\documentclass[runningheads]{llncs}

\usepackage{makeidx}  
\usepackage{graphicx}
\usepackage{subfigure}
\usepackage{multirow}
\usepackage{color,soul} 
\usepackage{ dsfont }
\usepackage{amssymb,amsfonts}
\usepackage{physics} 
\usepackage{cases}
\usepackage{hyperref}

\usepackage{graphicx}

\usepackage{amsmath}

\usepackage{subfigure}
\usepackage{algorithm}
\usepackage{algorithmic}

\newcommand{\mR}{{\mathbb R}}

\newcommand{\be}{\begin{equation}}
\newcommand{\ee}{\end{equation}}

\makeatletter
 \def\SOUL@hlpreamble{%
 \setul{}{2.4ex}
 \let\SOUL@stcolor\SOUL@hlcolor
 \SOUL@stpreamble
 }
\makeatother

\begin{document}
\mainmatter              

\title{Multimarginal Wasserstein Barycenter for Stain Normalization and Augmentation}
\titlerunning{Multimarginal Wasserstein Barycenter}  
%

\author{Saad Nadeem\inst{1}{\let\thefootnote\relax\footnote{{\hspace{-4mm} Email: nadeems@mskcc.org}}} \and Travis Hollmann\inst{2} \and Allen Tannenbaum\inst{3}}

\authorrunning{Nadeem \emph{et al}.} 

\tocauthor{Saad Nadeem, Travis Hollmann, Joseph O. Deasy and Allen Tannenbaum}

\institute{Department of Medical Physics, Memorial Sloan Kettering Cancer Center\\
\and
Department of Pathology, Memorial Sloan Kettering Cancer Center\\
\and
Departments of Computer Science \& Applied Mathematics, Stony Brook University\\
}

\maketitle              

\begin{abstract}
Variations in hematoxylin and eosin (H\&E) stained images (due to clinical lab protocols, scanners, etc) directly impact the quality and accuracy of clinical diagnosis, and hence it is important to control for these variations for a reliable diagnosis. In this work, we present a new approach based on the multimarginal Wasserstein barycenter to normalize and augment H\&E stained images given one or more references. Specifically, we provide a mathematically robust way of naturally incorporating additional images as intermediate references to drive stain normalization and augmentation simultaneously. The presented approach showed superior results quantitatively and qualitatively as compared to state-of-the-art methods for stain normalization. We further validated our stain normalization and augmentations in the nuclei segmentation task on a publicly available dataset, achieving state-of-the-art results against competing approaches.

\end{abstract}
\section{Introduction}
Histology is founded on the study of microscopic images to diagnose cell structures and arrangements for which staining is a critical part of the tissue preparation process. In particular, certain staining agents (mainly hematoxylin and eosin) transform the transparent tissue samples to become more distinguishable. Hematoxylin dyes the nuclei a dark purple color while eosin dyes other structures a pink color. A major problem is that results from the staining are inconsistent and prone to variability due to many factors including differing lab protocols, inter-patient variabilities, differences in raw materials, and variations in the slide scanners. These inconsistencies may cause major problems not only for pathologists, but may also degrade the performance of computer-aided diagnosis systems.

\begin{figure}[t!]
\begin{center}
\setlength{\tabcolsep}{1pt}
\tiny
\begin{tabular}{cccccc}
source & $t=0.4$ & $t=0.6$ & $t=0.8$ & $t=1$ & target\\
{\fboxsep=0mm
\fboxrule=1.5pt
\fcolorbox{blue}{white}{\includegraphics[width=0.16\textwidth]{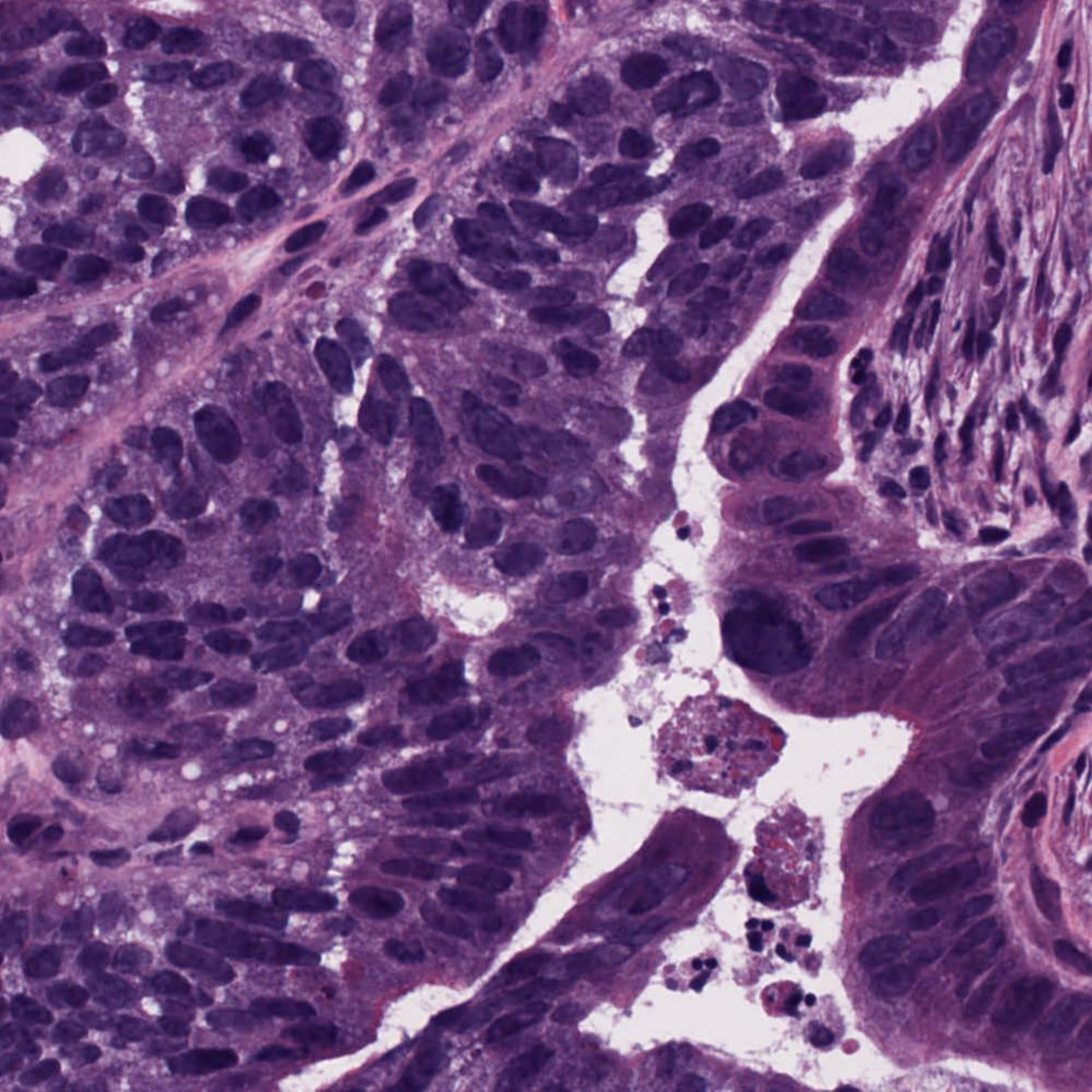}}}&
\includegraphics[width=0.16\textwidth]{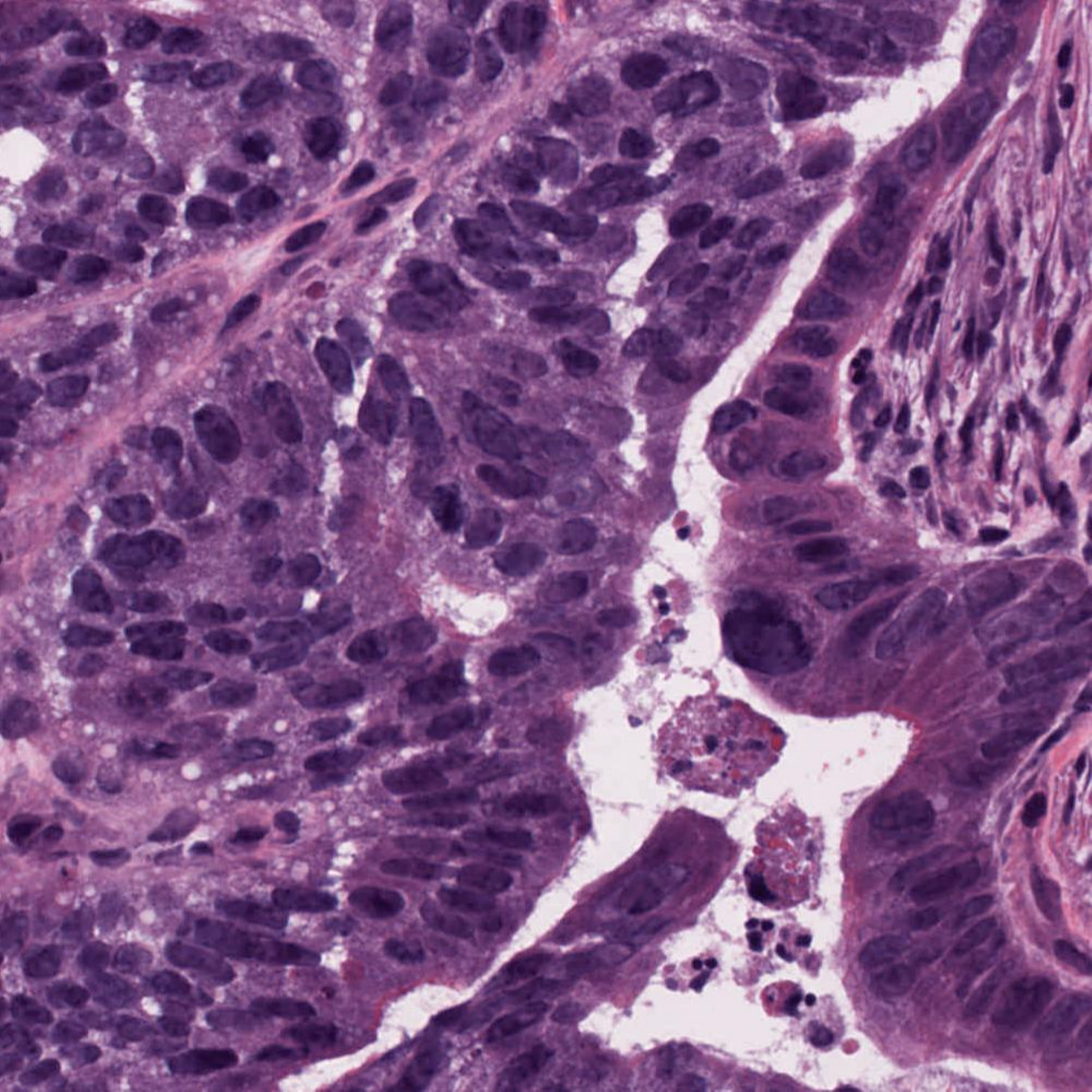}&
\includegraphics[width=0.16\textwidth]{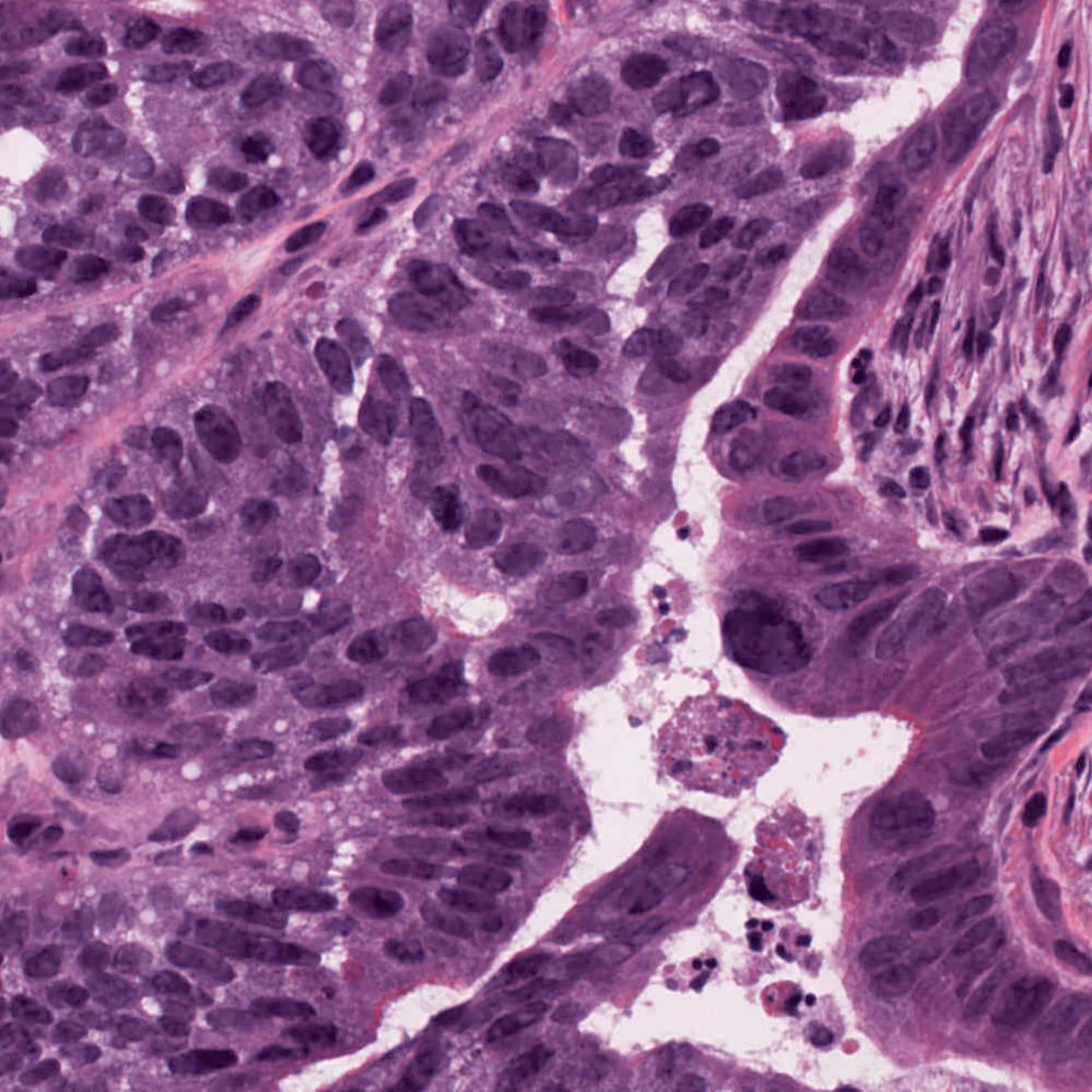}&
\includegraphics[width=0.16\textwidth]{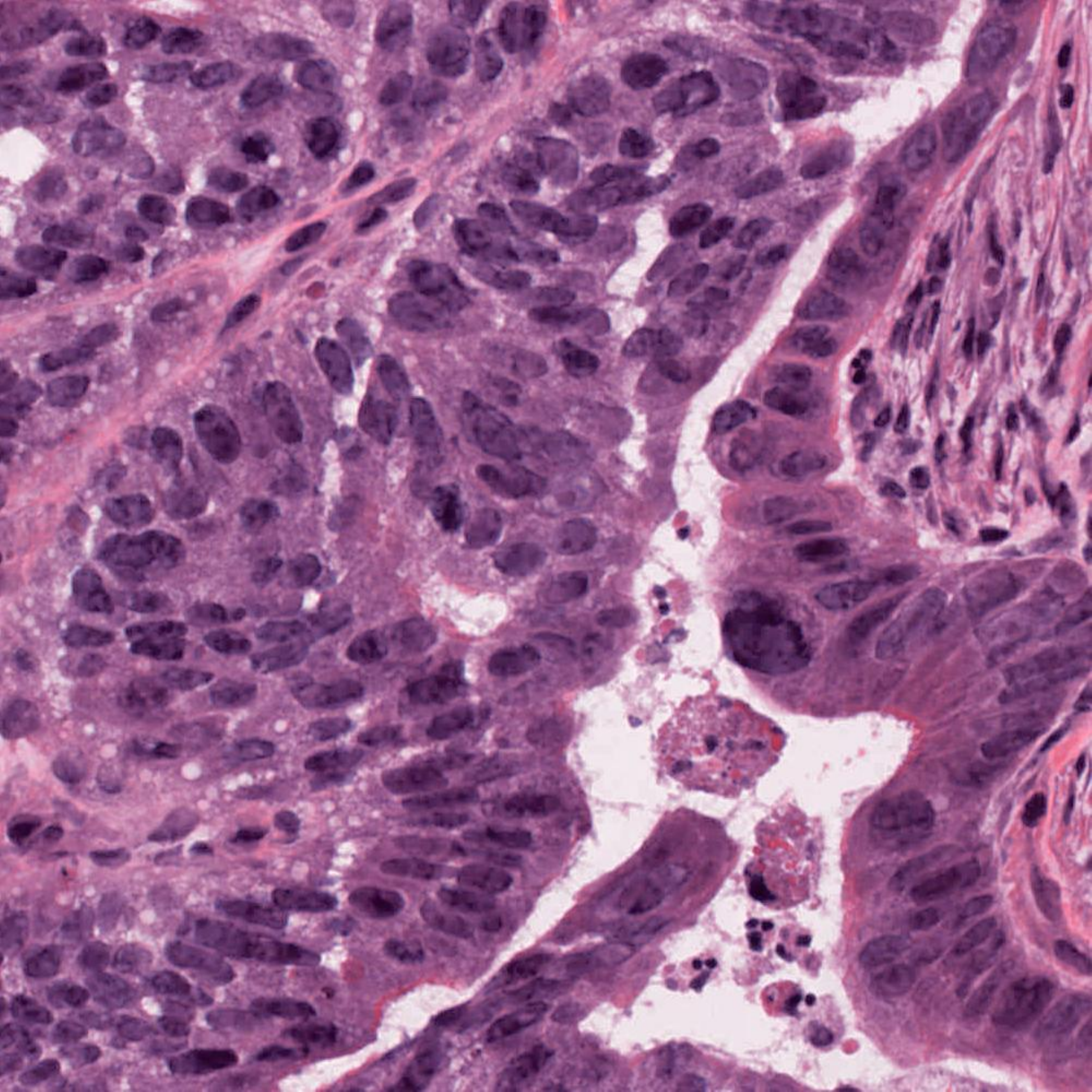}&
\includegraphics[width=0.16\textwidth]{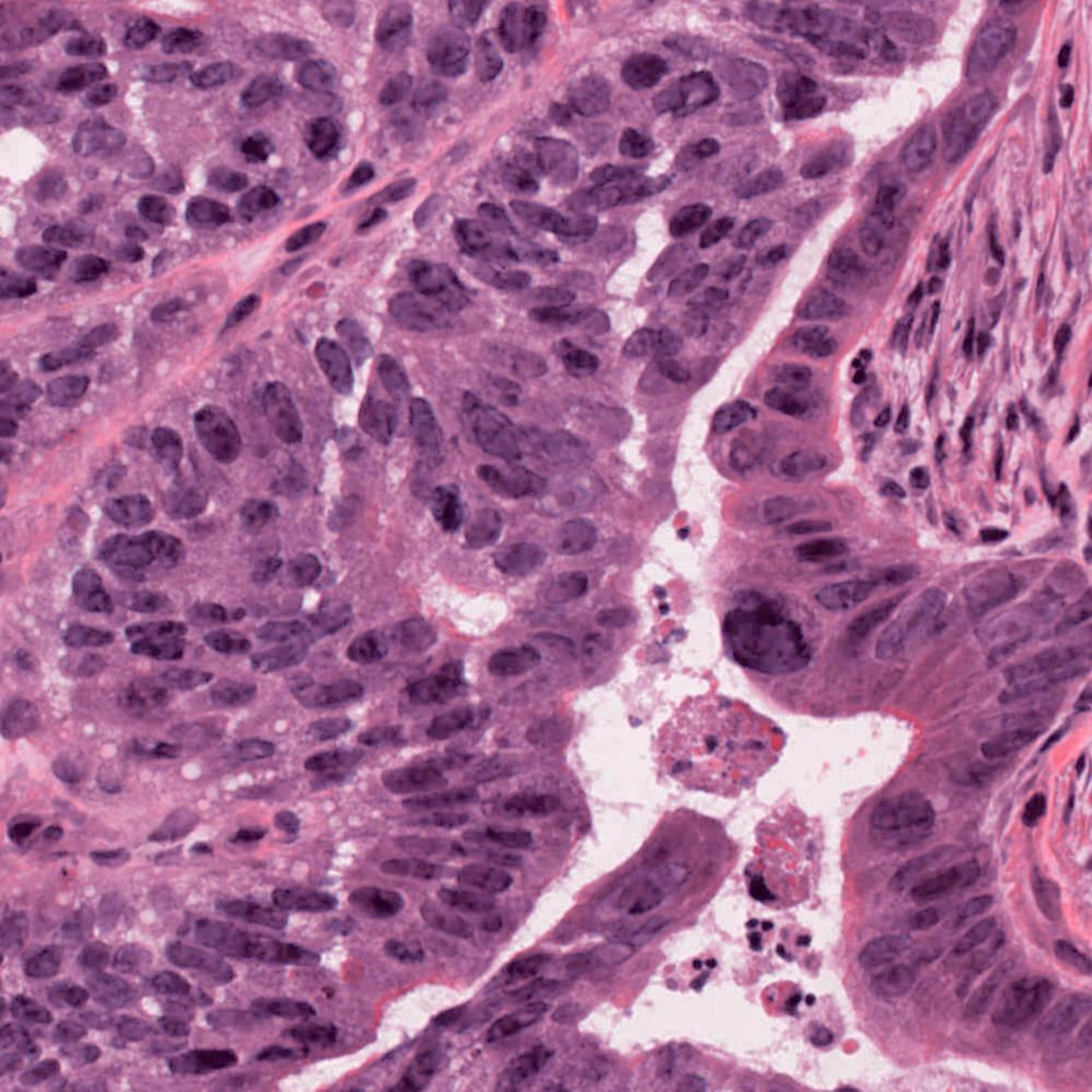}&
{\fboxsep=0mm
\fboxrule=1.5pt
\fcolorbox{red}{white}{\includegraphics[width=0.16\textwidth]{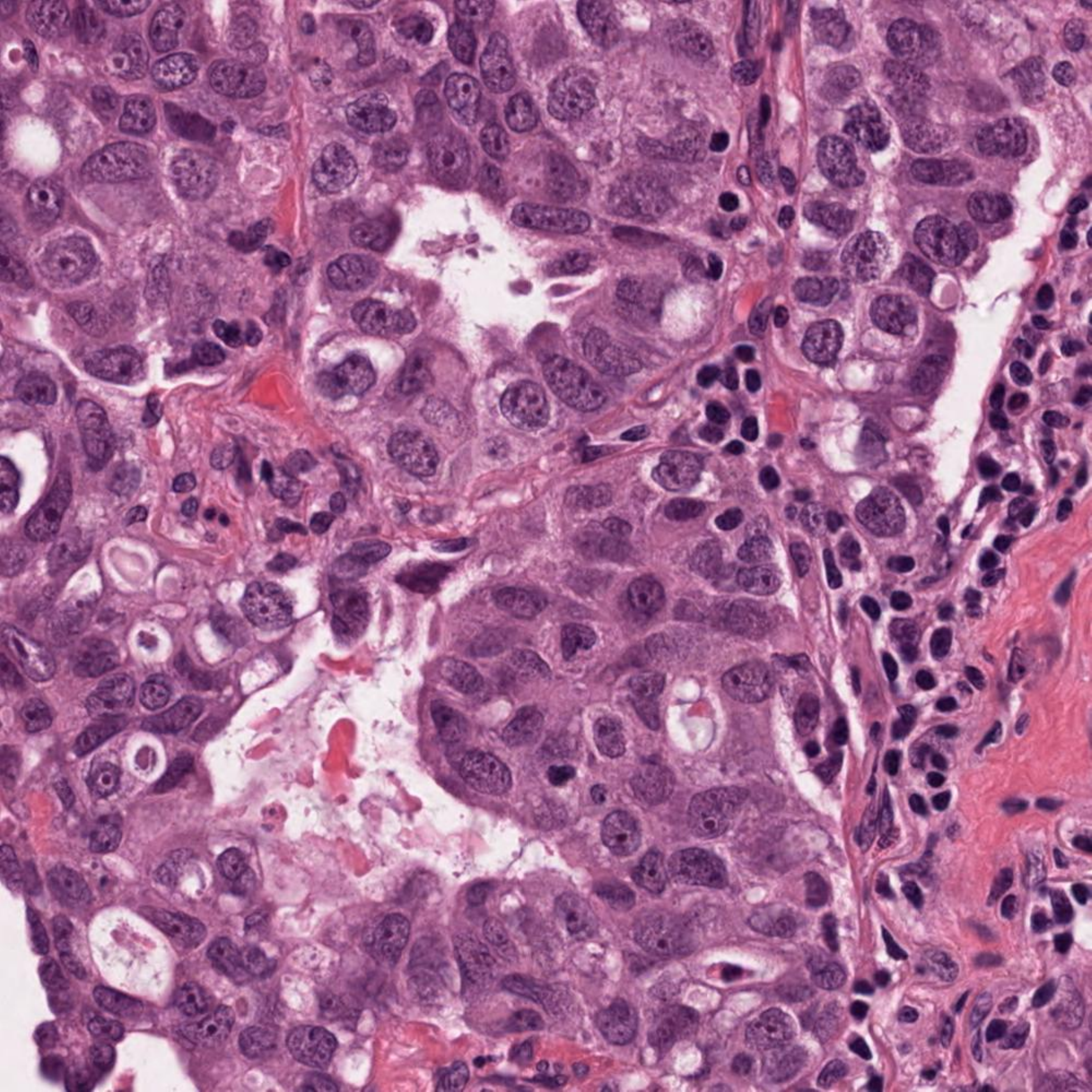}}}\\

{\fboxsep=0mm
\fboxrule=1.5pt
\fcolorbox{blue}{white}{\includegraphics[width=0.16\textwidth]{figures/MonuSeg_TCGA-38-6178-01Z-00-DX1-barycenter-unif-9.pdf}}}&
\includegraphics[width=0.16\textwidth]{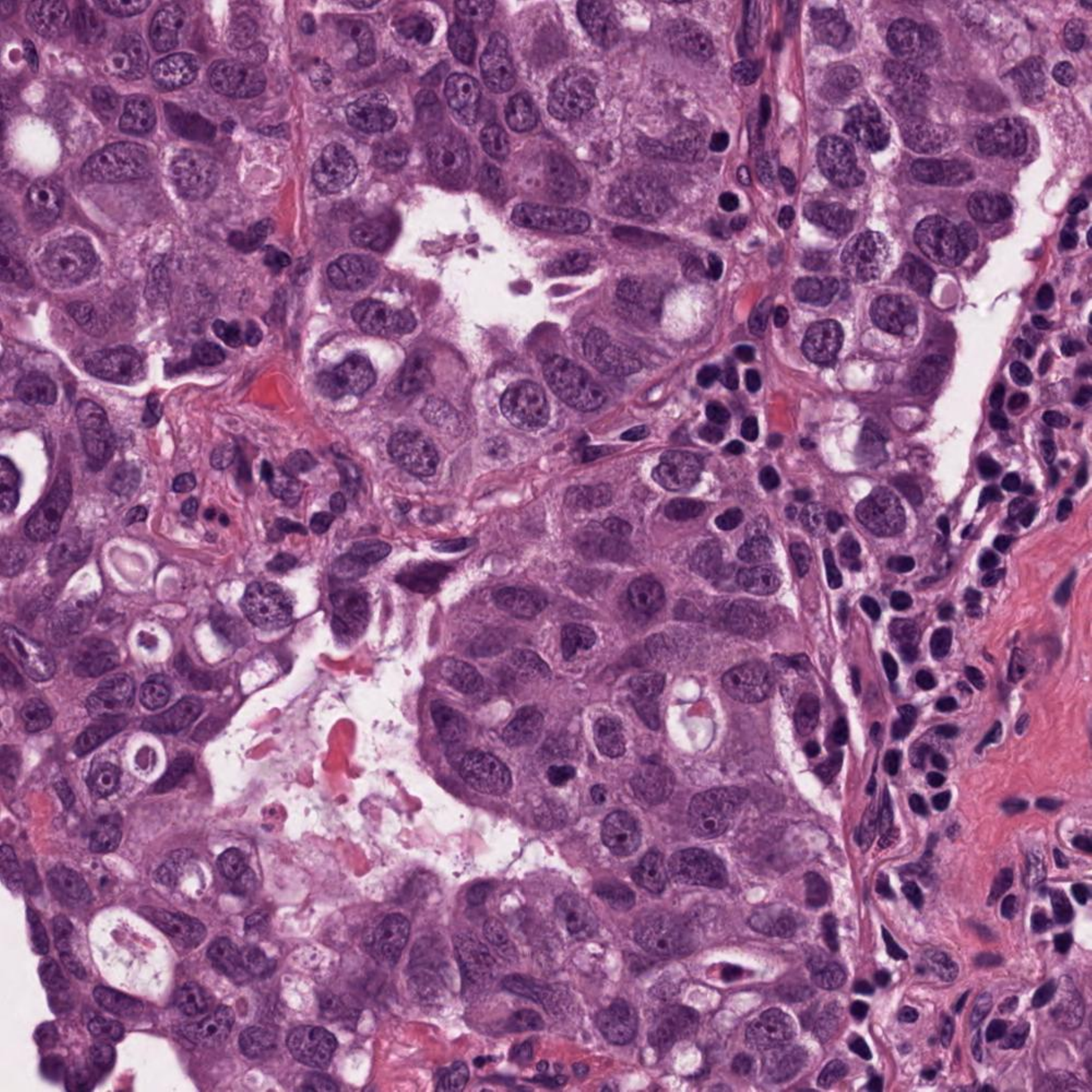}&
\includegraphics[width=0.16\textwidth]{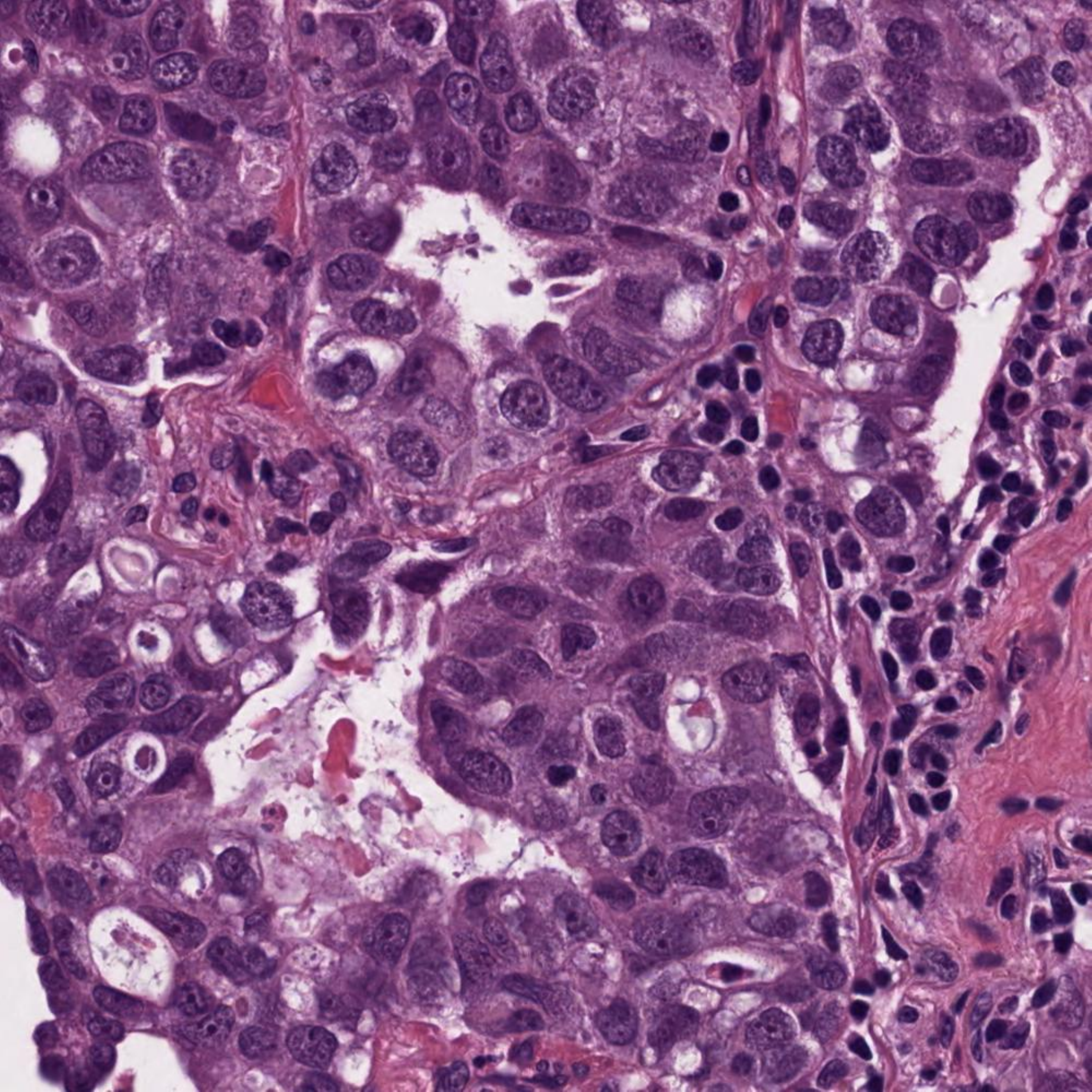}&
\includegraphics[width=0.16\textwidth]{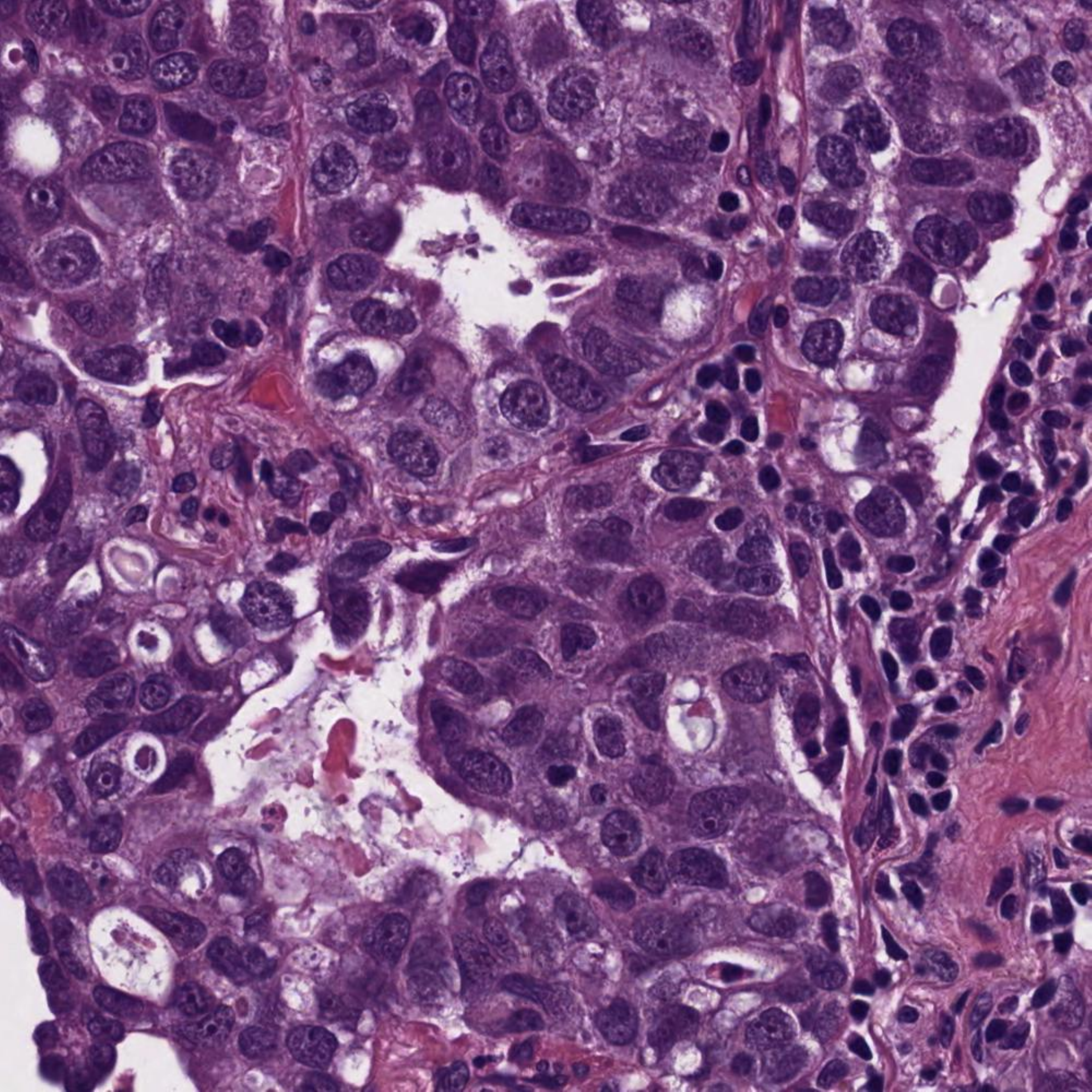}&
\includegraphics[width=0.16\textwidth]{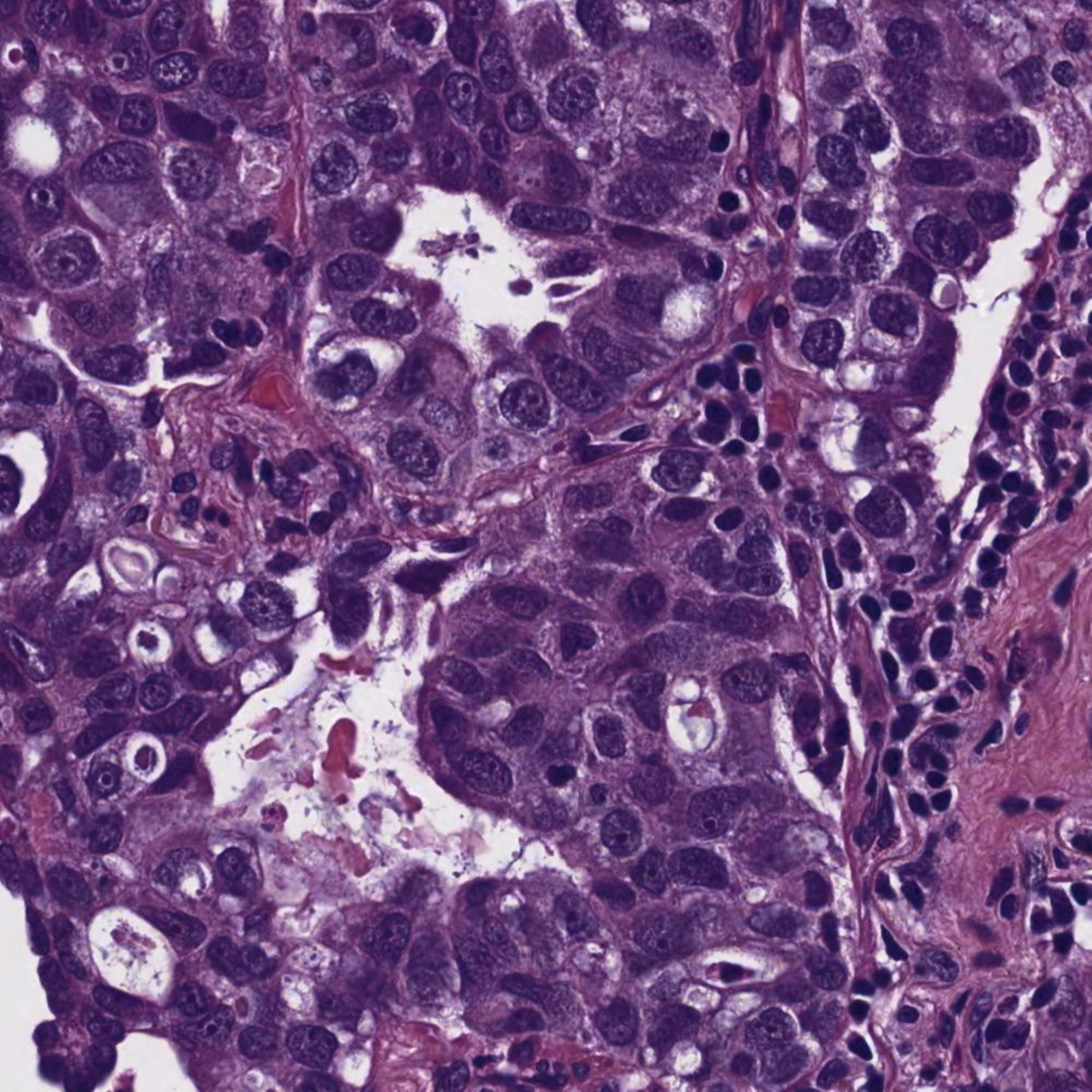}&
{\fboxsep=0mm
\fboxrule=1.5pt
\fcolorbox{red}{white}{\includegraphics[width=0.16\textwidth]{figures/MonuSeg_TCGA-NH-A8F7-01A-01-TS1-barycenter-unif-1.pdf}}}\\

\end{tabular}
\end{center}
\caption{Traditional ($N=2$) Wasserstein Barycenter. The source color distribution is interpolated towards the target color distribution in the Lab color space.}
\label{fig:traditional_barycenter}
\end{figure}

It is well-known that for downstream applications such as nuclei segmentation, stain normalization and augmentation are essential \cite{kumar2019multi,tellez2019quantifying}. In fact, augmentation might be more important than normalization, as pointed out by \cite{pontalba2019assessing}. Color jitter or random HSV shifts was identified as an important step for the top contenders of the MonuSeg competition \cite{kumar2017dataset}. Due to the coming telemedicine revolution, there will be a strong requirement for either matching to a reference distribution or incorporating stain invariant approaches using stain augmentation or training on diverse datasets.

Tellez \emph{et al.} \cite{tellez2019quantifying} showed that for certain specific tasks, stain augmentation (adding color jitters in HSV or HED color space), with and without normalization, may also improve performance while making the resultant models more robust; nuclei segmentation was not explored as one of the tasks in the latter paper. Further as shown in \cite{pontalba2019assessing}, deep learning stain normalization/transfer approaches, such as StainGAN \cite{shaban2019staingan}, are not particularly suitable for nuclei segmentation task given the lack of training data representing enough samples from same tissue, institution, preparation protocols, etc. Stain invariant models learnt via novel stain augmentation approaches can help achieve better performance. The work of Vahadane \emph{et al.} \cite{vahadane2016structure} has been the predominant choice in the past for stain normalization in the nuclei segmentation tasks.

\begin{figure}[t!]
\begin{center}
\setlength{\tabcolsep}{1pt}
\tiny
\begin{tabular}{cccccc}
& \multicolumn{4}{c}{$ \textcolor{blue}{\xrightarrow{\hspace*{7cm}}} $}\\
{\fboxsep=0mm
\fboxrule=2pt
\fcolorbox{blue}{white}{\includegraphics[width=0.16\textwidth]{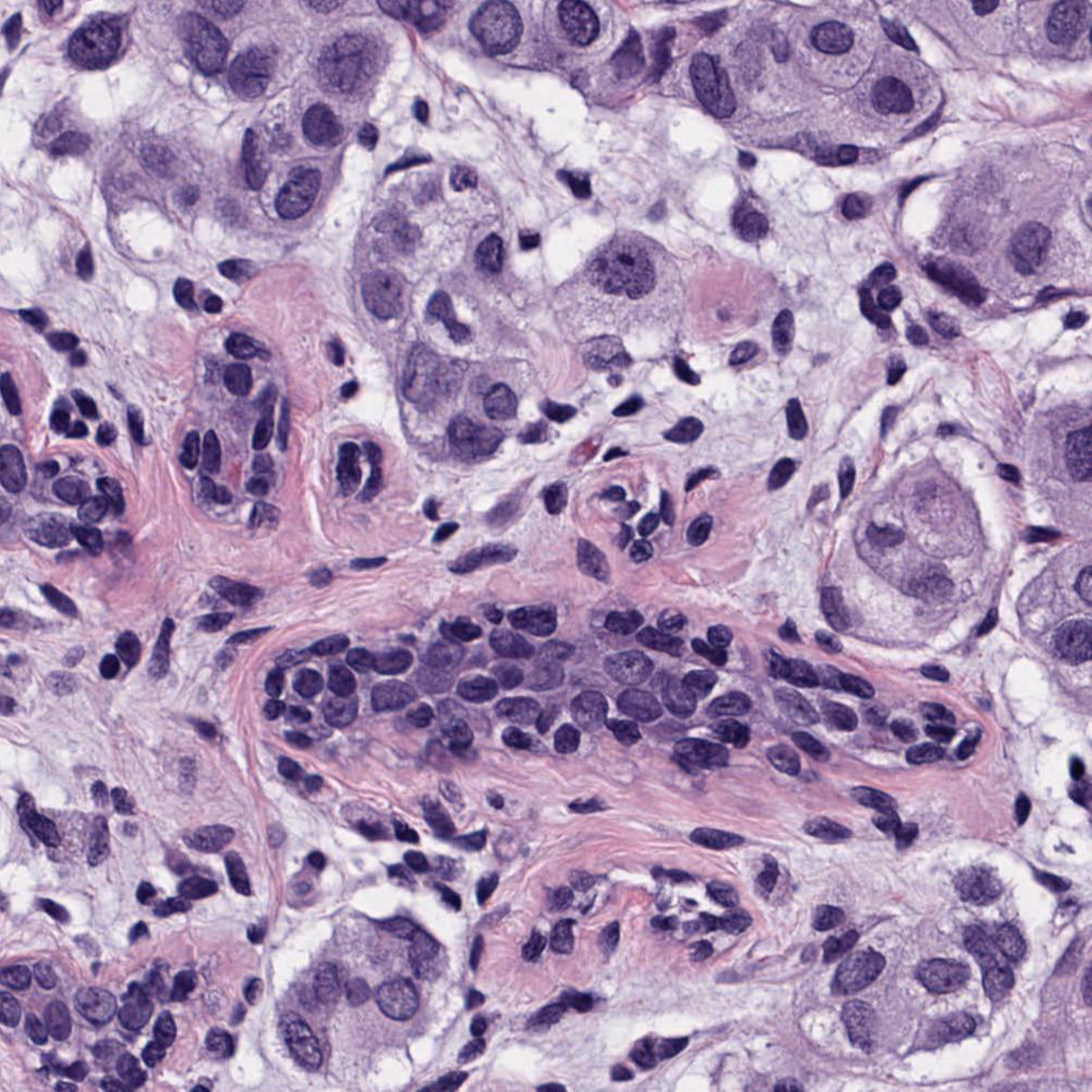}}}&
\includegraphics[width=0.16\textwidth]{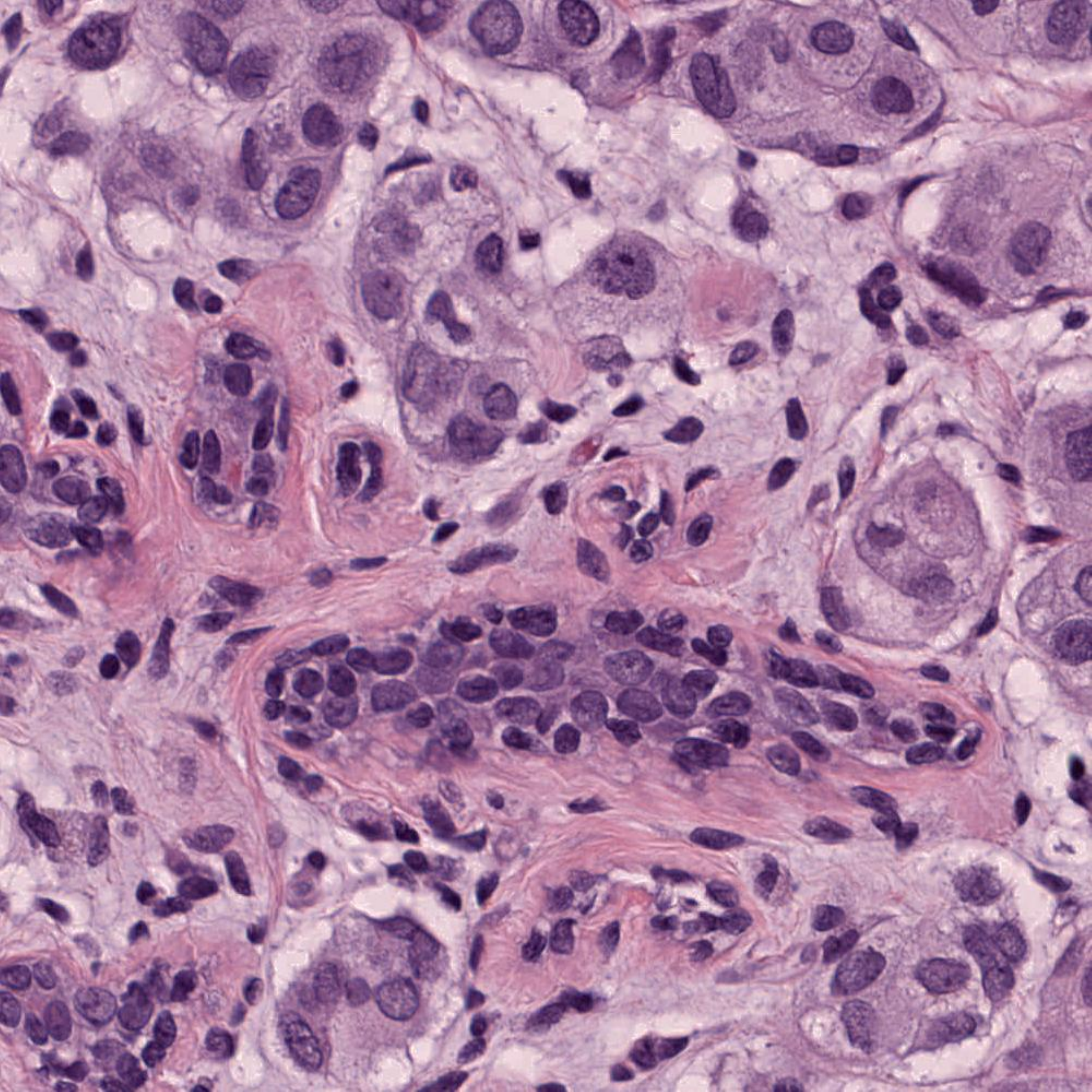}&
\includegraphics[width=0.16\textwidth]{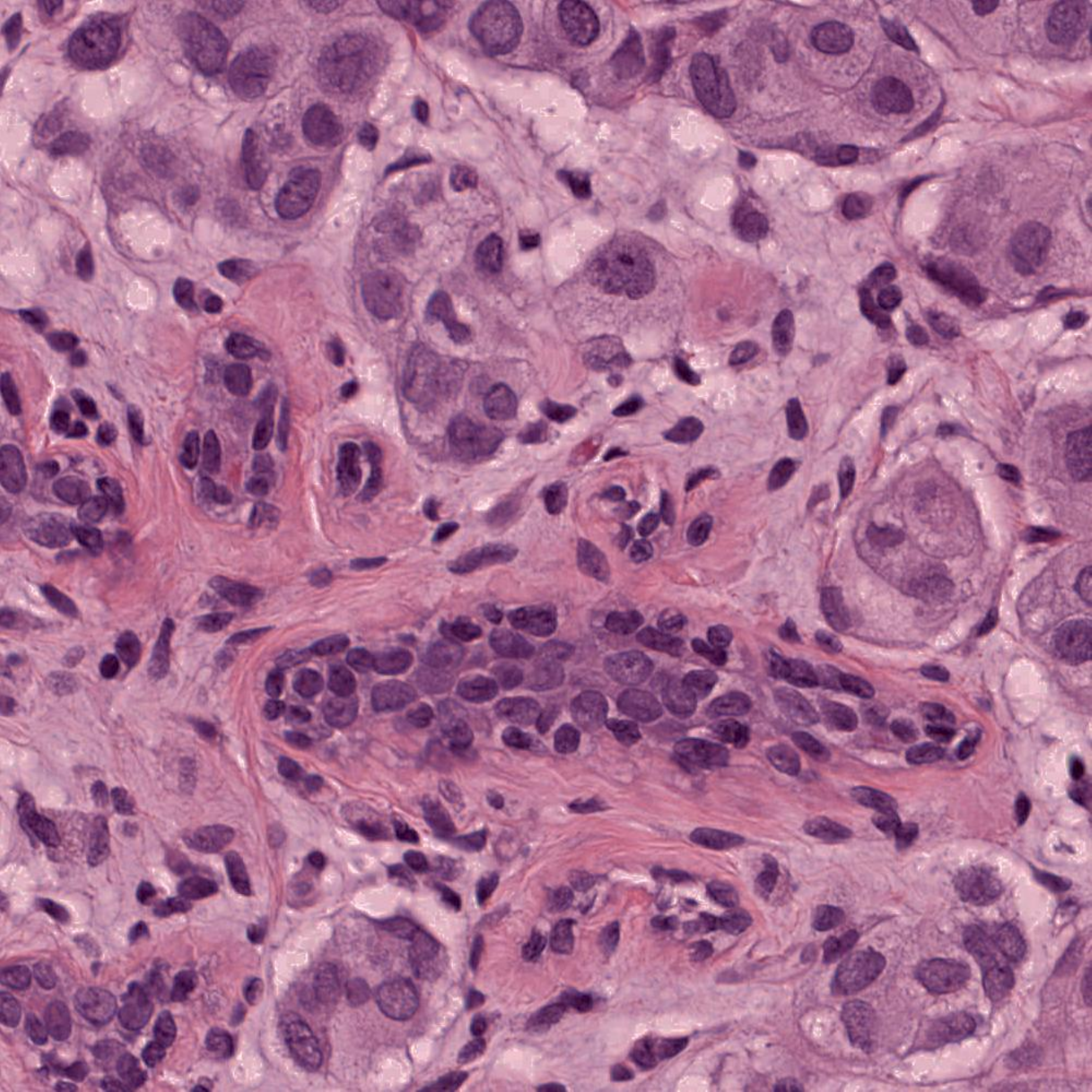}&
\includegraphics[width=0.16\textwidth]{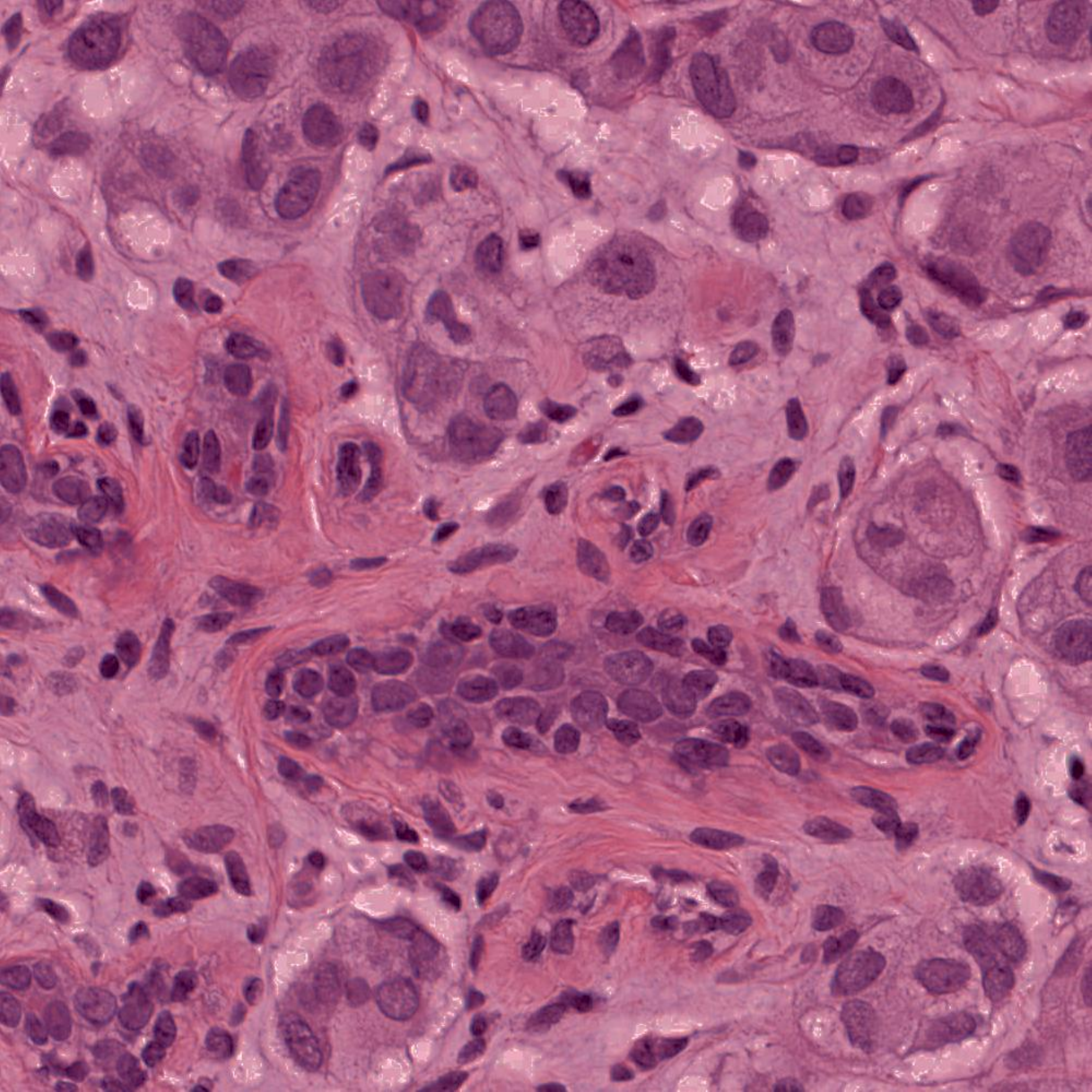}&
\includegraphics[width=0.16\textwidth]{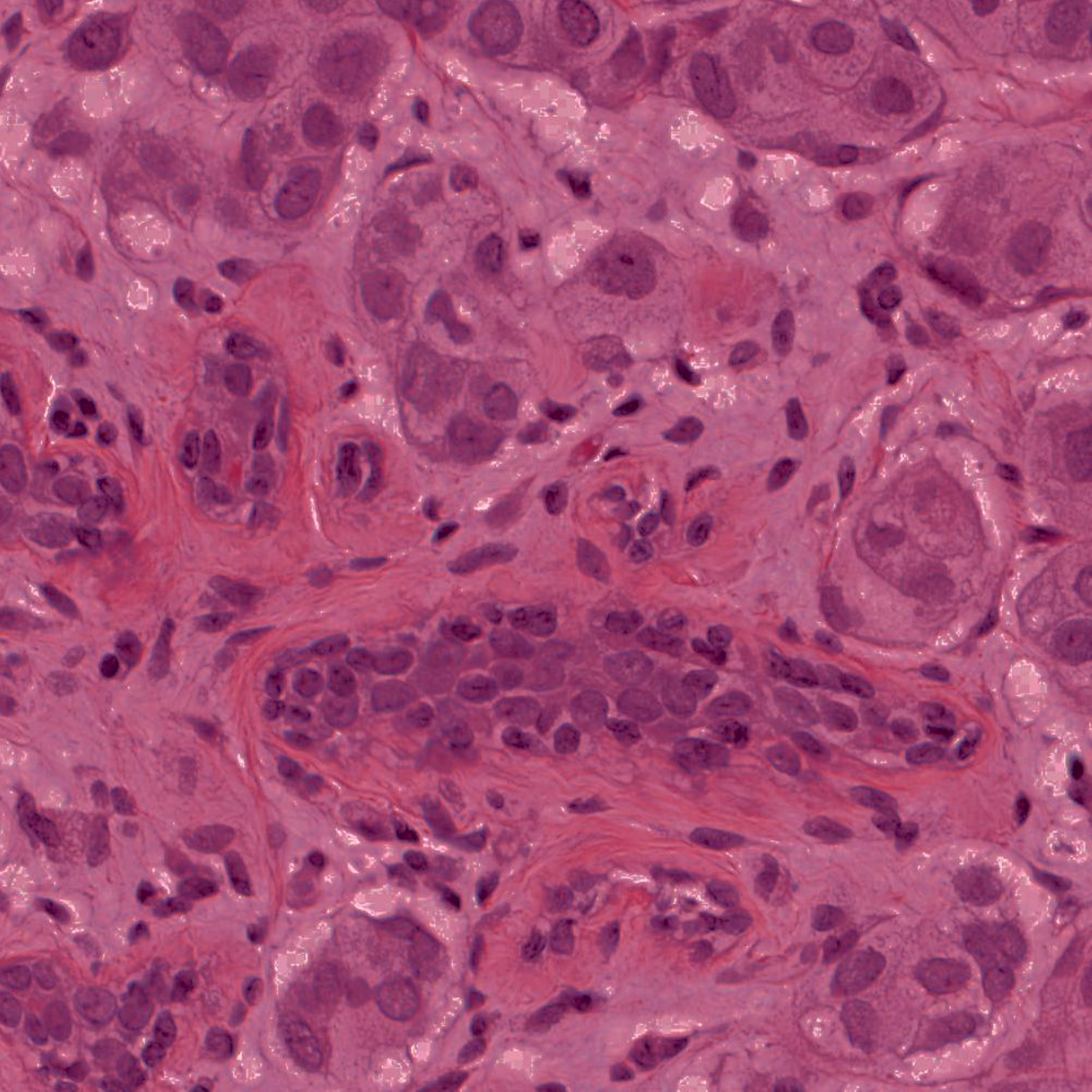}&
{\fboxsep=0mm
\fboxrule=2pt
\fcolorbox{red}{white}{\includegraphics[width=0.16\textwidth]{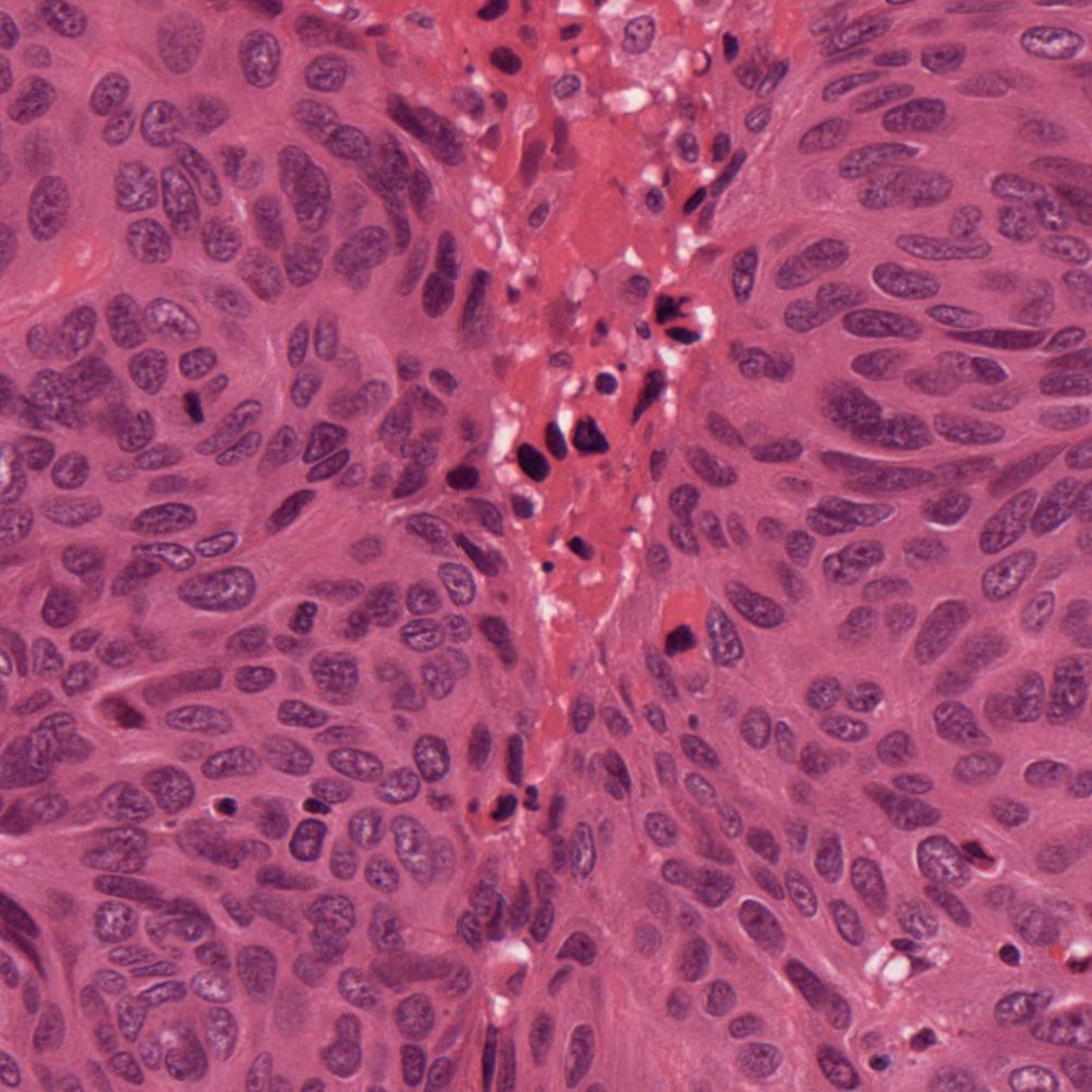}}}\\
\end{tabular}

\begin{tabular}{cccccc}
\hline
\hline
& \multicolumn{4}{c}{$ \textcolor{blue}{\xrightarrow{\hspace*{7cm}}} $} & \\
{\fboxsep=0mm
\fboxrule=2pt
\fcolorbox{blue}{white}{\includegraphics[width=0.16\textwidth]{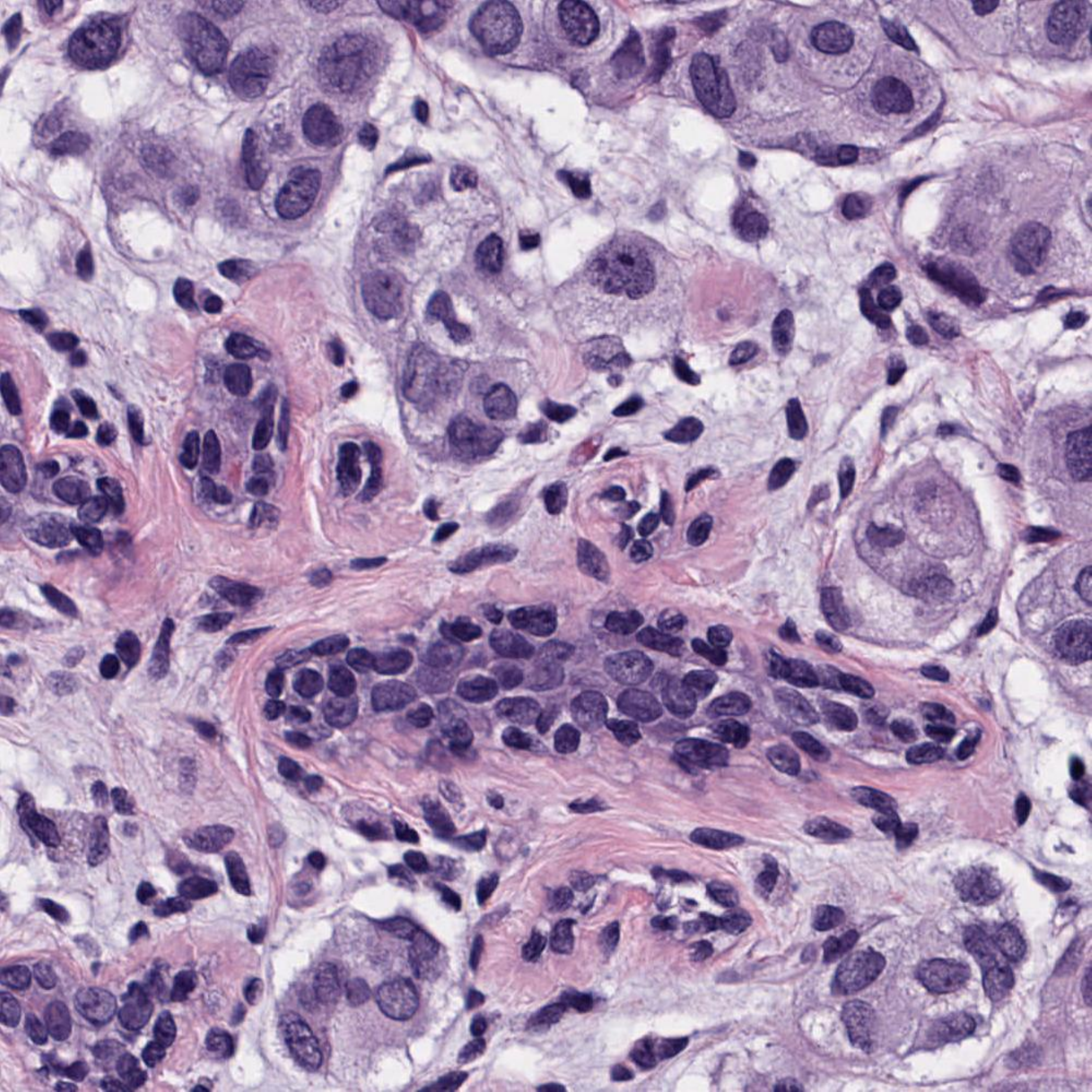}}}&
\includegraphics[width=0.16\textwidth]{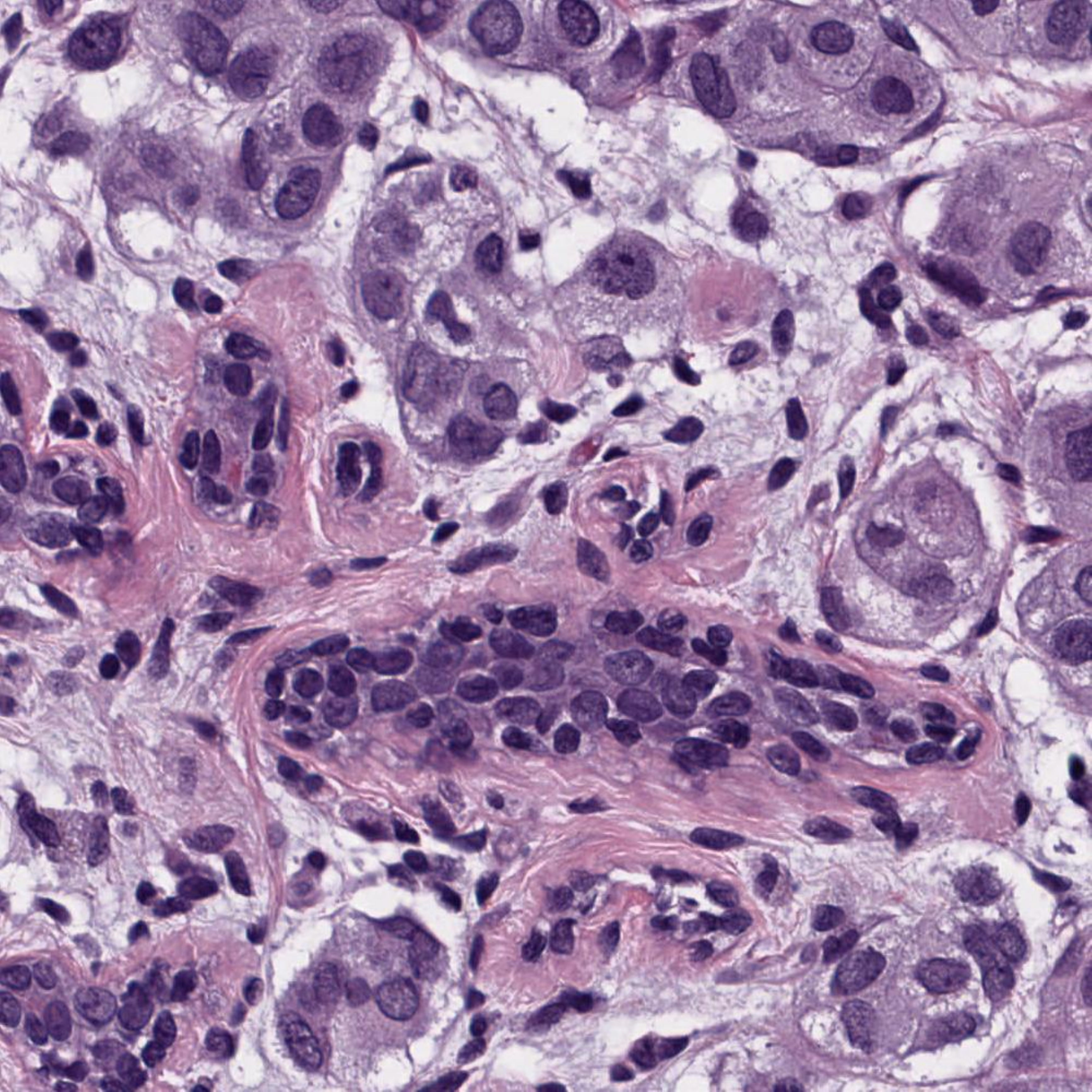}&
\includegraphics[width=0.16\textwidth]{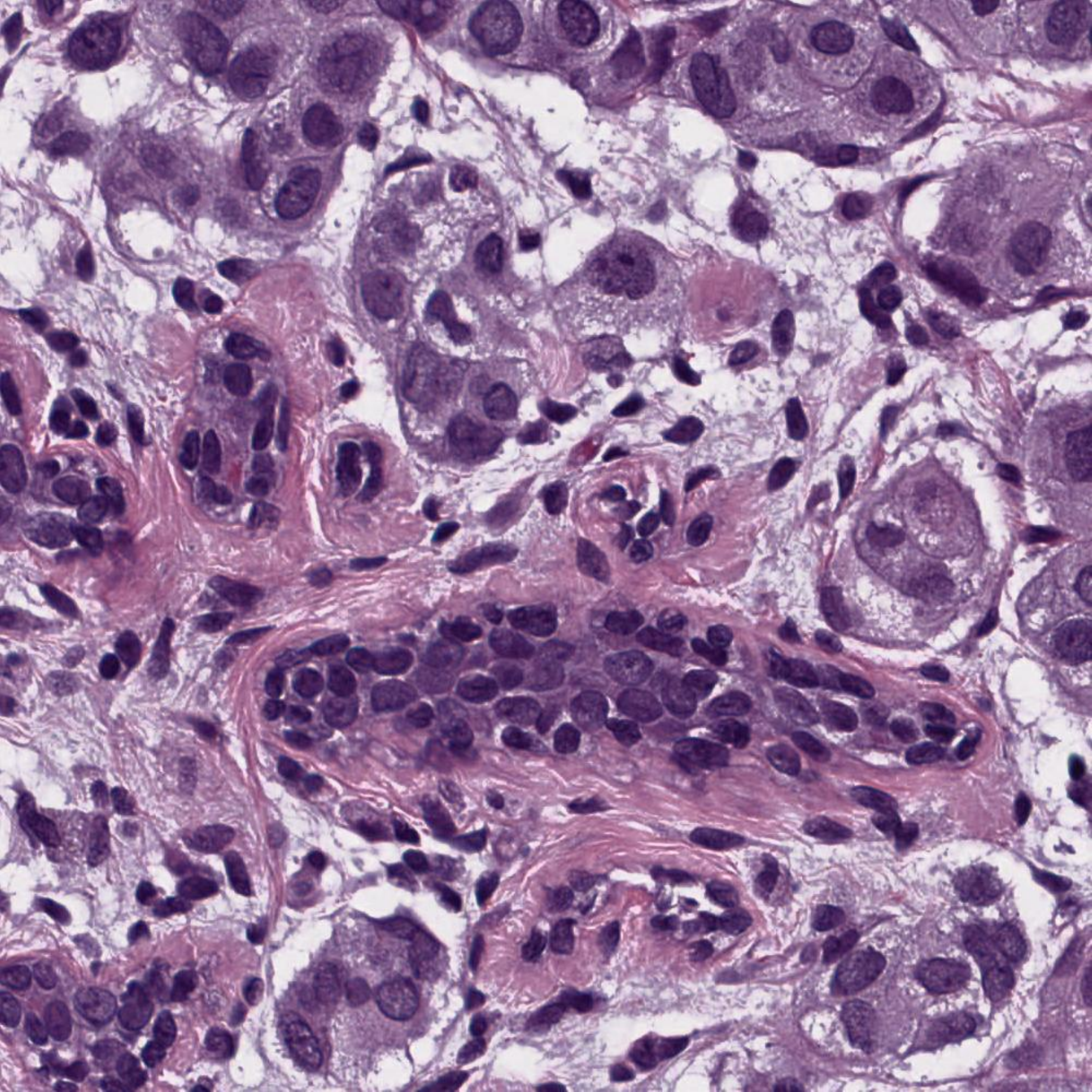}&
\includegraphics[width=0.16\textwidth]{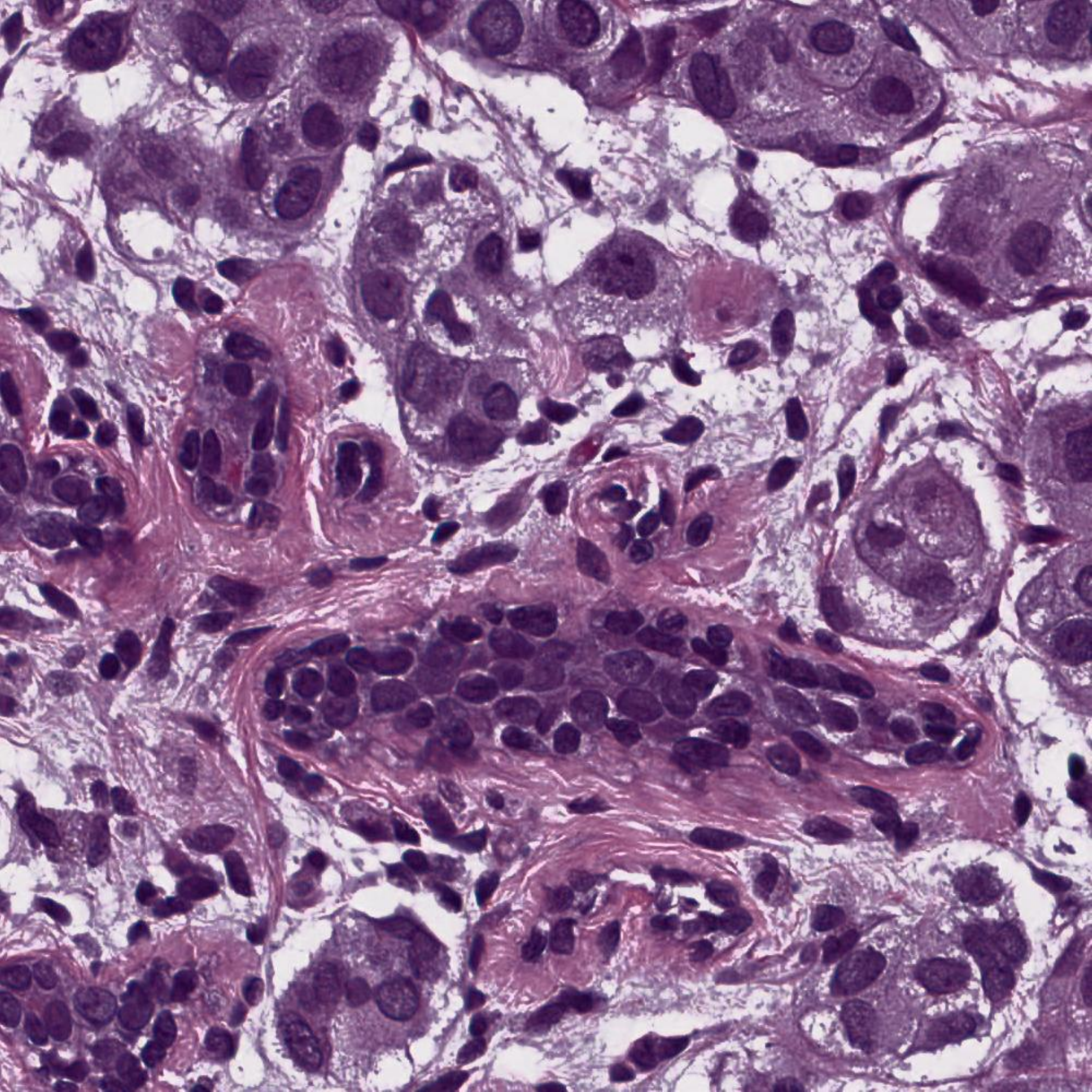}&
\includegraphics[width=0.16\textwidth]{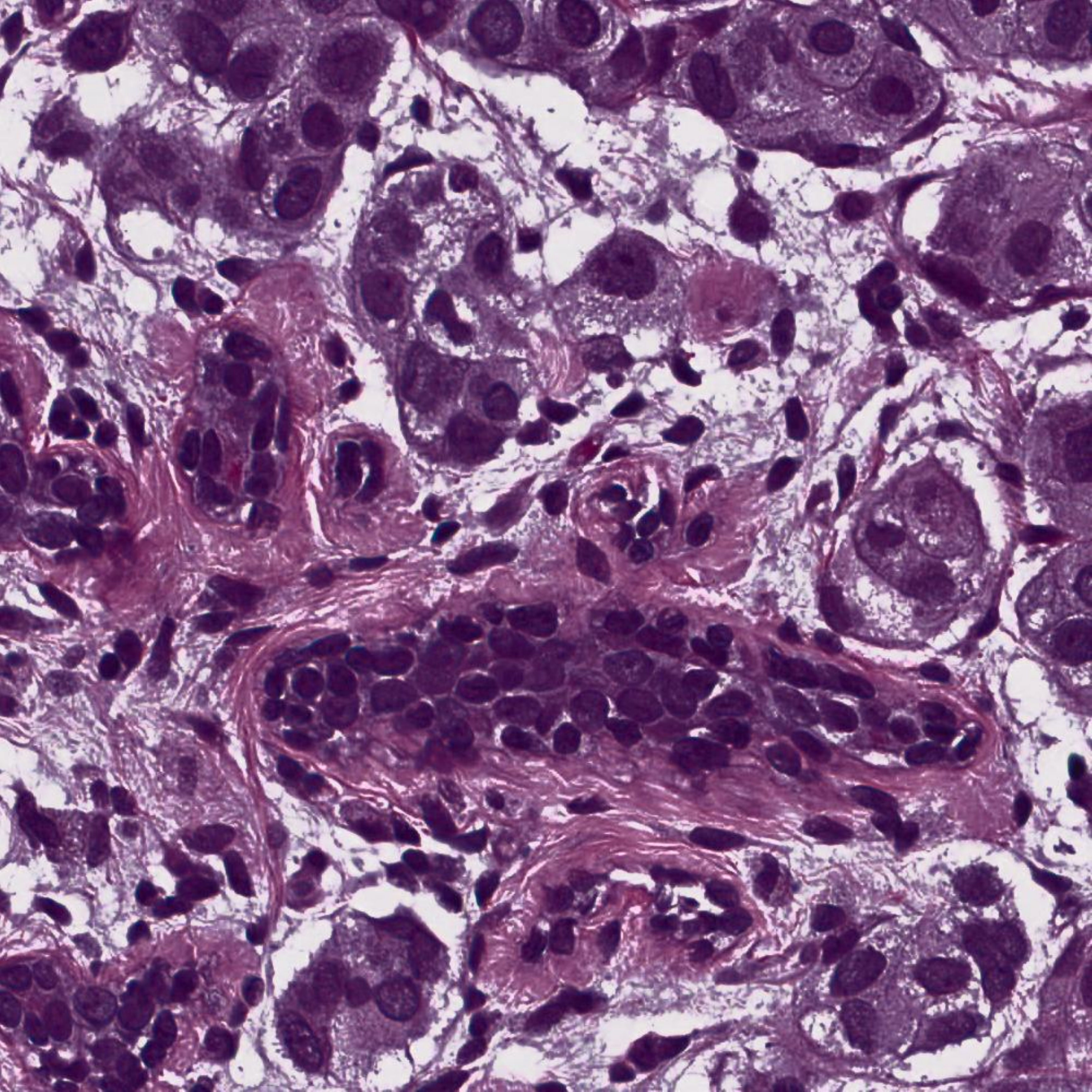}&
{\fboxsep=0mm
\fboxrule=1pt
\fcolorbox{green}{white}{\includegraphics[width=0.16\textwidth]{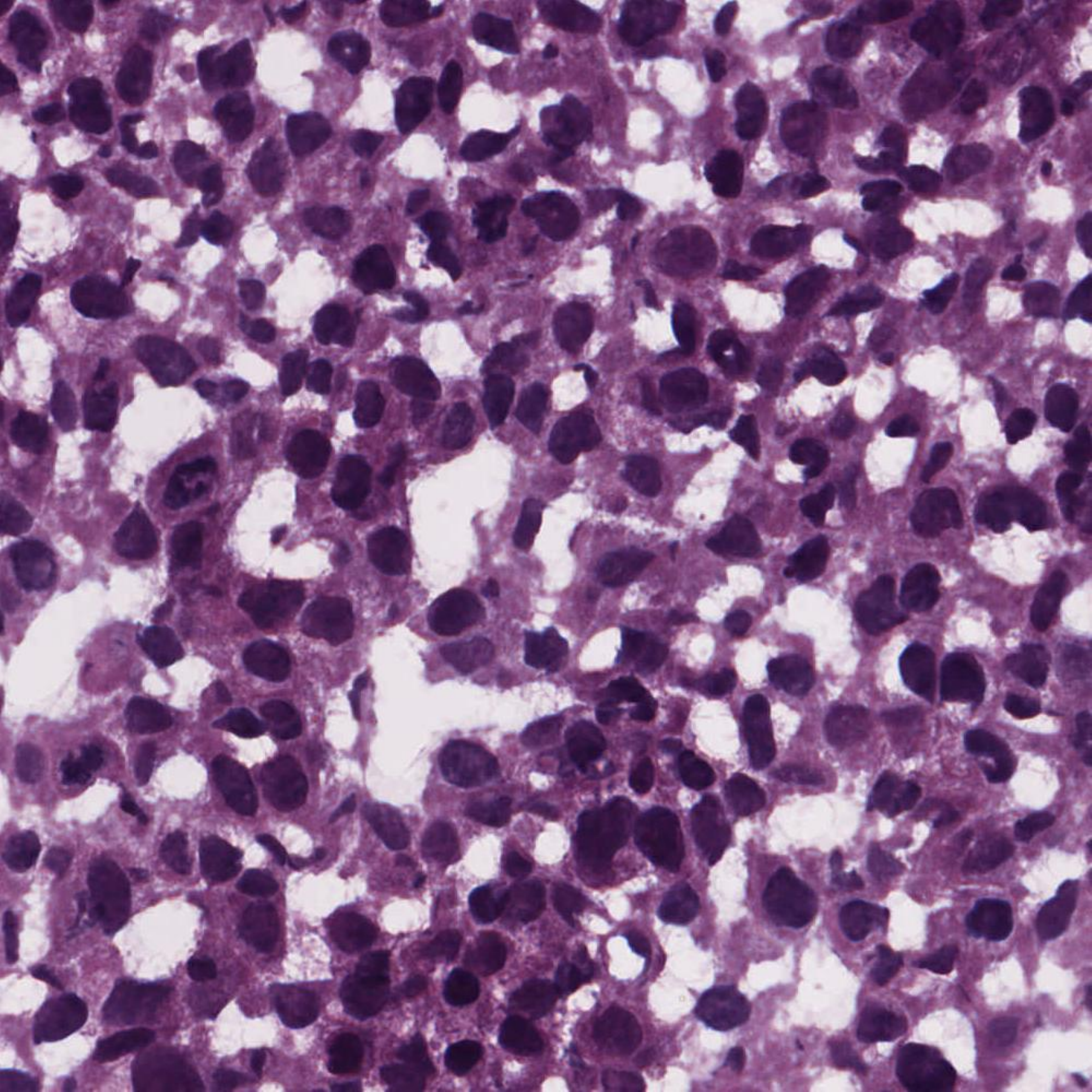}}}\\

\includegraphics[width=0.16\textwidth]{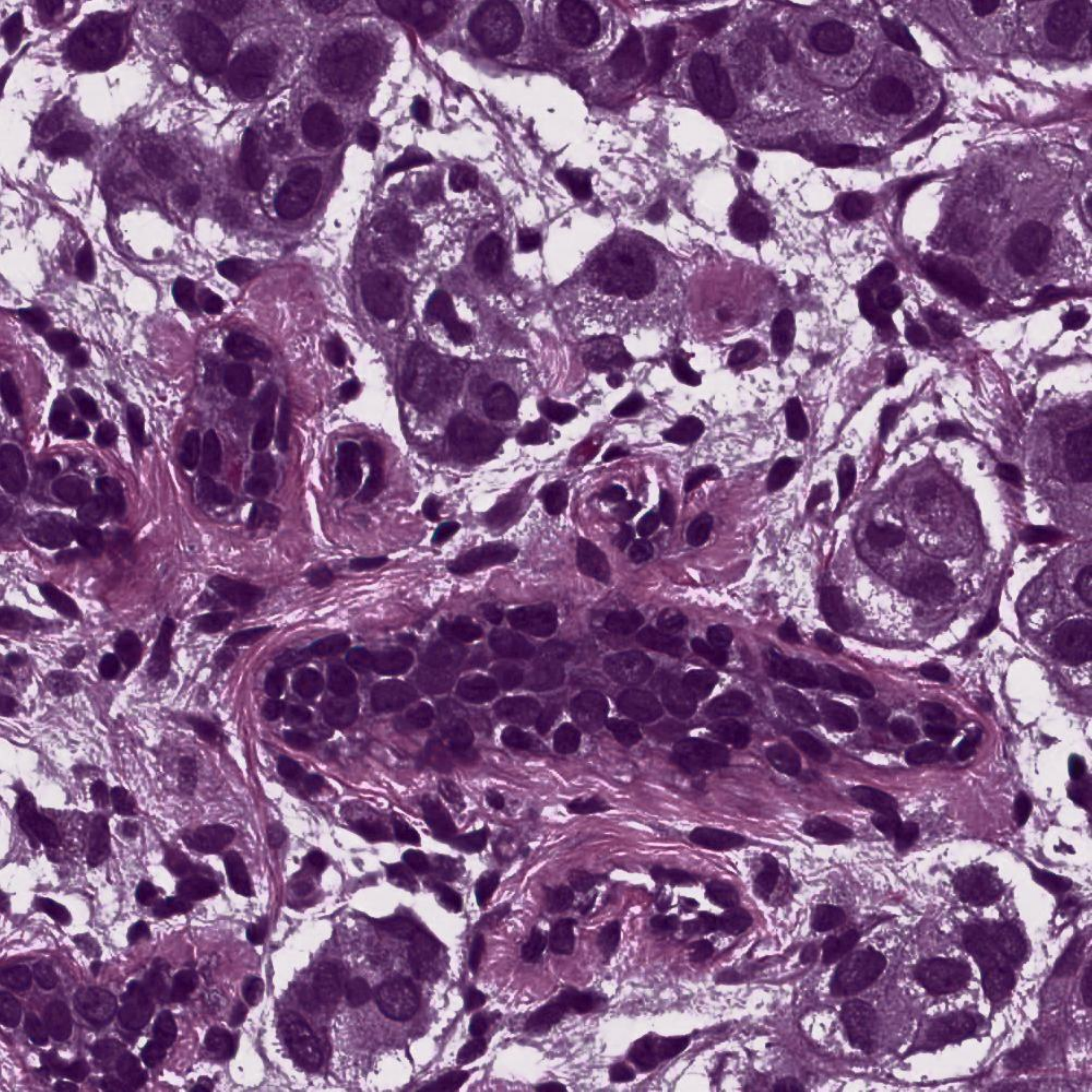}&
\includegraphics[width=0.16\textwidth]{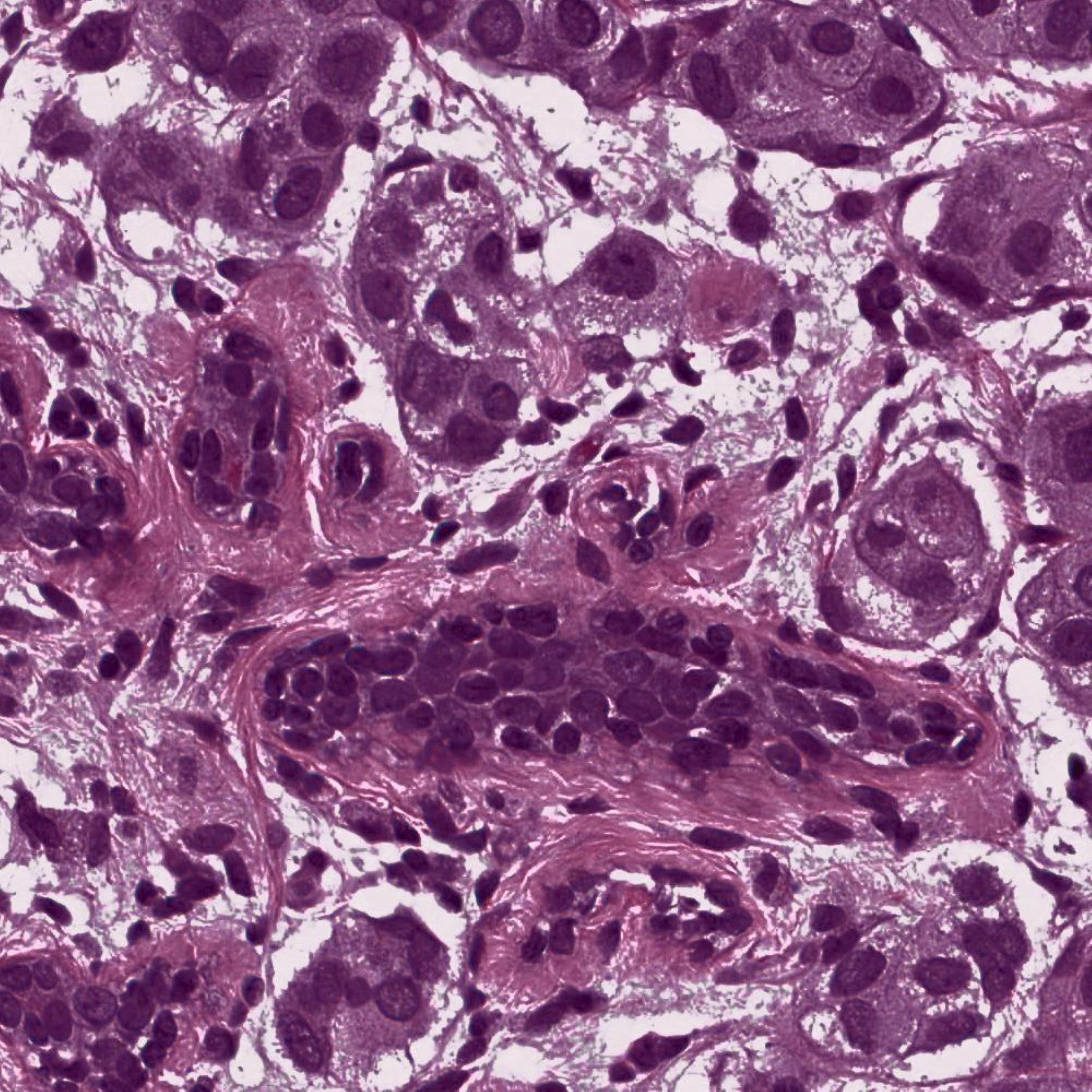}&
\includegraphics[width=0.16\textwidth]{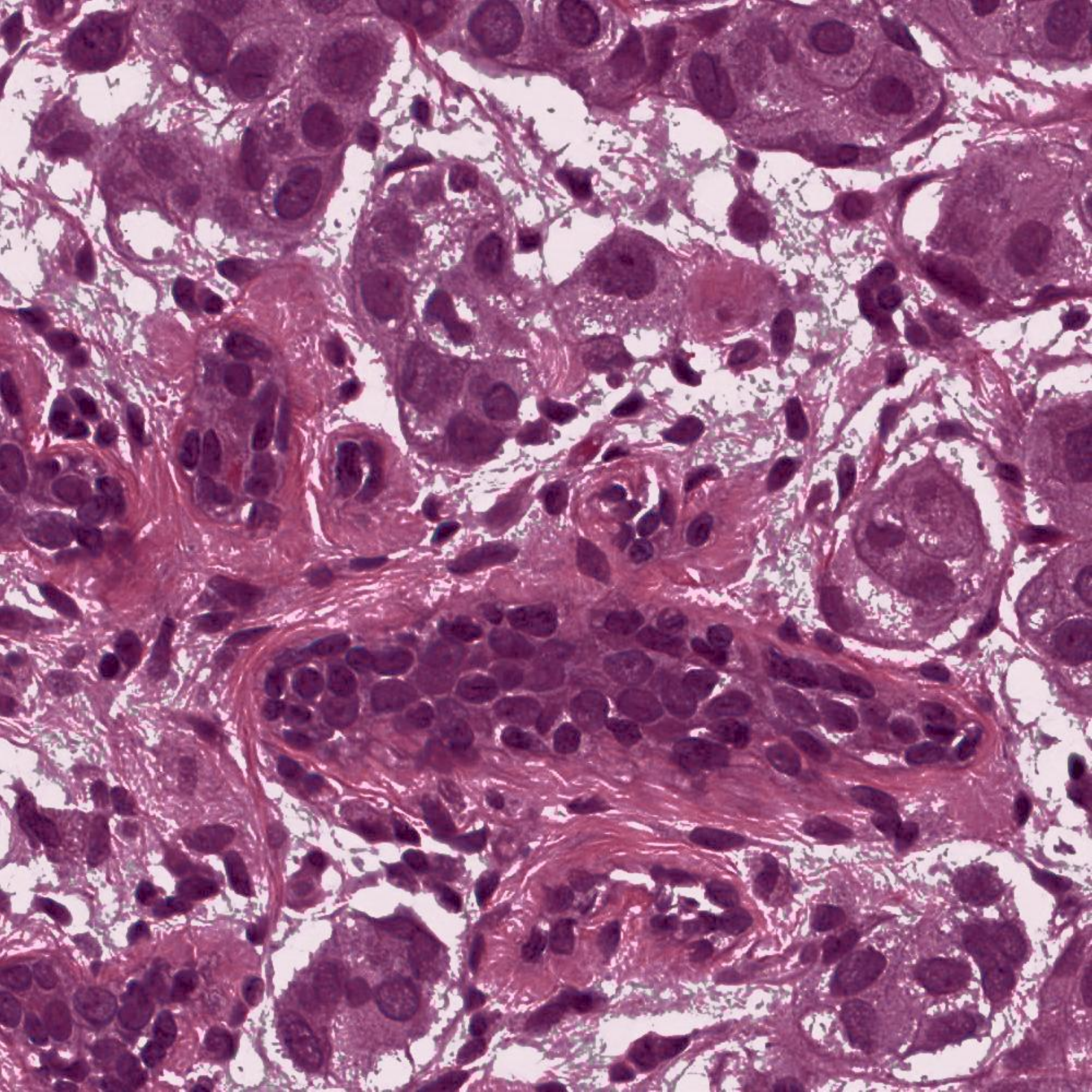}&
\includegraphics[width=0.16\textwidth]{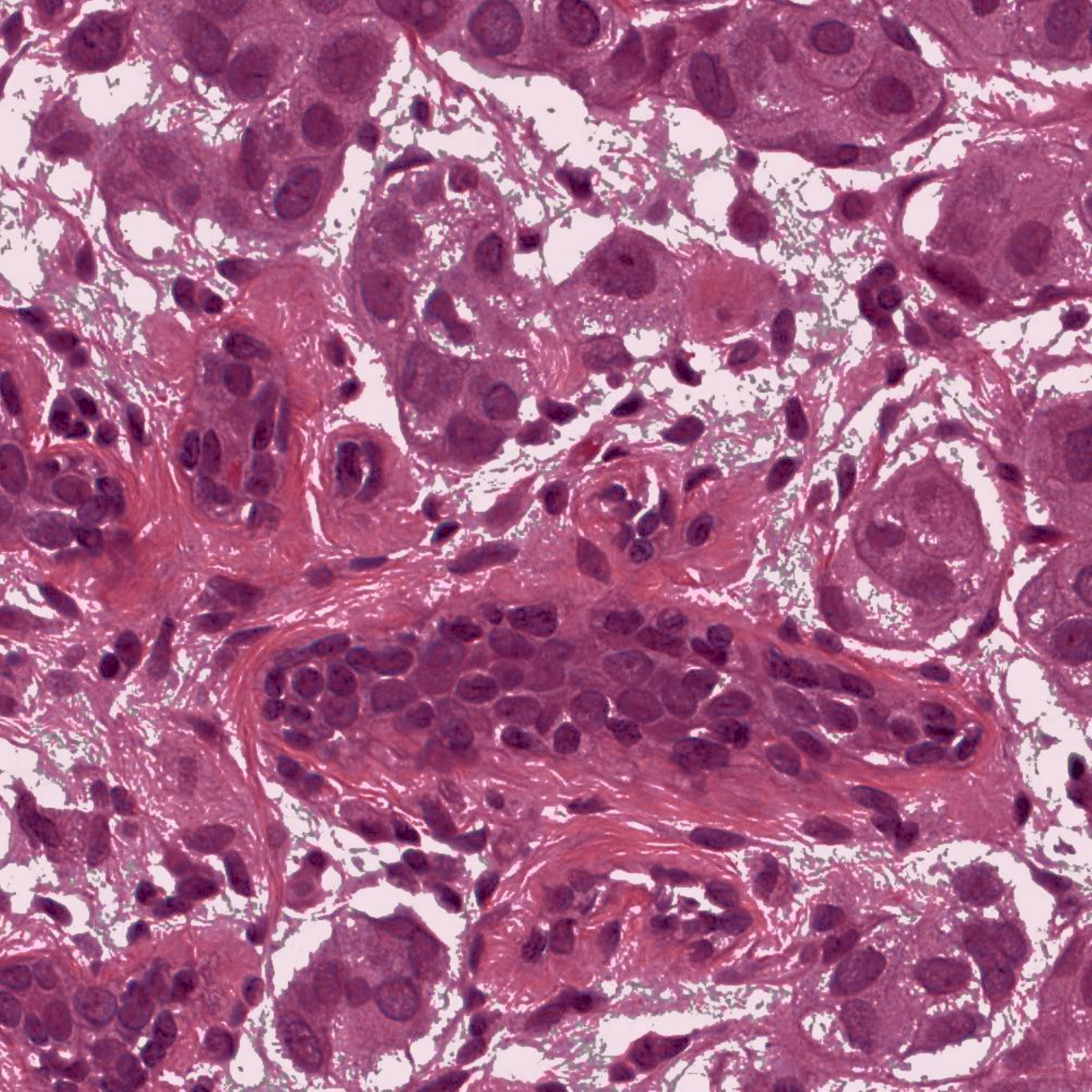}&
\includegraphics[width=0.16\textwidth]{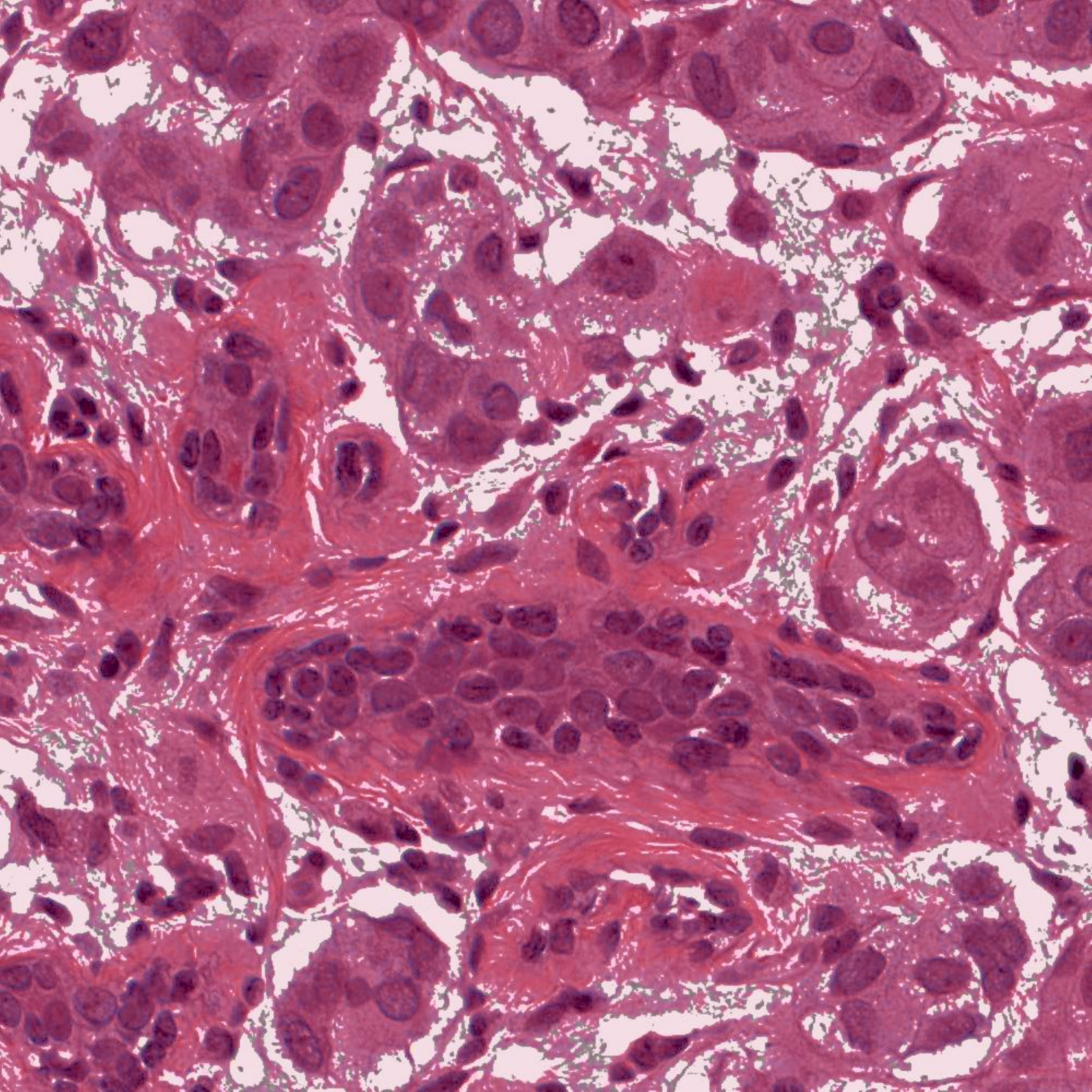}&
{\fboxsep=0mm
\fboxrule=1pt
\fcolorbox{red}{white}{\includegraphics[width=0.16\textwidth]{figures/MonuSeg_TCGA-18-5592-01Z-00-DX1-barycenter-unif-9.pdf}}}\\
\multicolumn{5}{c}{$ \textcolor{blue}{\xrightarrow{\hspace*{8.5cm}}} $} & \\

\end{tabular}

\begin{tabular}{cccccc}
\hline
\hline
& \multicolumn{4}{c}{$ \textcolor{blue}{\xrightarrow{\hspace*{7cm}}} $} & \\
{\fboxsep=0mm
\fboxrule=2pt
\fcolorbox{blue}{white}{\includegraphics[width=0.16\textwidth]{figures/MonuSeg_TCGA-A7-A13F-01Z-00-DX1-barycenter-unif-9.pdf}}}&
\includegraphics[width=0.16\textwidth]{figures/MonuSeg_TCGA-A7-A13F-01Z-00-DX1-barycenter-unif-7.pdf}&
\includegraphics[width=0.16\textwidth]{figures/MonuSeg_TCGA-A7-A13F-01Z-00-DX1-barycenter-unif-5.pdf}&
\includegraphics[width=0.16\textwidth]{figures/MonuSeg_TCGA-A7-A13F-01Z-00-DX1-barycenter-unif-3.pdf}&
\includegraphics[width=0.16\textwidth]{figures/MonuSeg_TCGA-A7-A13F-01Z-00-DX1-barycenter-unif-1.pdf}&
{\fboxsep=0mm
\fboxrule=1pt
\fcolorbox{green}{white}{\includegraphics[width=0.16\textwidth]{figures/MonuSeg_TCGA-DK-A2I6-01A-01-TS1-barycenter-unif-1.pdf}}}\\

\includegraphics[width=0.16\textwidth]{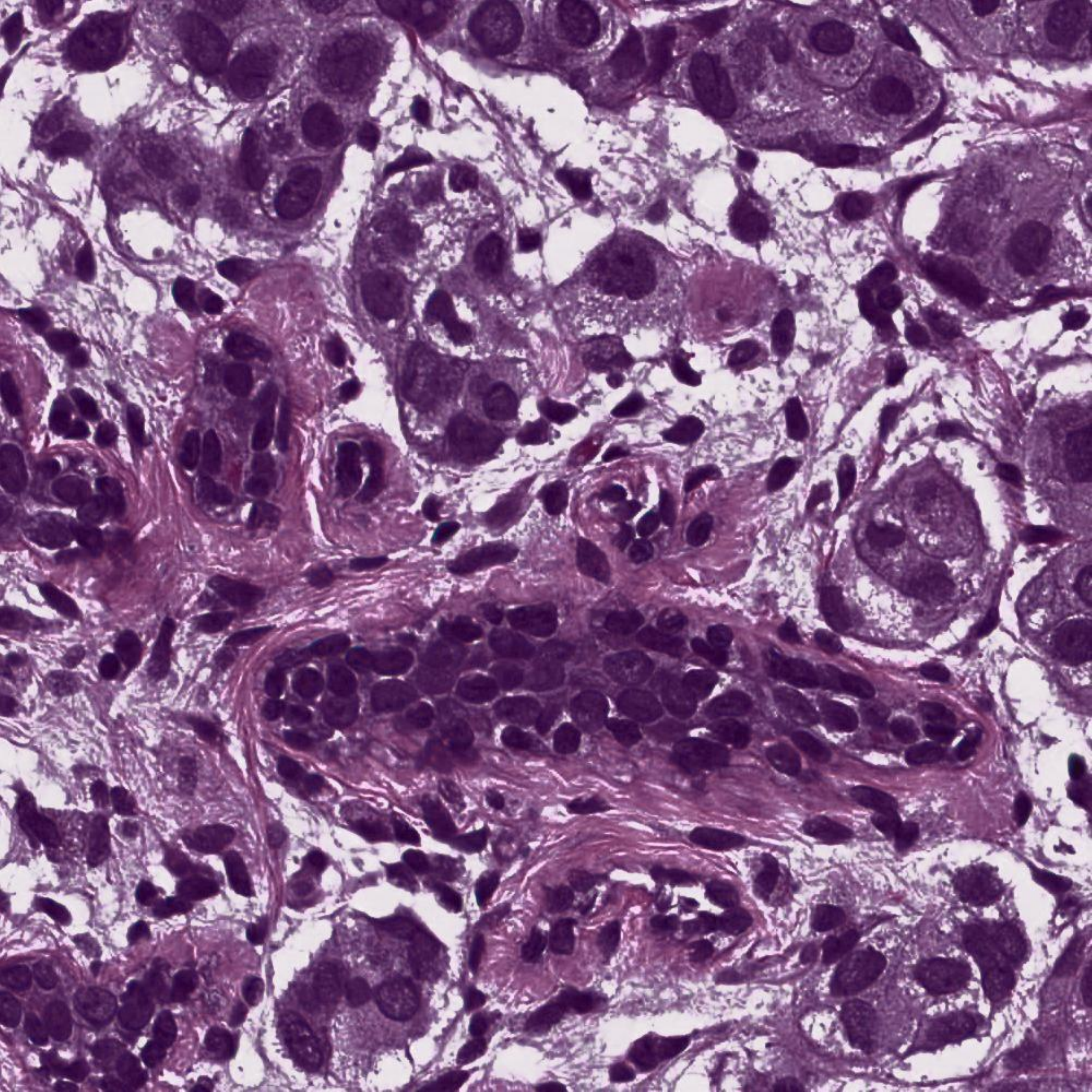}&
\includegraphics[width=0.16\textwidth]{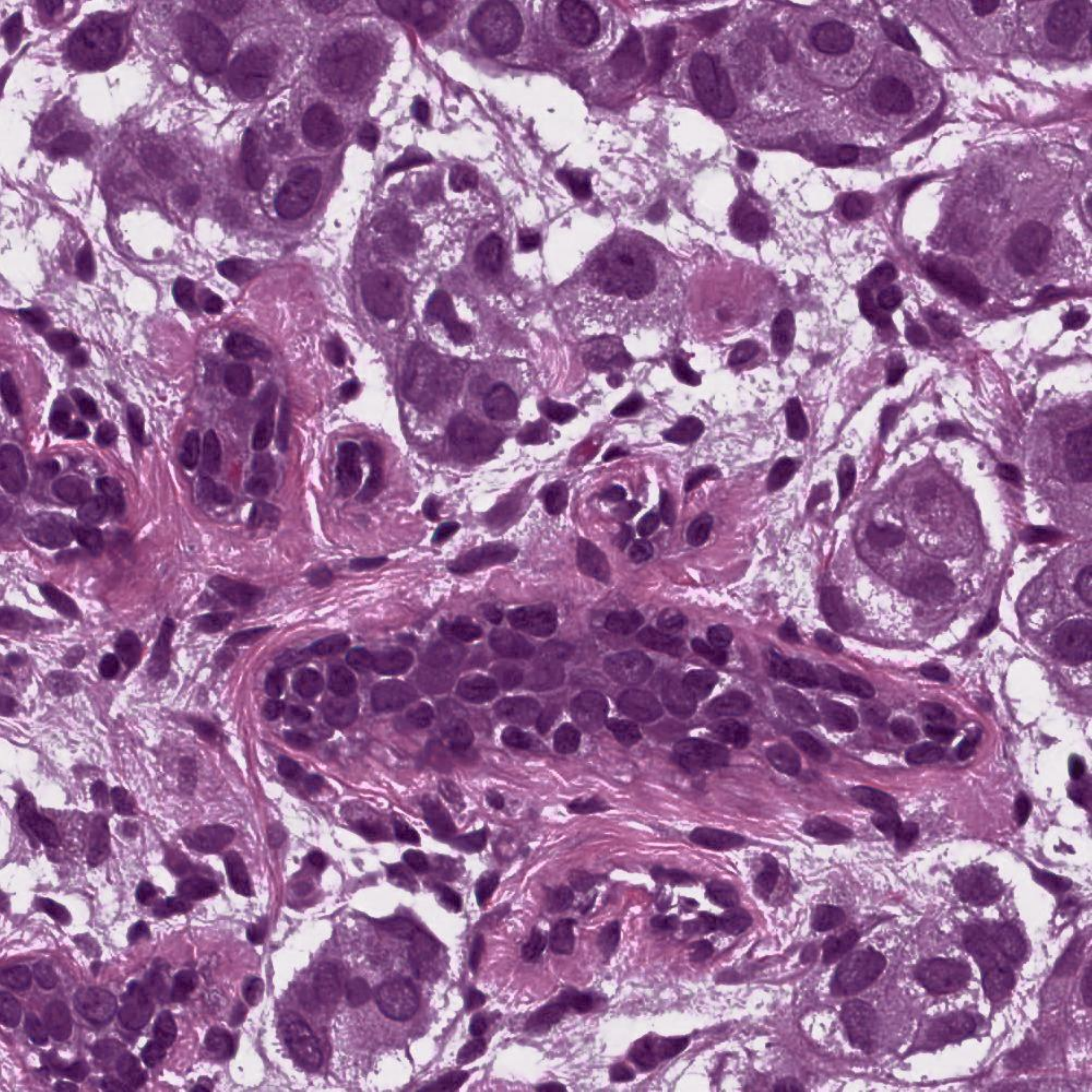}&
\includegraphics[width=0.16\textwidth]{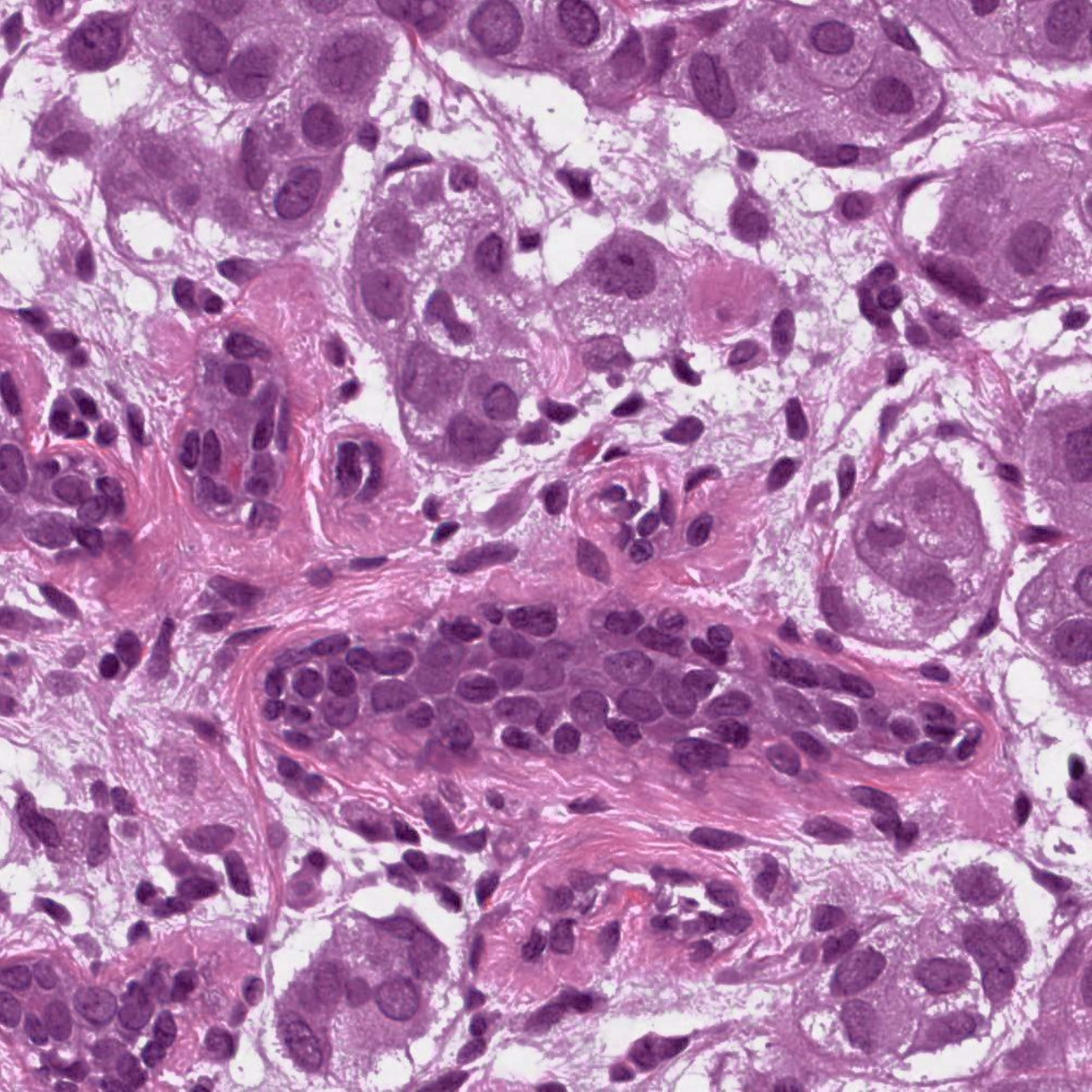}&
\includegraphics[width=0.16\textwidth]{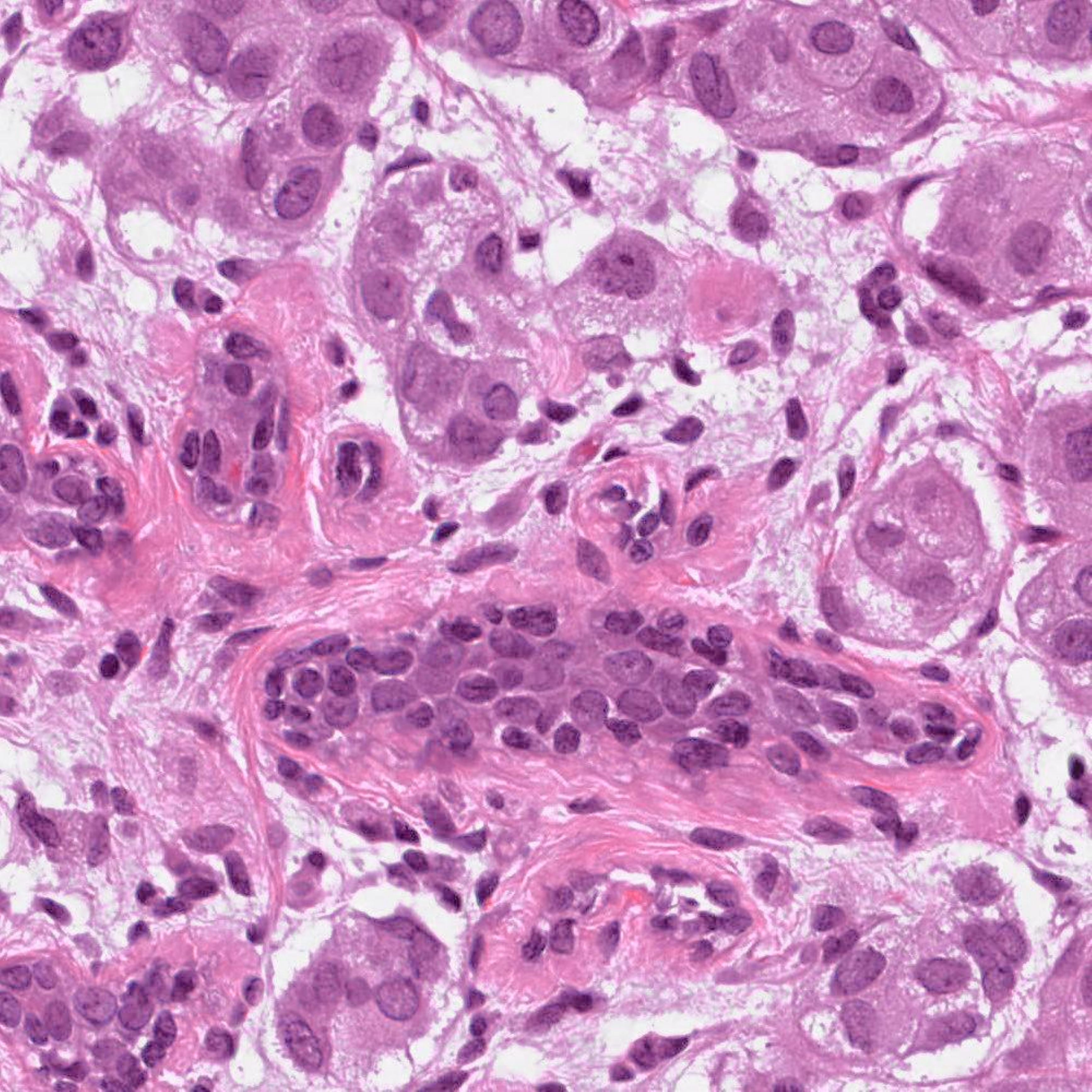}&
\includegraphics[width=0.16\textwidth]{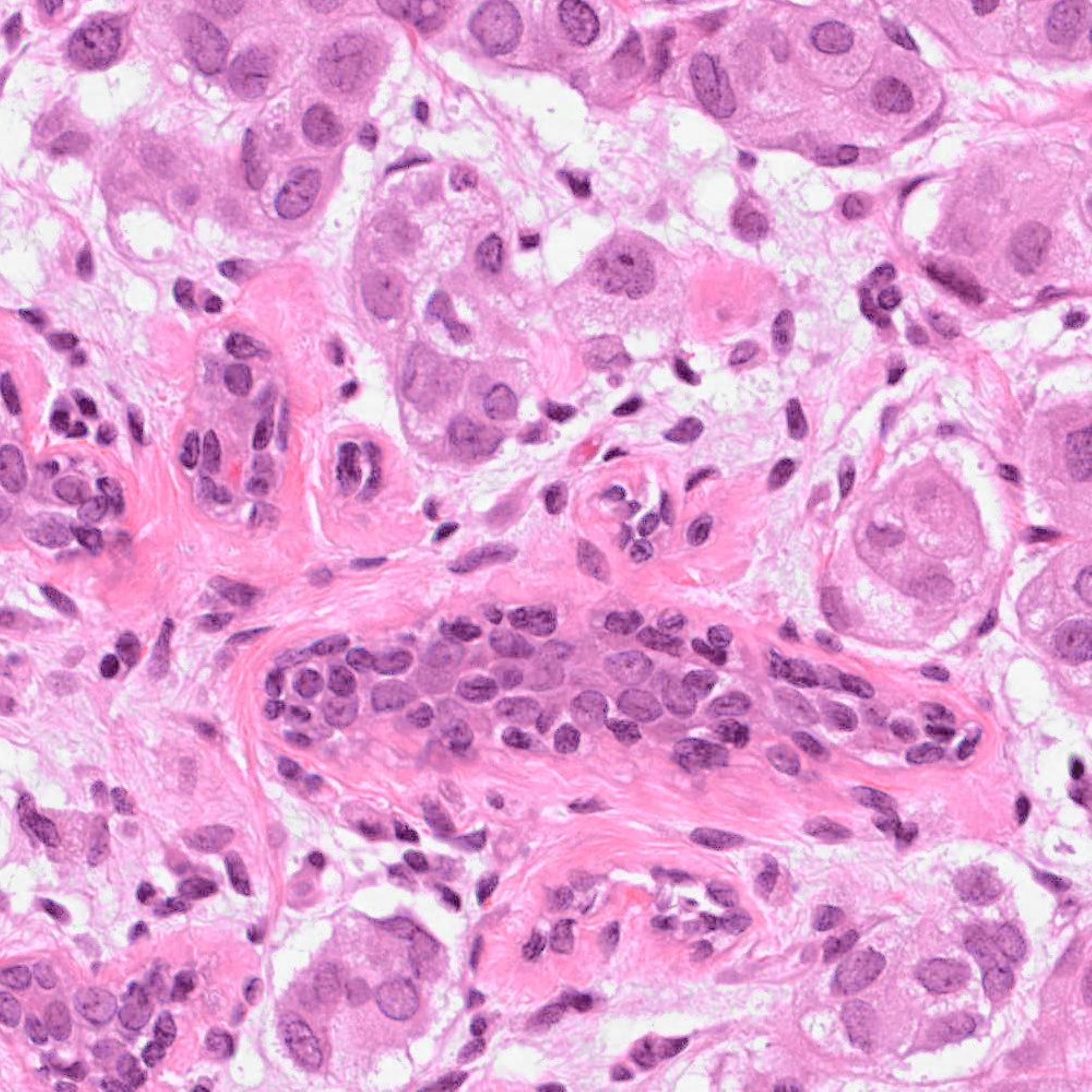}&
{\fboxsep=0mm
\fboxrule=1pt
\fcolorbox{green}{white}{\includegraphics[width=0.16\textwidth]{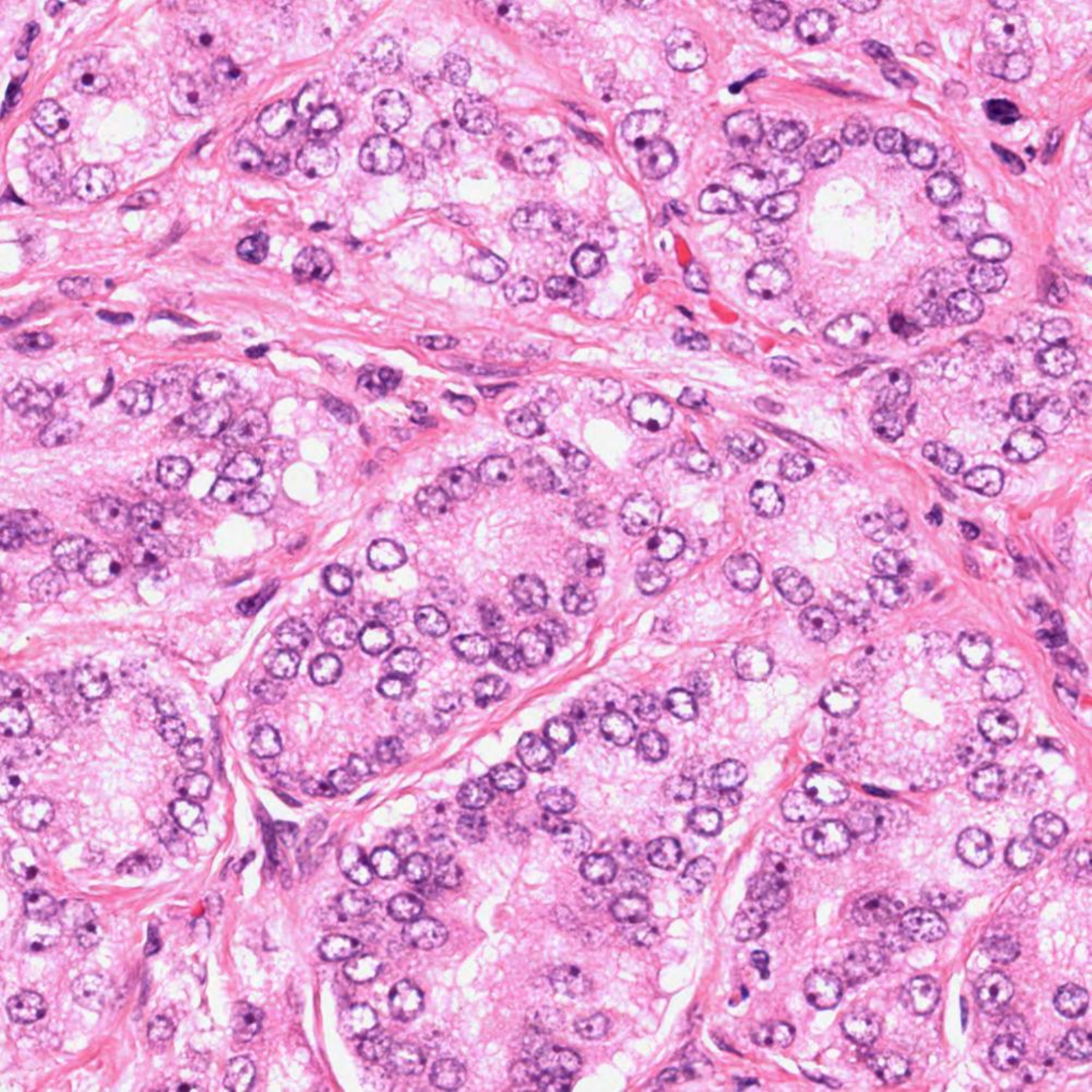}}}\\

\includegraphics[width=0.16\textwidth]{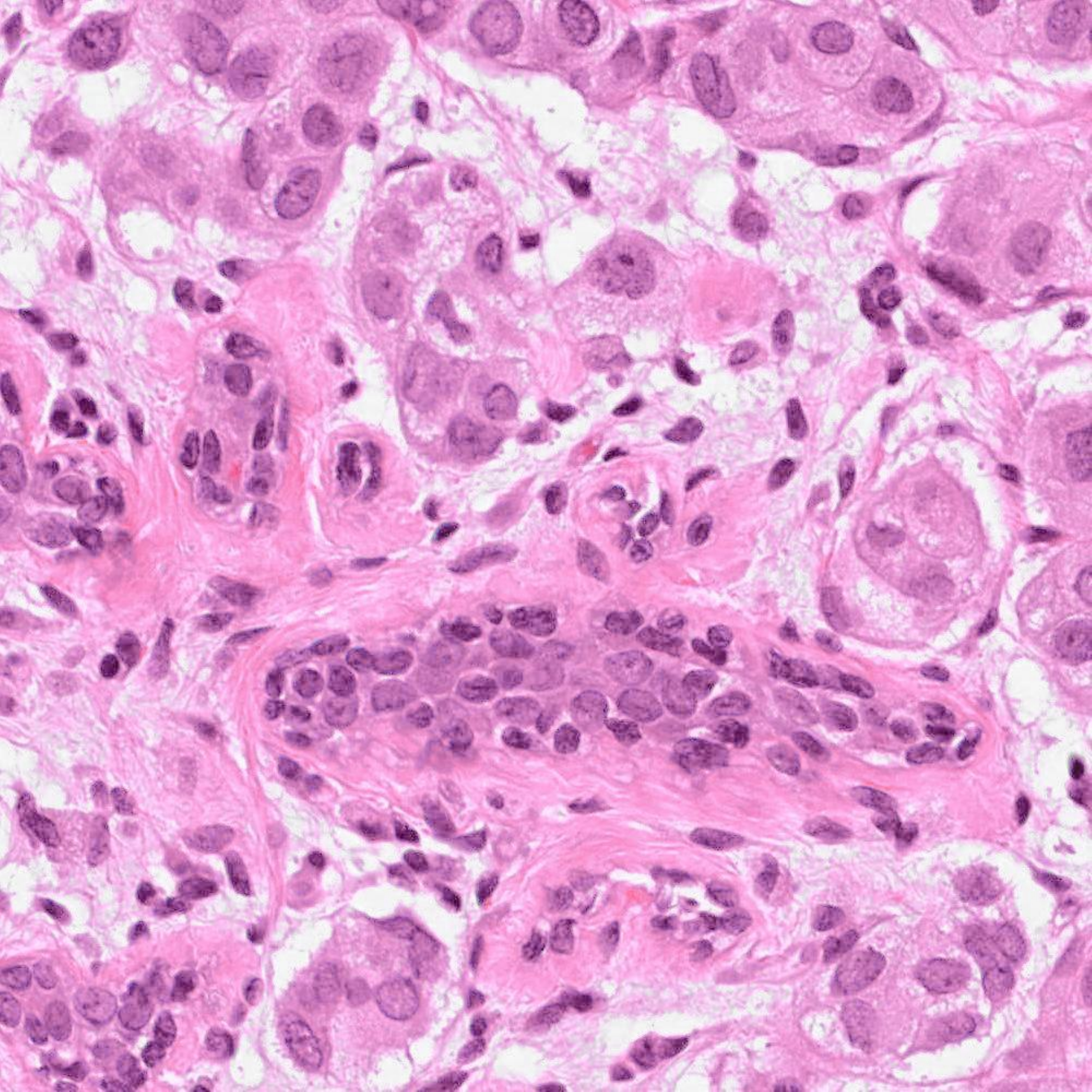}&
\includegraphics[width=0.16\textwidth]{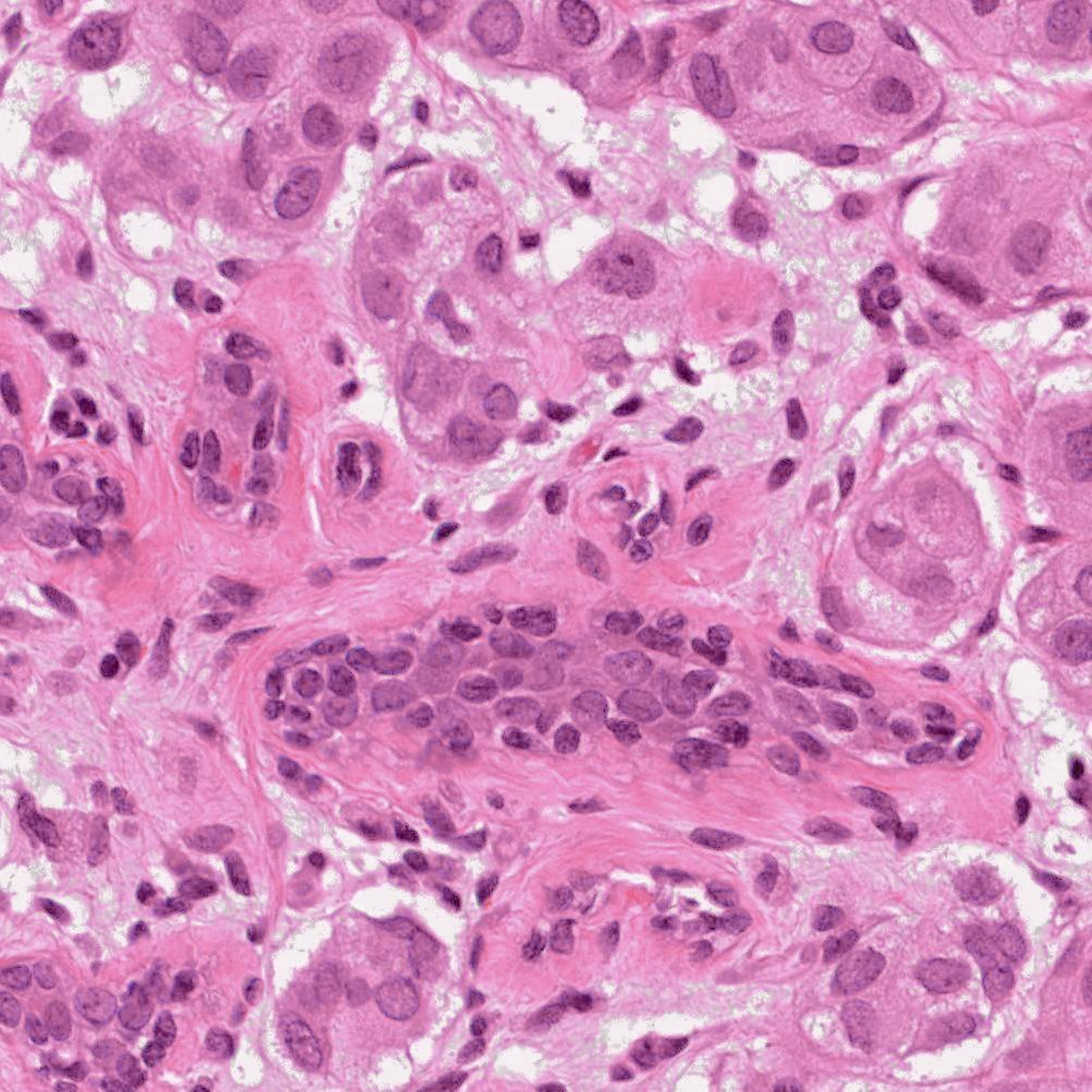}&
\includegraphics[width=0.16\textwidth]{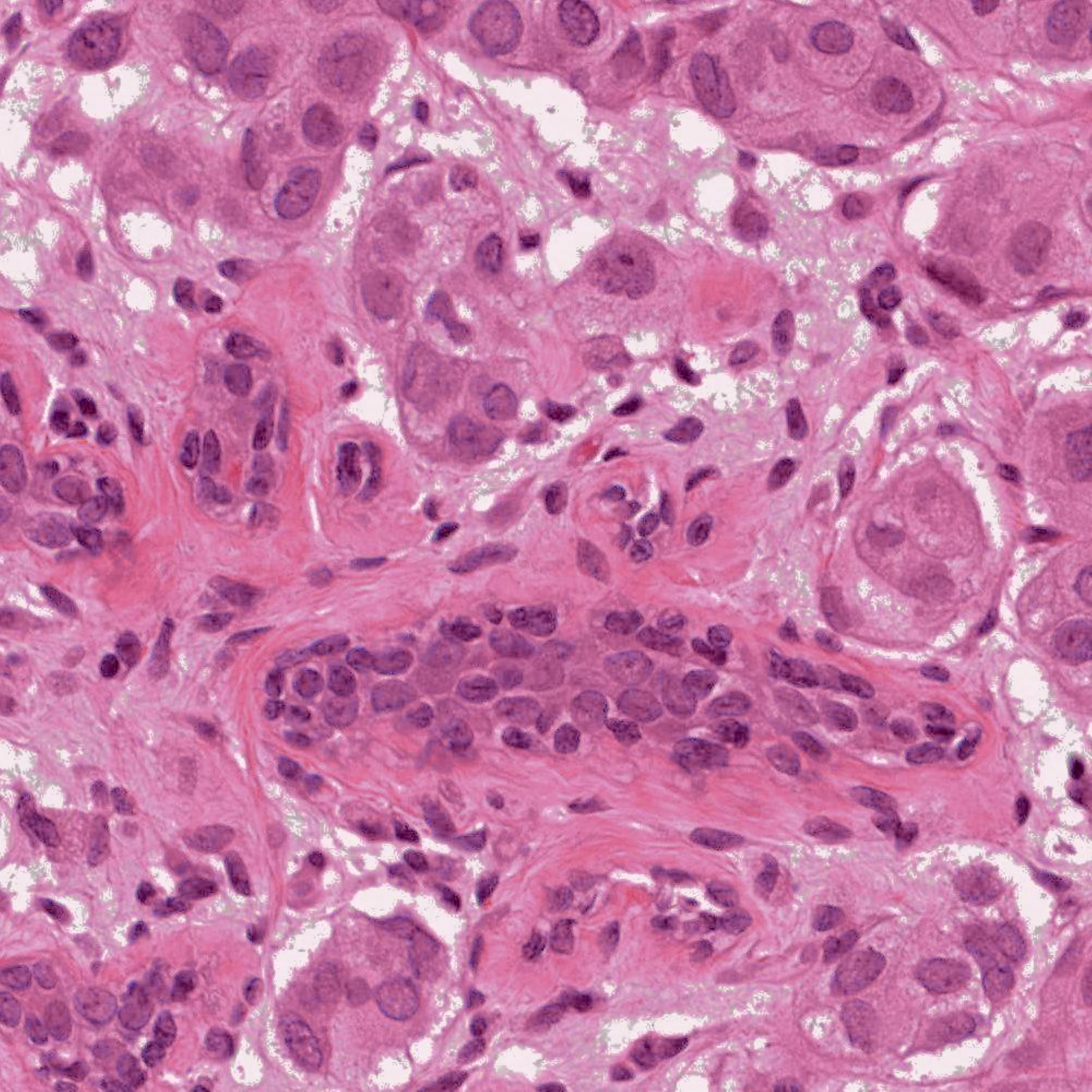}&
\includegraphics[width=0.16\textwidth]{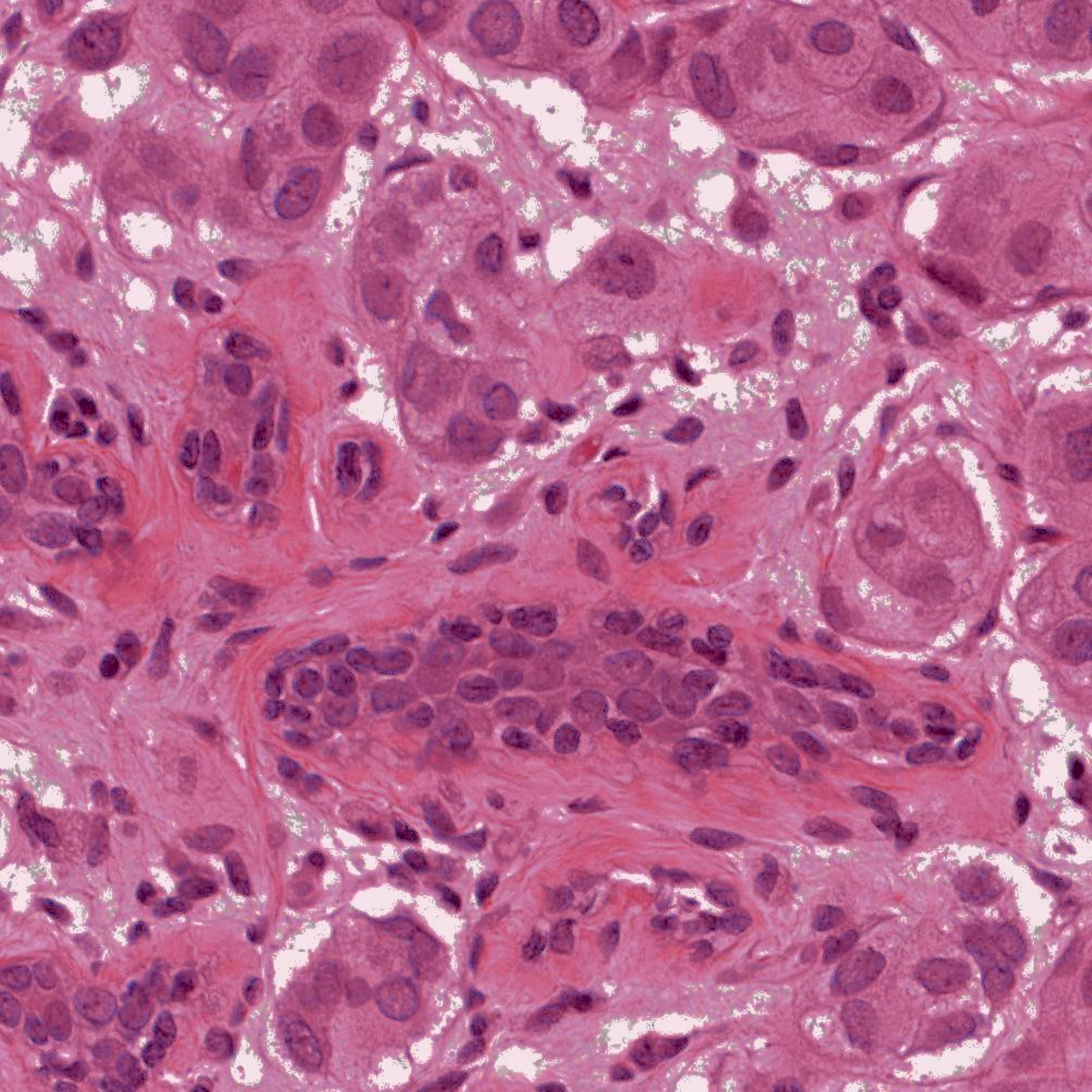}&
\includegraphics[width=0.16\textwidth]{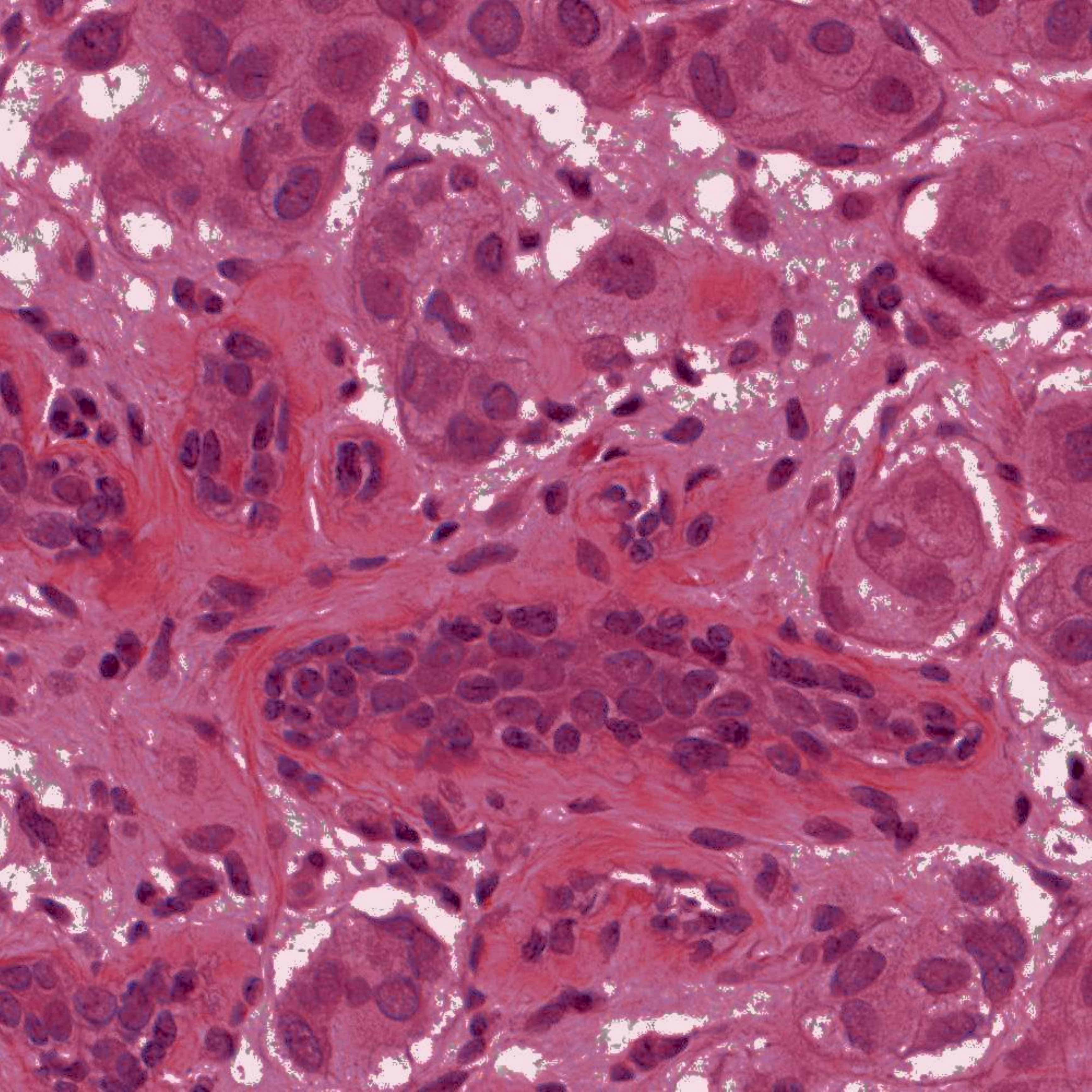}&
{\fboxsep=0mm
\fboxrule=1pt
\fcolorbox{red}{white}{\includegraphics[width=0.16\textwidth]{figures/MonuSeg_TCGA-18-5592-01Z-00-DX1-barycenter-unif-9.pdf}}}\\
\multicolumn{5}{c}{$ \textcolor{blue}{\xrightarrow{\hspace*{8.5cm}}} $} & \\

\end{tabular}

\end{center}
\caption{Multimarginal Wasserstein Barycenter with 1, 2, and 3 reference images, i.e., $N=2,3$ or $4$ in (\ref{eqn:bar}). The images with blue borders are input/source images, the ones with green are the (intermediate) references and the red ones are the final reference/target images. Reference images can be from the same or different domains. Ideally in the multimarginal case, one of the intermediate references should have some background.}
\label{fig:multi_barycenter}
\end{figure}

To deal with some of the aforementioned issues, \textit{we introduce a new approach for simultaneous H\&E stain normalization and augmentation based on the multimarginal Wasserstein barycenter approach}. Specifically, the novelty of the paper lies in first introducing the traditional Wasserstein barycenter approach for stain normalization/augmentation (Figure \ref{fig:traditional_barycenter}), and then introducing the multimarginal version \cite{Agueh,Pass} to overcome the limitations of the traditional approach in this context (Figure \ref{fig:multi_barycenter}). Note that the traditional Wasserstein barycenter (1 source and 1 reference), although widely employed in computer vision, to the best of our knowledge has never been used for stain normalization/augmentation and the more general multimarginal Wasserstein barycenter (1 source and multiple references) has hardly ever been used in computer vision or medical imaging communities. For more accurate stain normalization, the multimarginal version allows one to incorporate additional distributions by utilizing one or more intermediate reference images (Figure \ref{fig:multi_barycenter}). The resultant interpolations span a broad spectrum of stain variations allowing for simultaneous stain normalization and augmentation.

With respect to the pipeline, we convert given source and target images from RGB to the Lab color space, interpolate between the given color distributions using the Wasserstein barycenter, and then convert back to the RGB space. The Lab color space was chosen based on its general effectiveness (decorrelated color channels, etc.) in color transfer applications; see Reinhard and Pouli \cite{reinhard2011colour} and the references therein.  Finally, we quantified our results on stain normalization with respect to a publicly available dataset and obtained state-of-the-art results. Similarly, we augment and normalize stain images in a publicly available nuclei segmentation dataset and again achieved state-of-the-art results using our method.

\section{Background on Wasserstein Distance} \label{sec:background}

We first very briefly review some basic material on optimal mass transport (OMT) theory and the Wasserstein distance that we will need in the sequel. We refer the reader to \cite{Vil08} for a more detailed development of the subject and references.

Consider two probability measures $\mu_0, \mu_1$ on ${\mathbb R}^n$.
In the original formulation of OMT due to Gaspard Monge \cite{Vil03,Vil08}, one seeks a transport map
\[
T\;:\;{\mathbb R}^n\to{\mathbb R}^n\;:\;x\mapsto T(x)
\]
which specifies where the initial mass $\mu_0(dx)$ at $x$ should be transported in order match the final distribution. This means that $T_\sharp \mu_0=\mu_1$ where
$\mu_1$ is the ``push-forward'' of $\mu_0$ under $T$:
\[
\mu_1(B)=\mu_0(T^{-1}(B))
\]
for every Borel set $B$ in $\mR^n$. Moreover, given the transportation cost $c(x,y)$, the map should minimize
\begin{equation}\label{eq:monge}
\int_{\mR^n} c(x,T(x))d\mu_0(x).
\end{equation}
In this paper, we will only consider the case $c(x,y)=\|x-y\|^2$. To ensure finite cost, we will assume that  $\mu_0$ and $\mu_1$ lie in the space of probability densities with finite second moments, denoted by $P_2(\mR^n)$.

The dependence of the transportation cost on $T$ is highly nonlinear and a minimum may not exist for general costs $c$. In order to handle this problem, Leonid Kantorovich proposed a relaxed formulation \cite{Vil03,Vil08}, in which one seeks a joint distribution $\pi\in\Pi(\mu_0,\mu_1)$ on $\mR^n\times\mR^n$, referred to as a {\em coupling} of $\mu_0$ and $\mu_1$, i.e., the  marginals along the two coordinate directions should coincide with $\mu_0$ and $\mu_1$, respectively. More precisely, in this setting, we consider
    \begin{equation}\label{eq:OptTrans}
      K:=  \inf_{\pi\in\Pi(\mu_0,\mu_1)}\int_{\mR^n\times\mR^n}\|x-y\|^2d\pi(x,y).
    \end{equation}

For the case where $\mu_0,\mu_1$ are absolutely continuous with respect to the Lebesgue measure, it is a standard result that OMT (\ref{eq:OptTrans}) has a unique solution \cite{Vil03,Vil08}. This is of the form
	\[
		\pi=({\rm Id}\times T)_\sharp \mu_0,
	\]
where ${\rm Id}$ stands for the identity map, and $T$ is the unique minimizer of \eqref{eq:monge}. One may also show that the unique optimal transport $T$ is the gradient of a convex function $\omega$, i.e.,
	\begin{equation}\label{eq:optimalmap}
		y=T(x)=\nabla\omega(x).
	\end{equation}

\noindent
\textbf{Wasserstein metric\label{sec:wass}}: The square root of the optimal cost formally defines a Riemannian metric on $P_2(\mR^n)$, known as the Wasserstein metric $W_2$ \cite{Vil03,Vil08}, i.e., $W_2(\mu_0,\mu_1):=\sqrt{K}$
with $K$ in \eqref{eq:OptTrans}.

Naturally $P_2(\mR^n)$ is a geodesic space: a geodesic between $\mu_0$ and $\mu_1$ is of the form
	\begin{equation}\label{eq:displacementinterp1}
		\mu_t=(T_t)_\sharp \mu_0,~~~T_t(x)=(1-t)x+tT(x).
	\end{equation}
A geodesic path is also known as \emph{displacement interpolation} \cite{McCann}. It holds that
	\begin{equation}\label{eq:W2geodesic}
		W_2(\mu_s,\mu_t) = (t-s) W_2(\mu_0,\mu_1),\quad 0\le s< t\le 1.
	\end{equation}
$\mu_t$ also solves the Wasserstein barycenter problem in the case of two probability measures as we will now describe below.

\section{Wasserstein Barycenter}

We follow the theory described in \cite{Agueh,Pass} to which we refer the interested reader to all the relevant references. We follow the notation and set-up from Section~\ref{sec:background}. In the case of images of interest in the present work, we take $n=2$ or $n=3$ in $P_2(\mR^n)$.

Then the \textbf{\emph{Wasserstein barycenter}} of $N$ probability measures $\mu_1, \ldots , \mu_N \in P_2(\mR^n)$ is the minimizer of the functional
\be \label{eqn:bar} f(\mu) =  \sum_{i=1}^N \lambda_i W_p^p (\mu_i, \mu), \quad \lambda_i \ge 0, \quad \sum_{i=1}^N \lambda_i =1. \ee
This is a special case of the Multimarginal Optimal Transport (MOMT) problem \cite{Agueh,Pass}.

The case $N=2$ is classical due to McCann \cite{McCann} has been sketched in Section~\ref{sec:wass}. For our purposes for stain normalization and augmentation in histological data, we may regard $\mu_1$ as the source distribution and $\mu_2$ as the reference distribution. Then for
$t \in [0,1]$, we can consider minimizing the family of functionals
$$f_t(\mu) = (1-t) W_2^2 (\mu_1, \mu) + t W_2^2 (\mu_2, \mu),$$ and hence get a continuous family of interpolations which form a geodesic in the space of probability distributions as described in Section~\ref{sec:wass}. See equation \eqref{eq:displacementinterp1}.

In the present work, we also employ one source distribution $\mu_1$ and either $1,2$ or $3$ reference distributions $\mu_i$ ($2\le i \le 4$). For $N \ge 3$ in (\ref{eqn:bar}), i.e., two or more references, we use the term ``multimarginal OMT,'' to emphasize the fact that we are considering more than two measures. For the application to images, one can always normalize to make sure that the total mass (intensity) is 1.
Further in the examples below (see Section~\ref{sec:normalization}), we choose $\lambda_i=1/N$, $1 \le i \le N$. Let $\mu_{opt}$ denote the optimal solution of (\ref{eqn:bar}), that is, the barycenter. Notice that taking $\mu_1$ and $\mu_{opt}$ as the marginals, we also find the optimal transport map $T_\# \mu_1 = \mu_{opt}.$ We also set the parameter $m=N-2$, which is the number of intermediate reference images. Thus, $m=0$ refers to the usual Wasserstein barycenter computed with respected to a reference and a source, with no intermediate reference images.\\

\noindent
\textbf{Computation of Wasserstein Barycenters:} There are a number of algorithms for the computation of the multimarginal Wasserstein barycenter; see the very recent paper \cite{Ruschen} and the references therein. In our implementation, we used the approach developed in Cuturi and Doucet \cite{Cuturi_fast}, which we briefly sketch.

In the latter work, the authors propose (sub)gradient descent framework based on a modification of the functional (\ref{eqn:bar}); see in particular Section 4 of \cite{Cuturi_fast}. Because of the computational complexity, they first smooth the Wasserstein distance via an entropic regularizer. This allows them to employ the {\em Sinkhorn algorithm} \cite{cuturi2013sinkhorn}, which is an iterative rescaling descent procedure that converges to the desired regularized distance. The procedure of \cite{Cuturi_fast} leads to a strictly convex objective function whose gradients can be computed in a fast, efficient manner. We employed the algorithm for the case in which we want the barycenter for $N$ distributions ($N=2,3$ or $4$) where $N$ is the total number of masses in the weighted sum defining the Wasserstein barycenter (Eq. \ref{eqn:bar}). As mentioned above, $m=N-2$ is the number of intermediate reference images.

\section{Experiments and Results} \label{sec:normalization}
We implemented our algorithm using the Python Optimal Transport (POT) library \footnote{https://github.com/rflamary/POT} which include GPU-accelerated versions of Sinkhorn regularization. We used Nvidia GeForce RTX 2080 Ti for our experiments. Pytorch framework was used for StainGAN \footnote{\url{https://github.com/xtarx/StainGAN}} and CNN3 \footnote{\url{https://github.com/neerajkumarvaid/Nuclei_Segmentation}} implementations. We evaluated our approach against Reinhard \emph{et al.} \cite{reinhard2001color}, Macenko \emph{et al.}\cite{macenko2009method}, Khan \emph{et al.} \cite{khan2014nonlinear}, Vahadne \emph{et al.} \cite{vahadane2016structure}, and StainGAN \cite{shaban2019staingan}.

\begin{figure}[t!]
\begin{center}
\setlength{\tabcolsep}{0.5pt}
\tiny
\begin{tabular}{ccccccccc}

Source & Reinhard\cite{reinhard2001color} & Macenko\cite{macenko2009method} & Khan\cite{khan2014nonlinear} & Vahadane\cite{vahadane2016structure} & StainGAN\cite{shaban2019staingan} & $m=0$(Ours) & $m=1$(Ours) & Target\\

{\fboxsep=0mm
\fboxrule=1pt
\fcolorbox{blue}{white}{\includegraphics[width=0.11\textwidth]{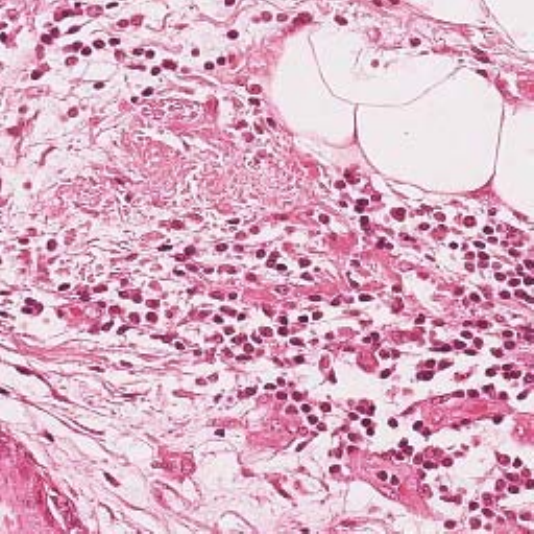}}}&
\includegraphics[width=0.11\textwidth]{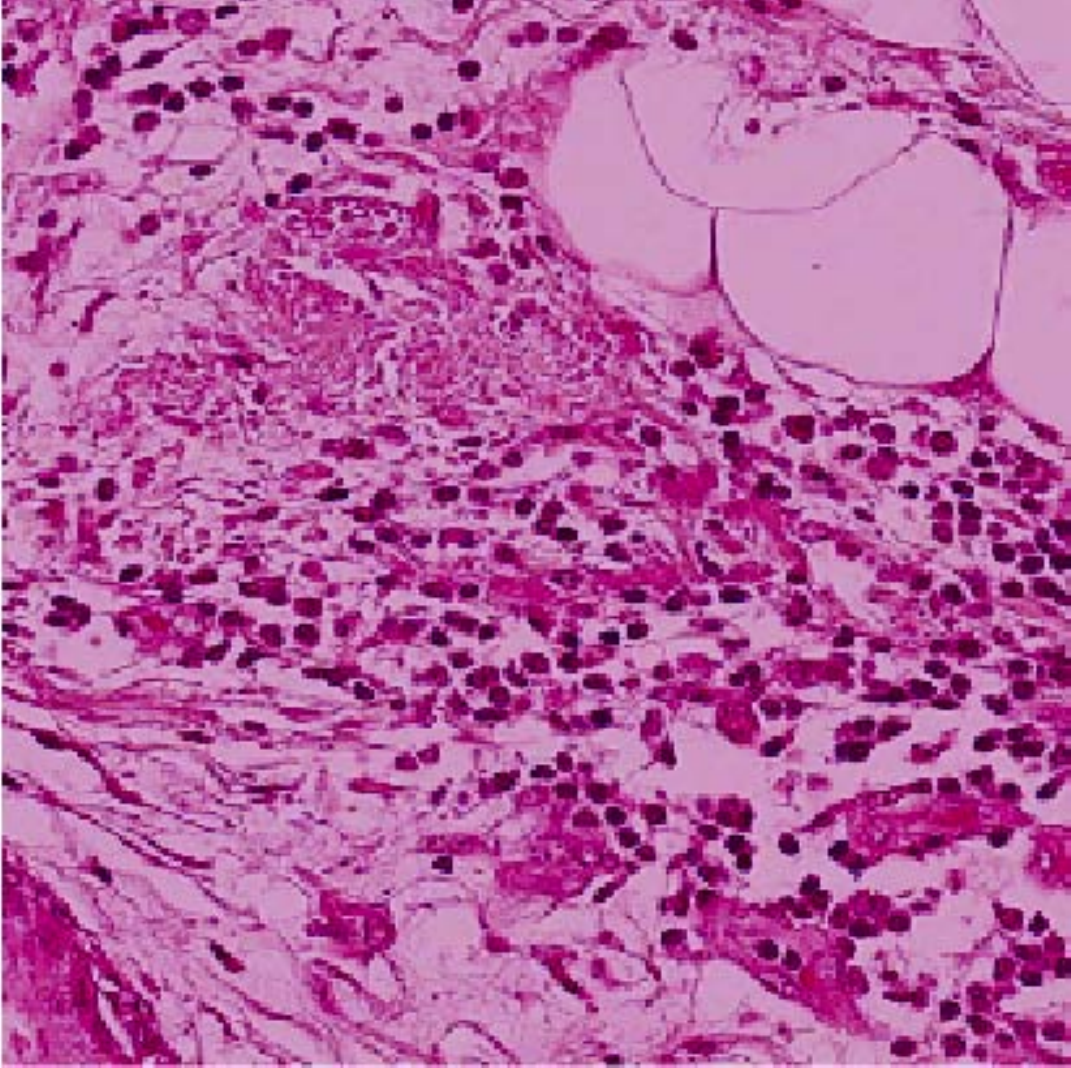}&
\includegraphics[width=0.11\textwidth]{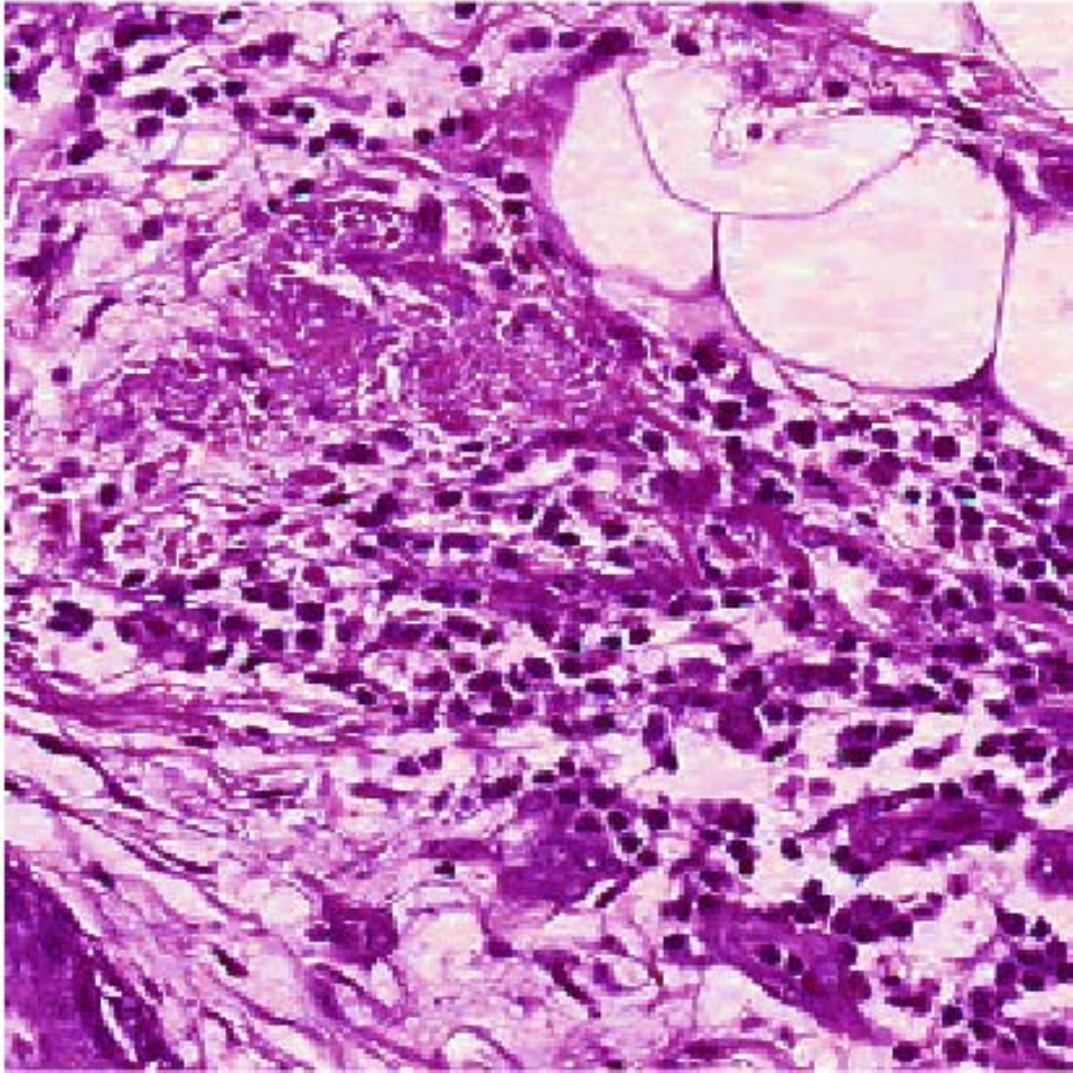}&
\includegraphics[width=0.11\textwidth]{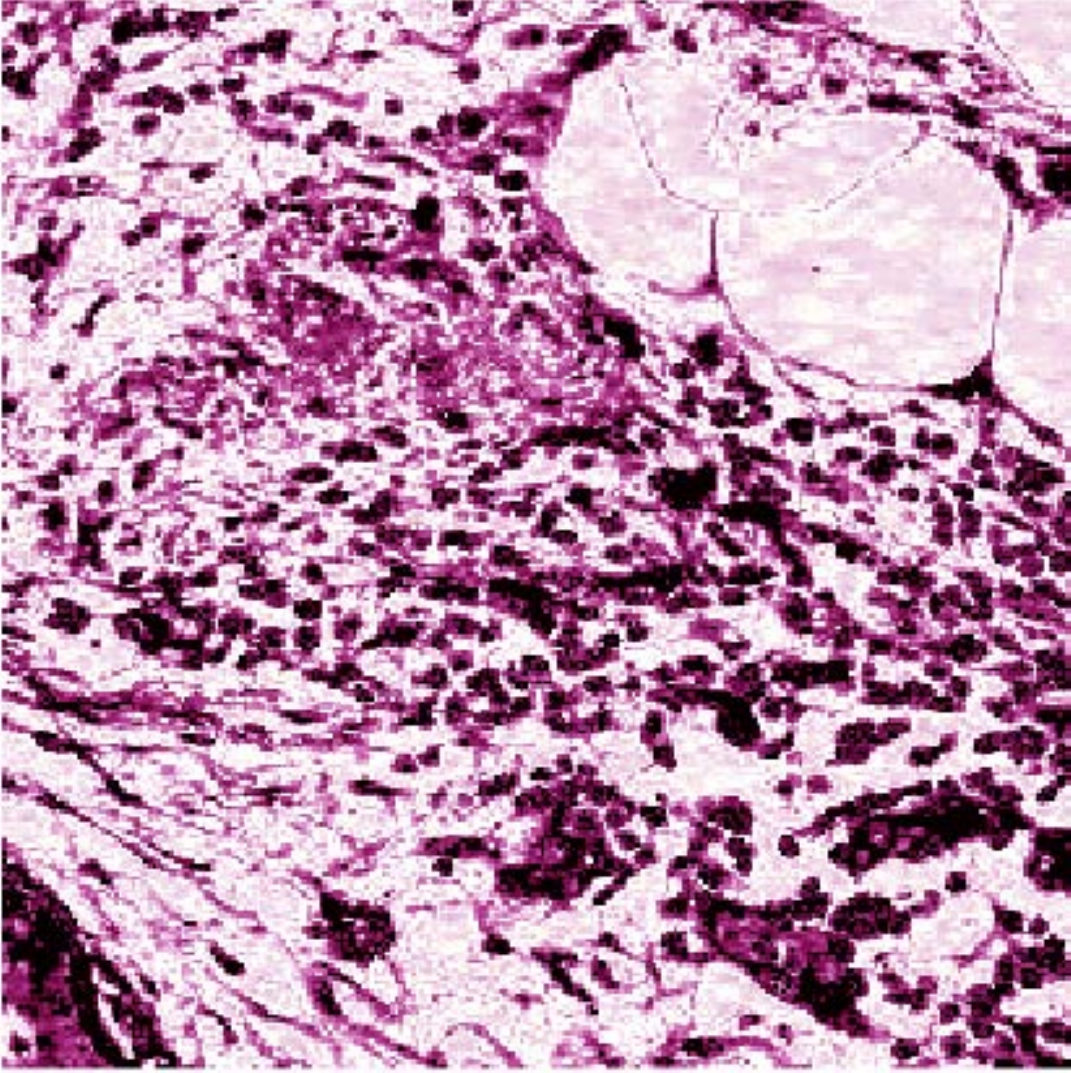}&
\includegraphics[width=0.11\textwidth]{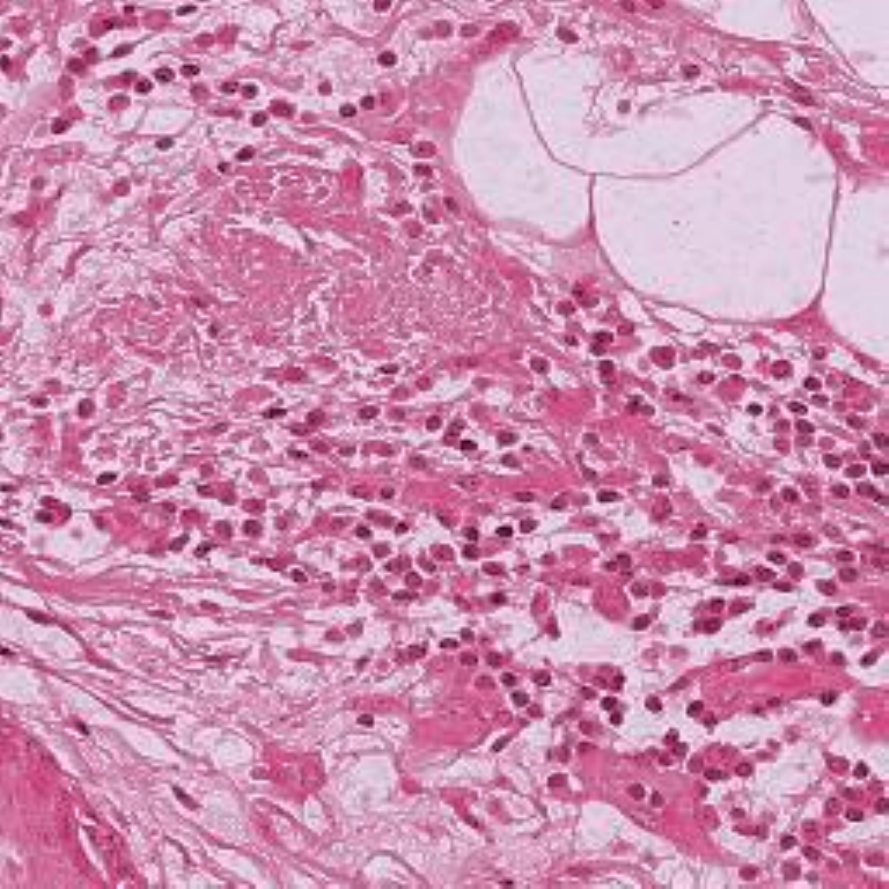}&
\includegraphics[width=0.11\textwidth]{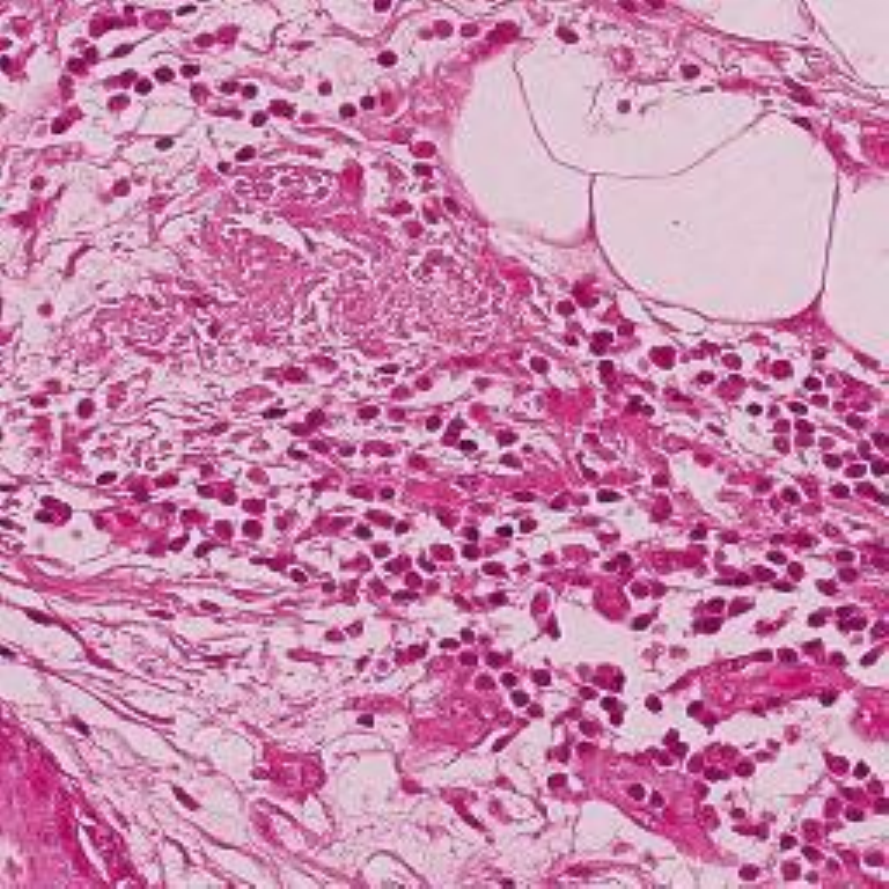}&
\includegraphics[width=0.11\textwidth]{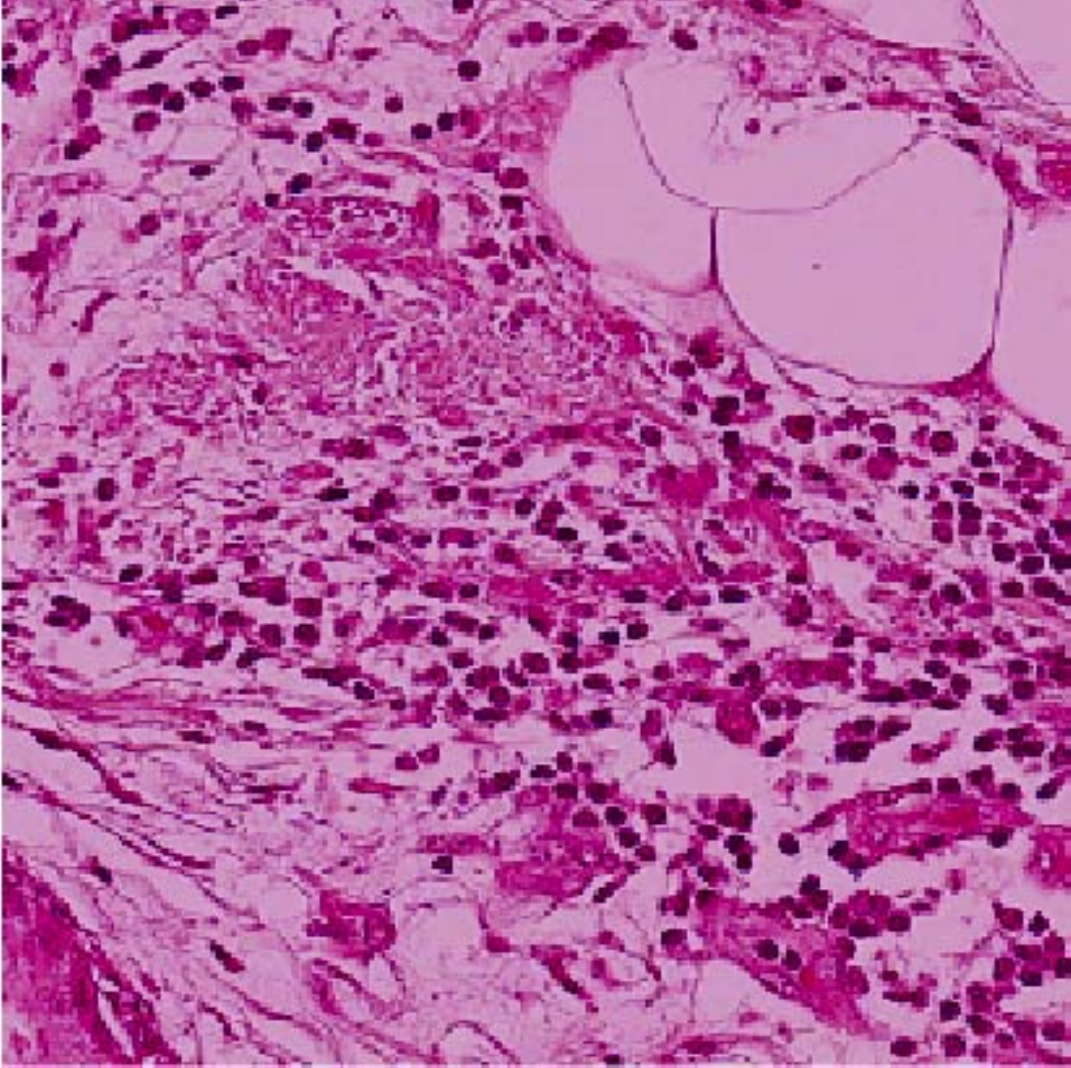}&
\includegraphics[width=0.11\textwidth]{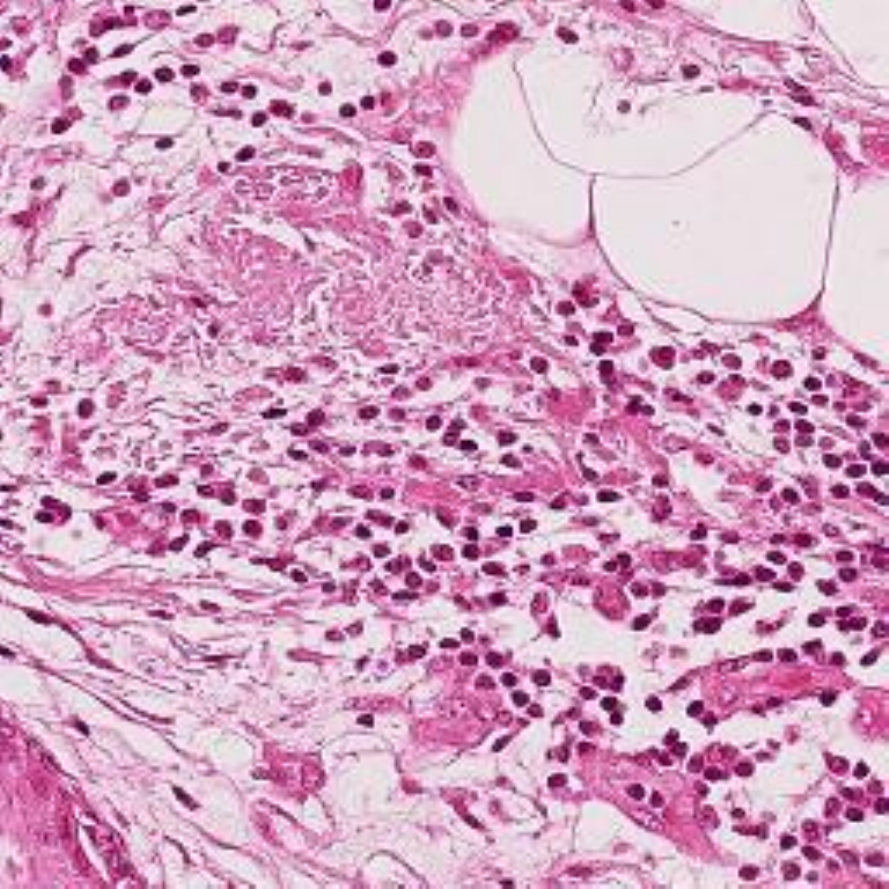}&
{\fboxsep=0mm
\fboxrule=1pt
\fcolorbox{red}{white}{\includegraphics[width=0.11\textwidth]{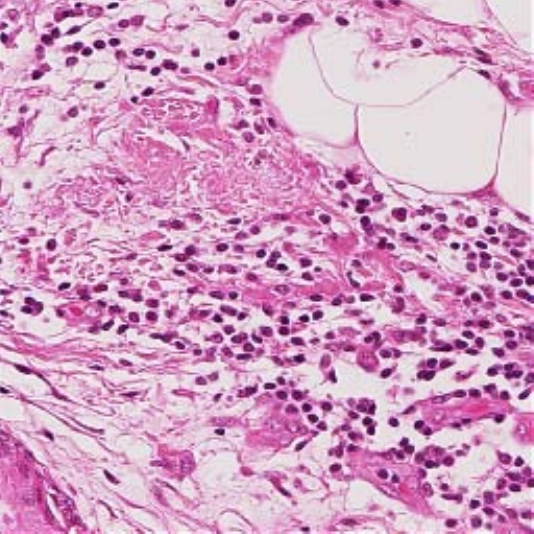}}}\\

{\fboxsep=0mm
\fboxrule=1pt
\fcolorbox{blue}{white}{\includegraphics[width=0.11\textwidth]{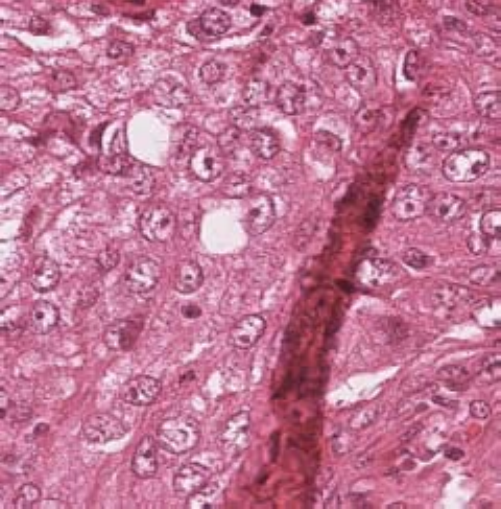}}}&
\includegraphics[width=0.11\textwidth]{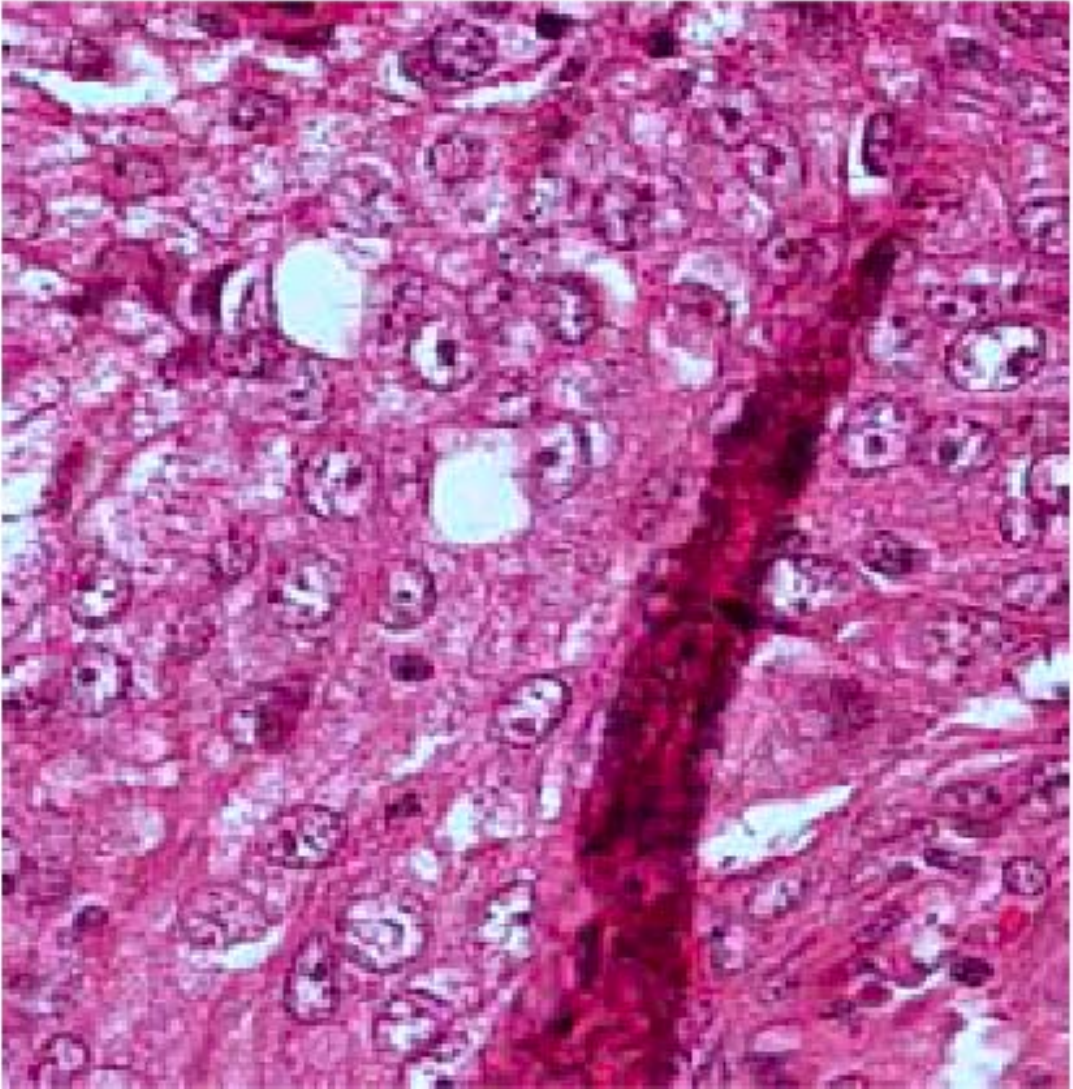}&
\includegraphics[width=0.11\textwidth]{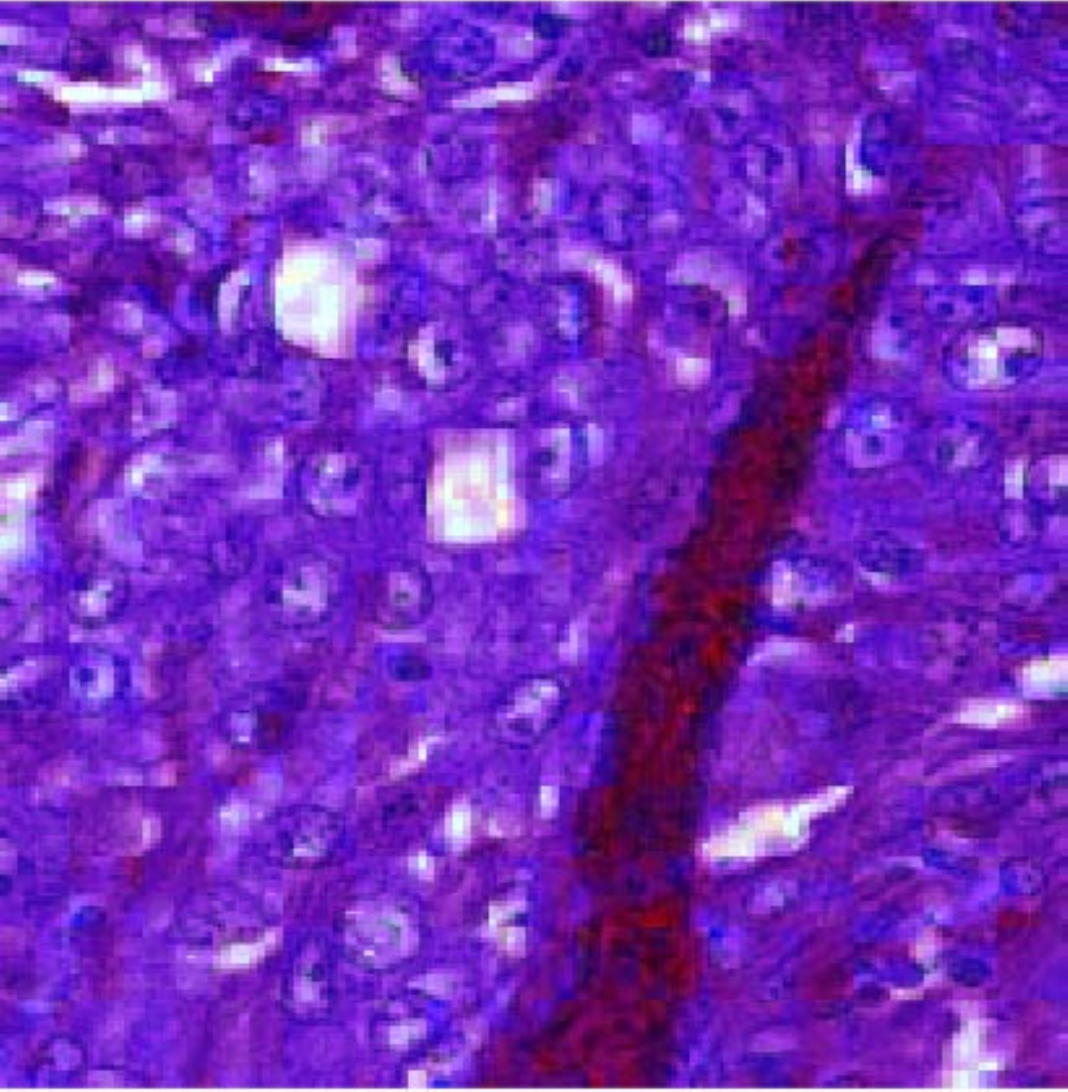}&
\includegraphics[width=0.11\textwidth]{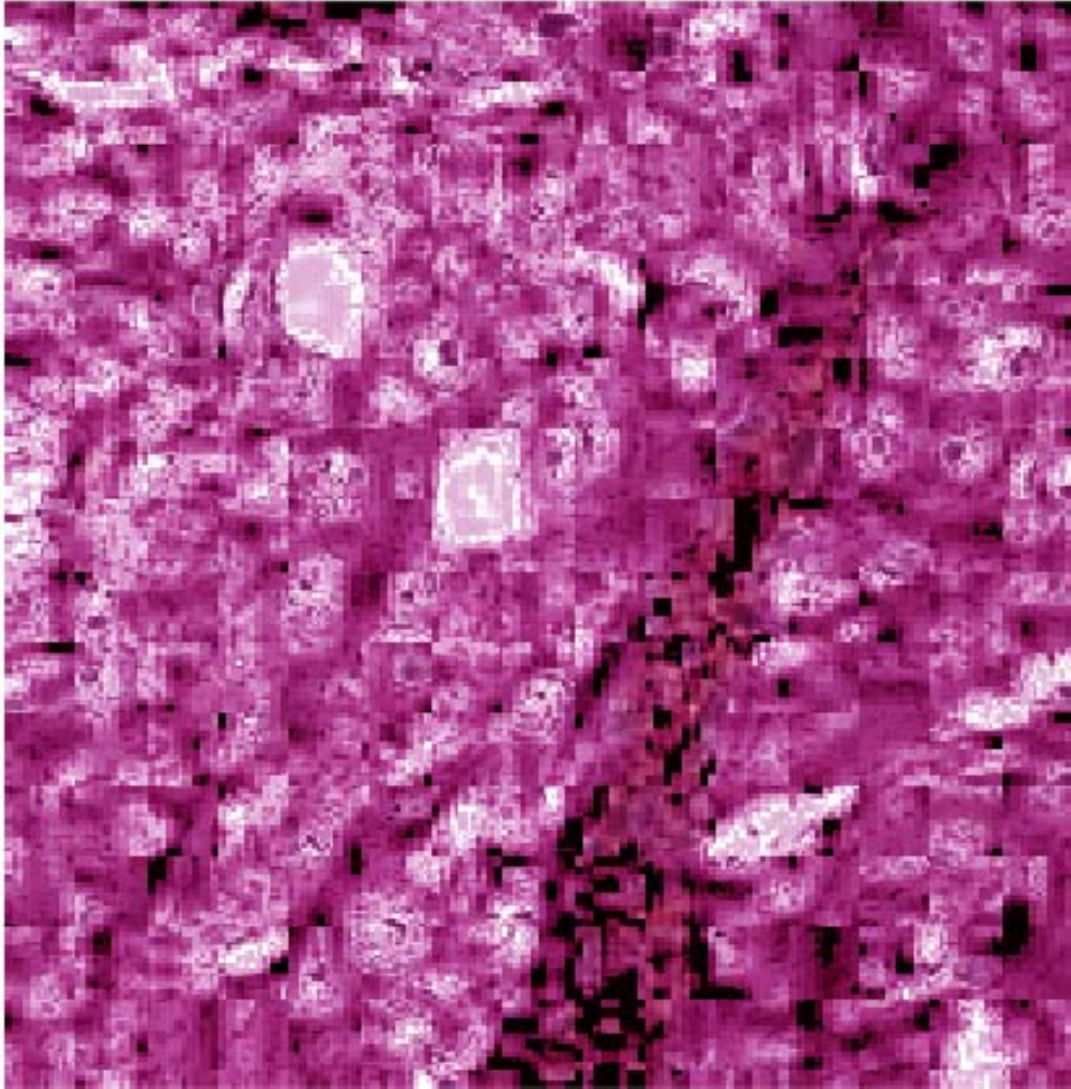}&
\includegraphics[width=0.11\textwidth]{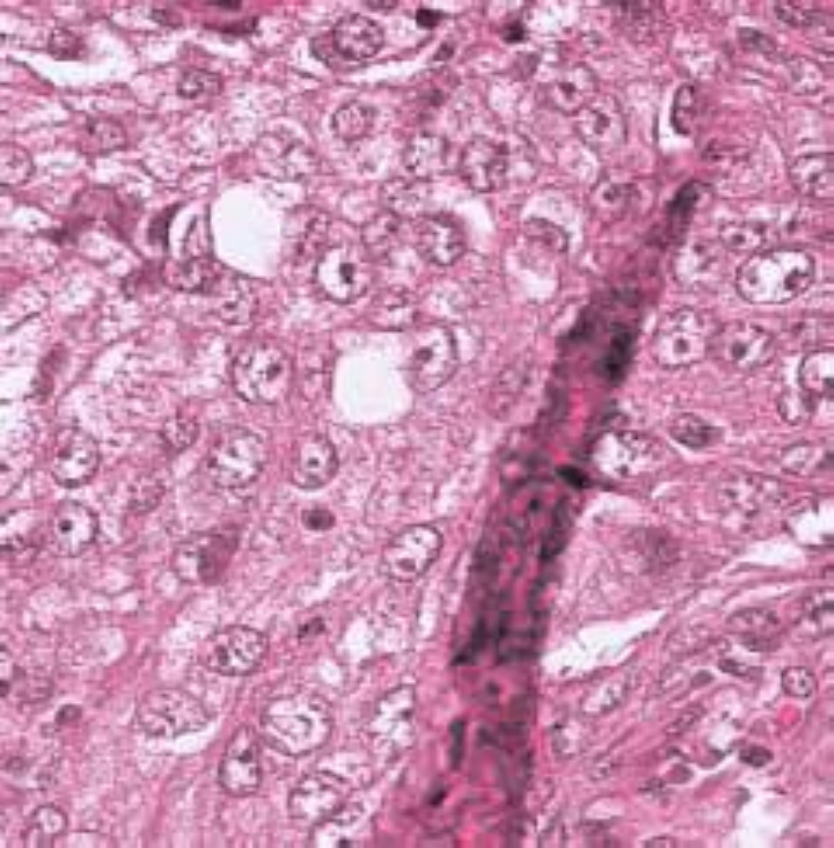}&
\includegraphics[width=0.11\textwidth]{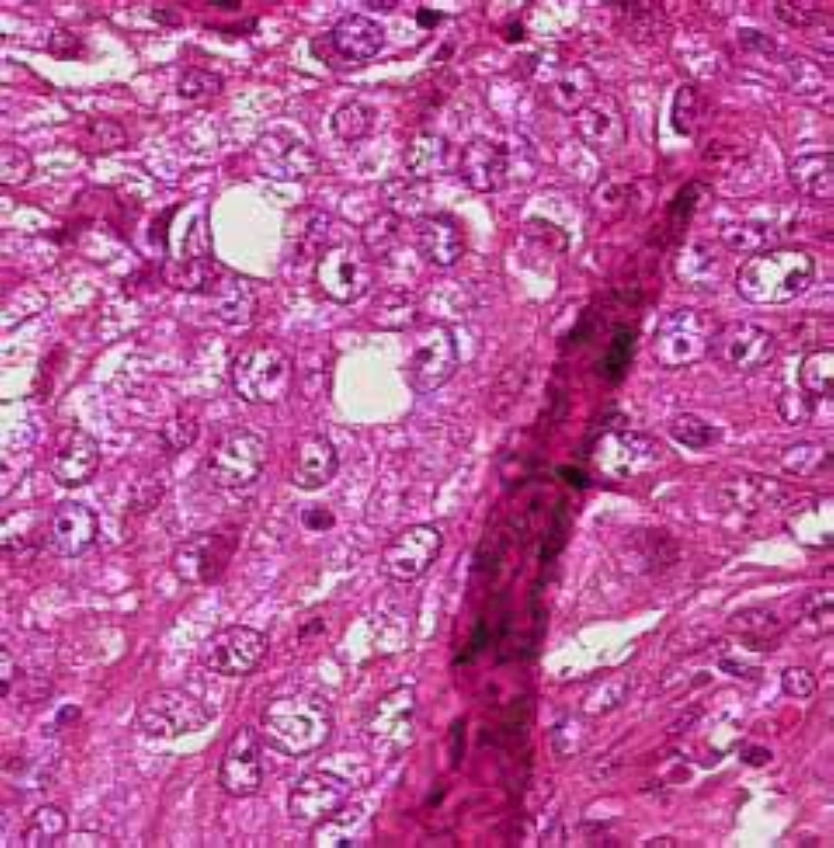}&
\includegraphics[width=0.11\textwidth]{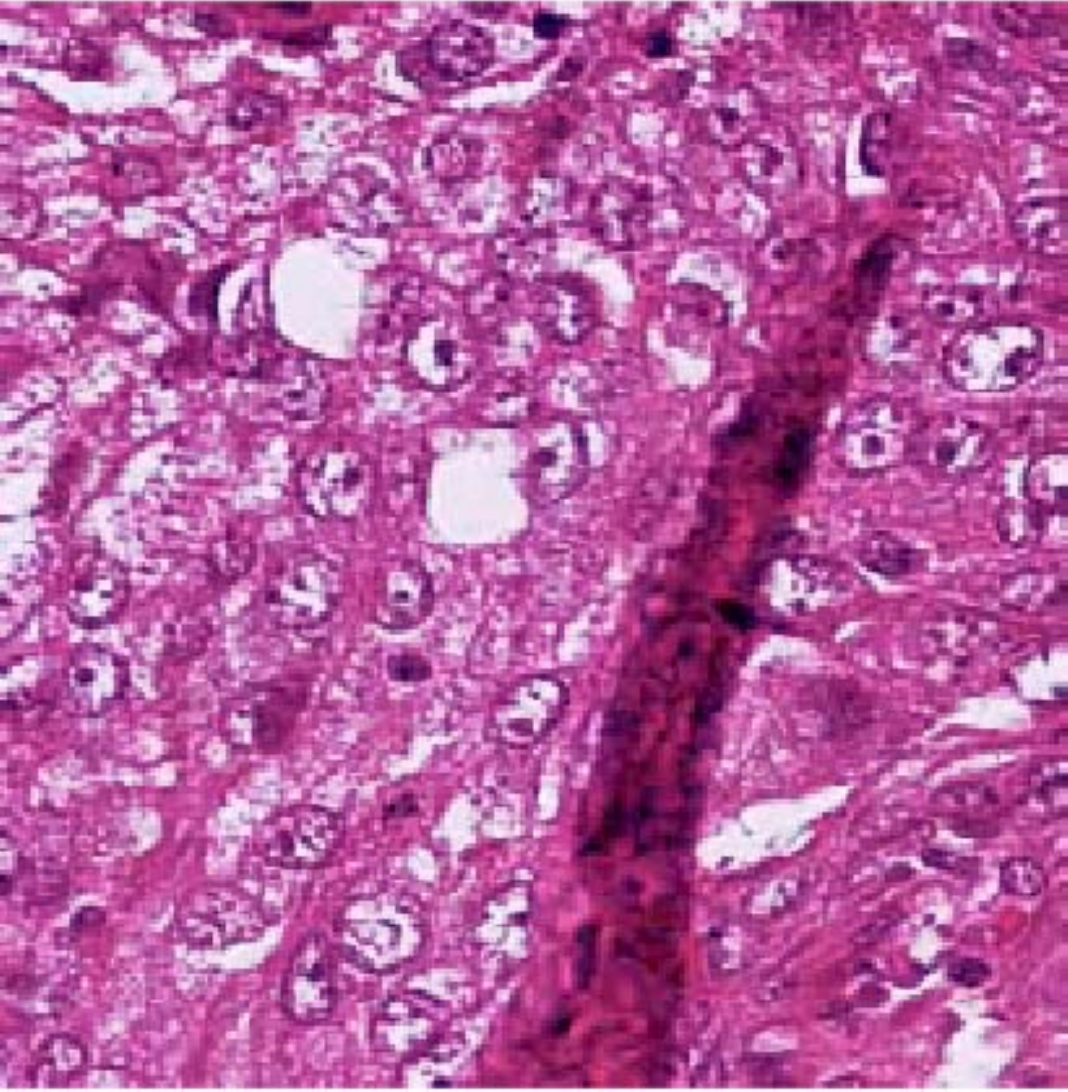}&
\includegraphics[width=0.11\textwidth]{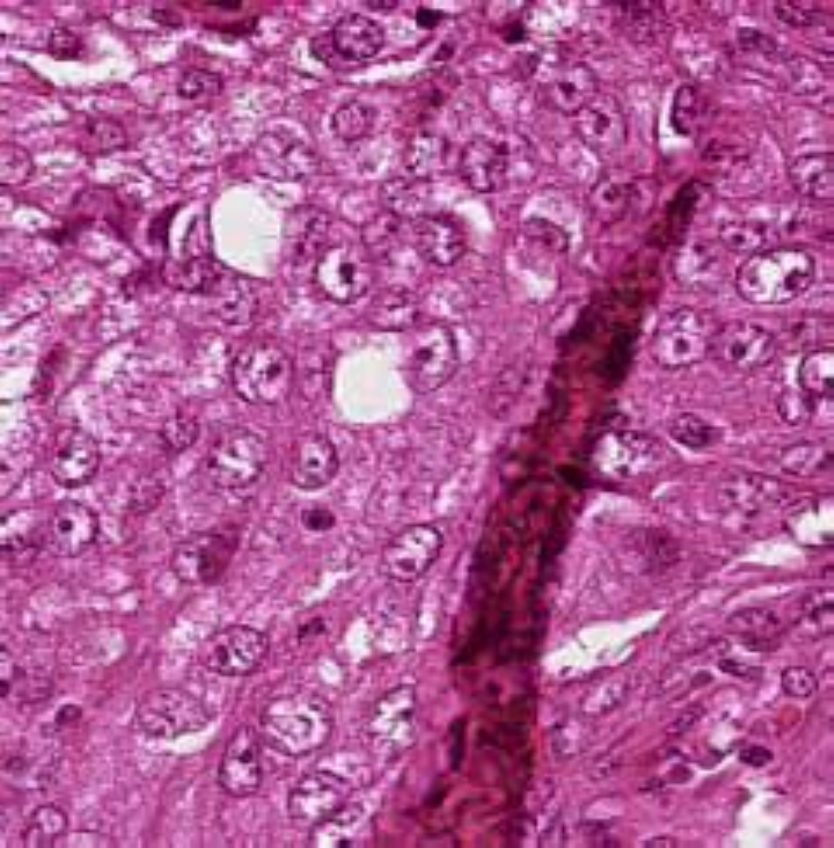}&
{\fboxsep=0mm
\fboxrule=1pt
\fcolorbox{red}{white}{\includegraphics[width=0.11\textwidth]{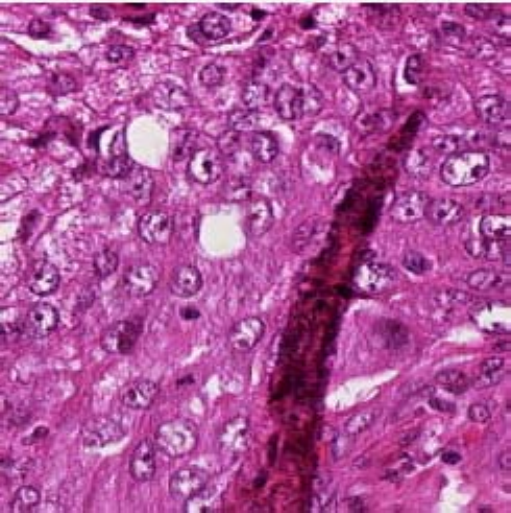}}}\\

{\fboxsep=0mm
\fboxrule=1pt
\fcolorbox{blue}{white}{\includegraphics[width=0.11\textwidth]{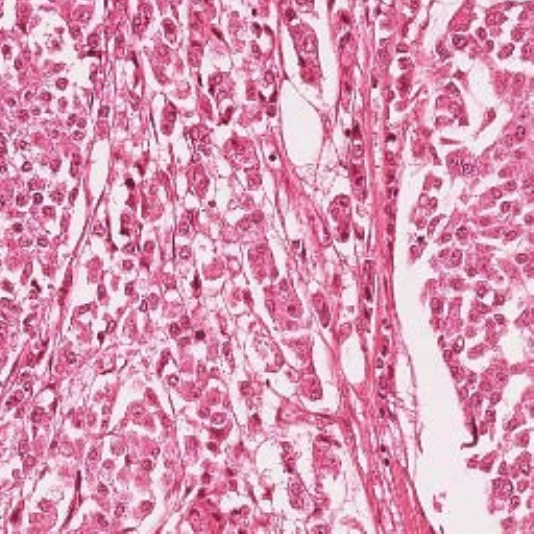}}}&
\includegraphics[width=0.11\textwidth]{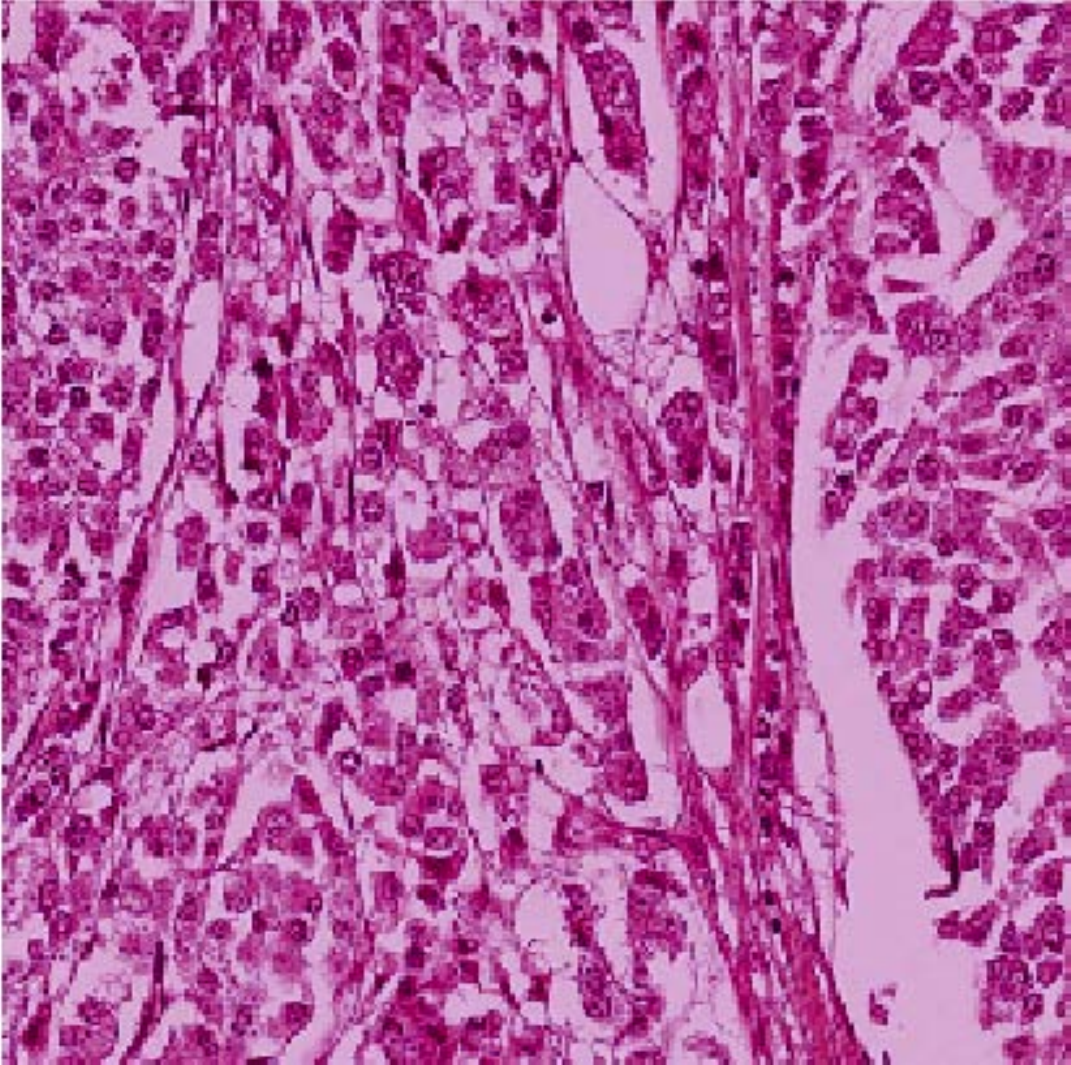}&
\includegraphics[width=0.11\textwidth]{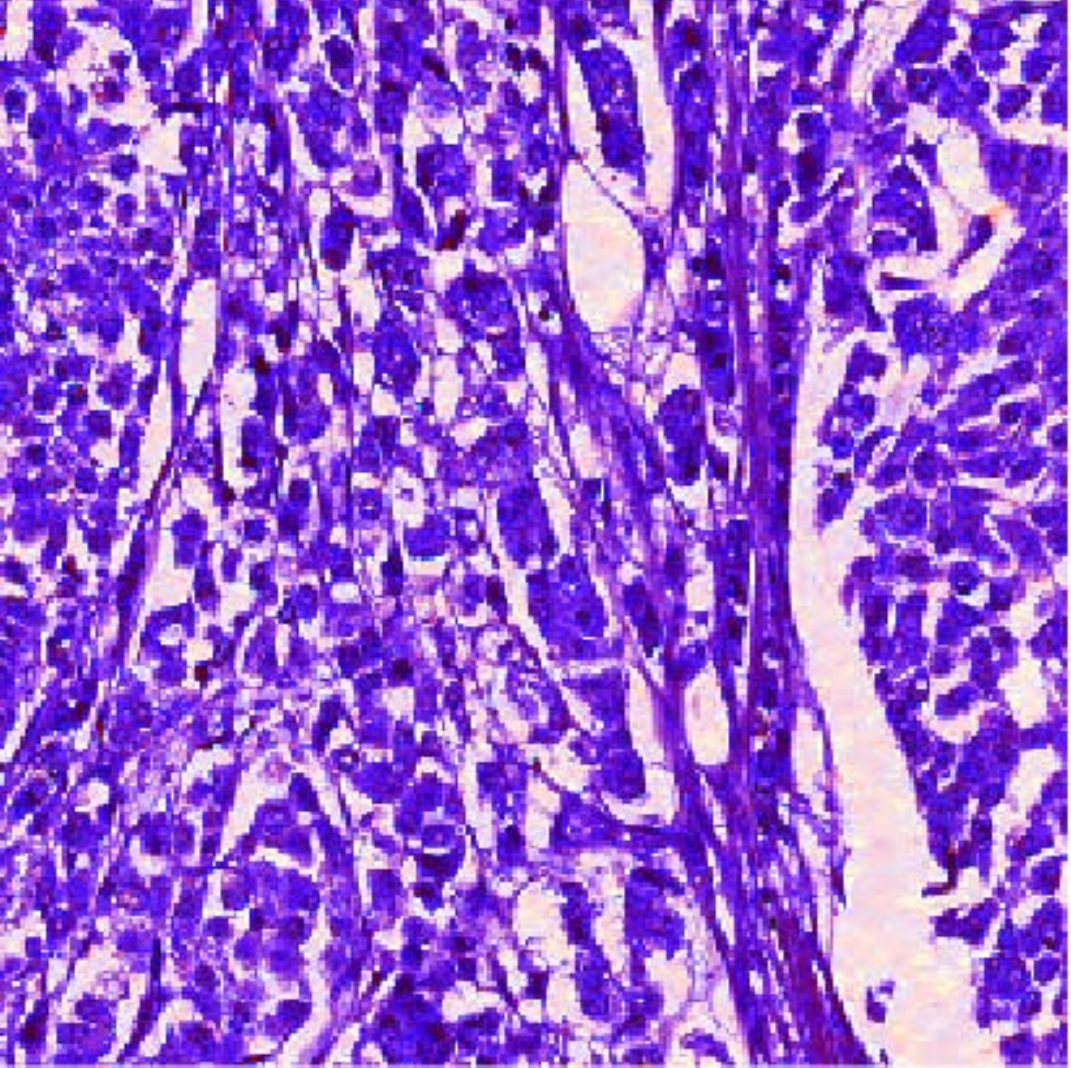}&
\includegraphics[width=0.11\textwidth]{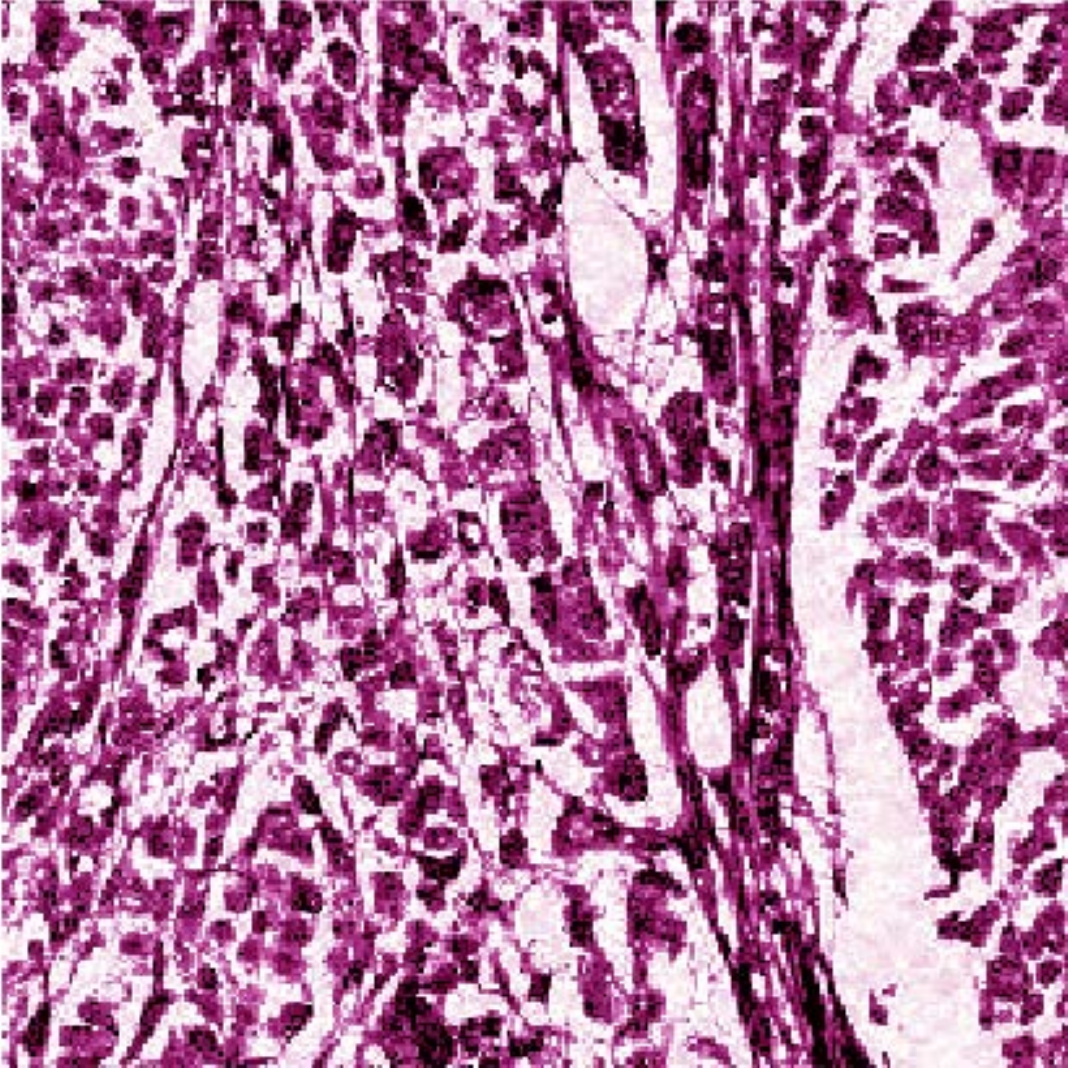}&
\includegraphics[width=0.11\textwidth]{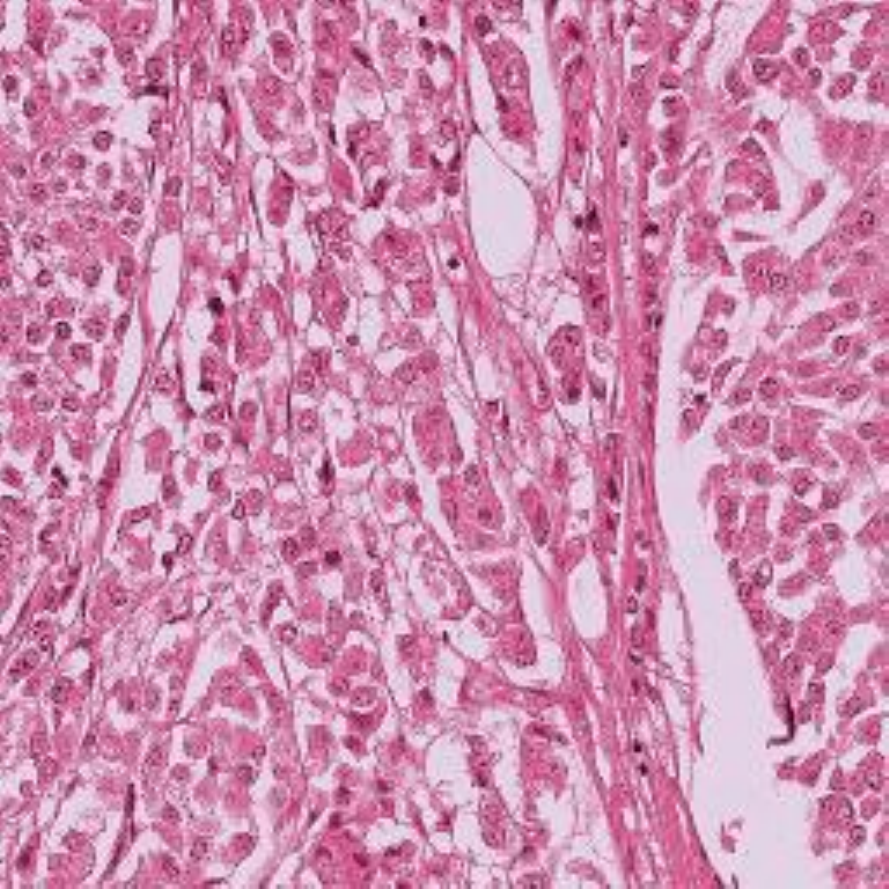}&
\includegraphics[width=0.11\textwidth]{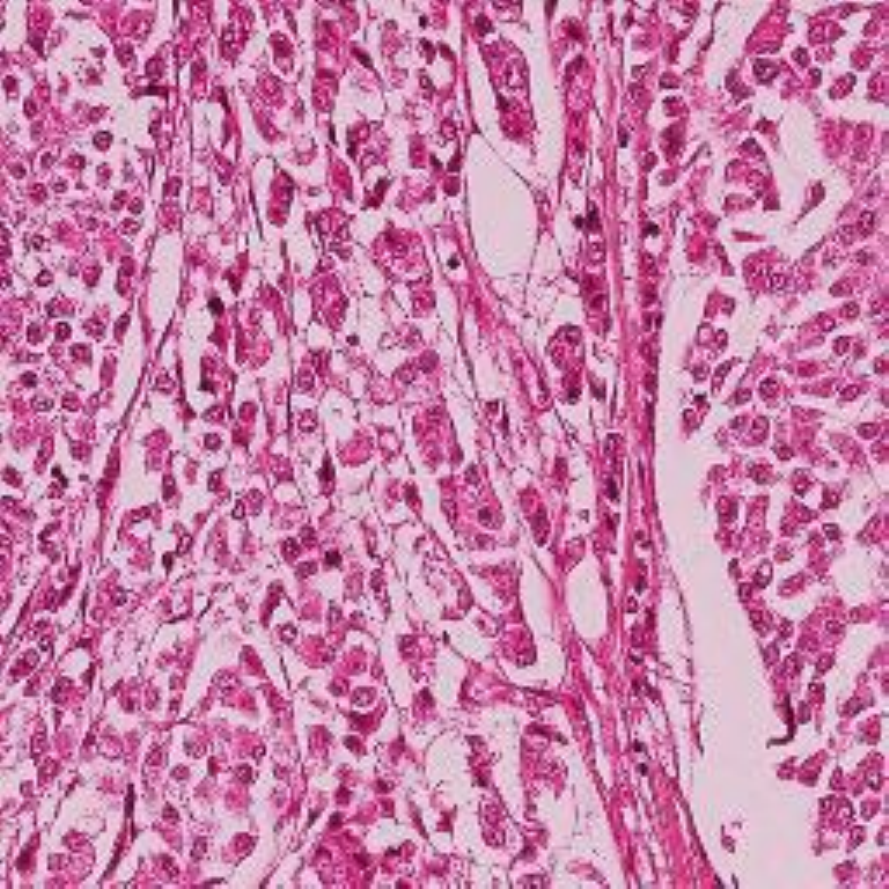}&
\includegraphics[width=0.11\textwidth]{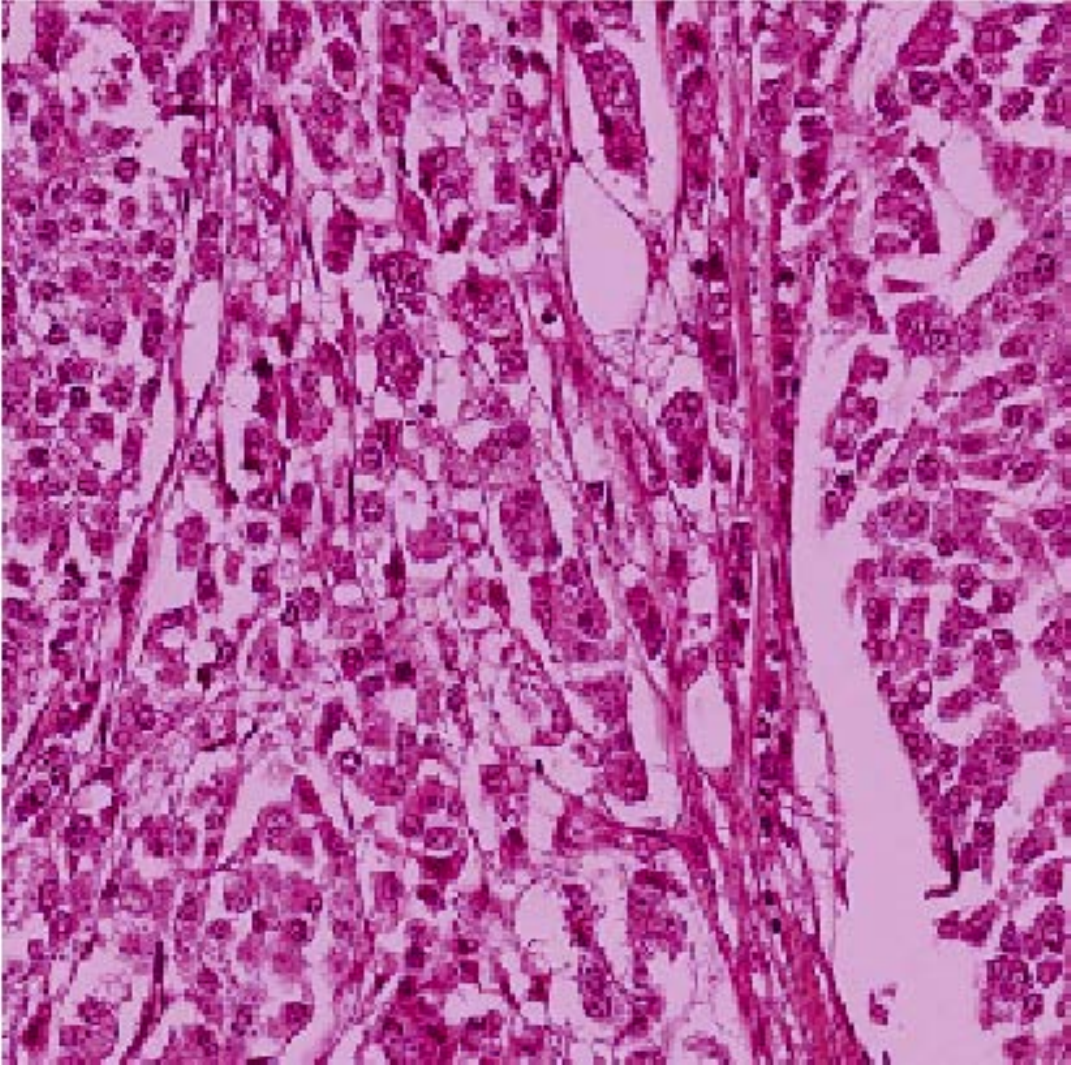}&
\includegraphics[width=0.11\textwidth]{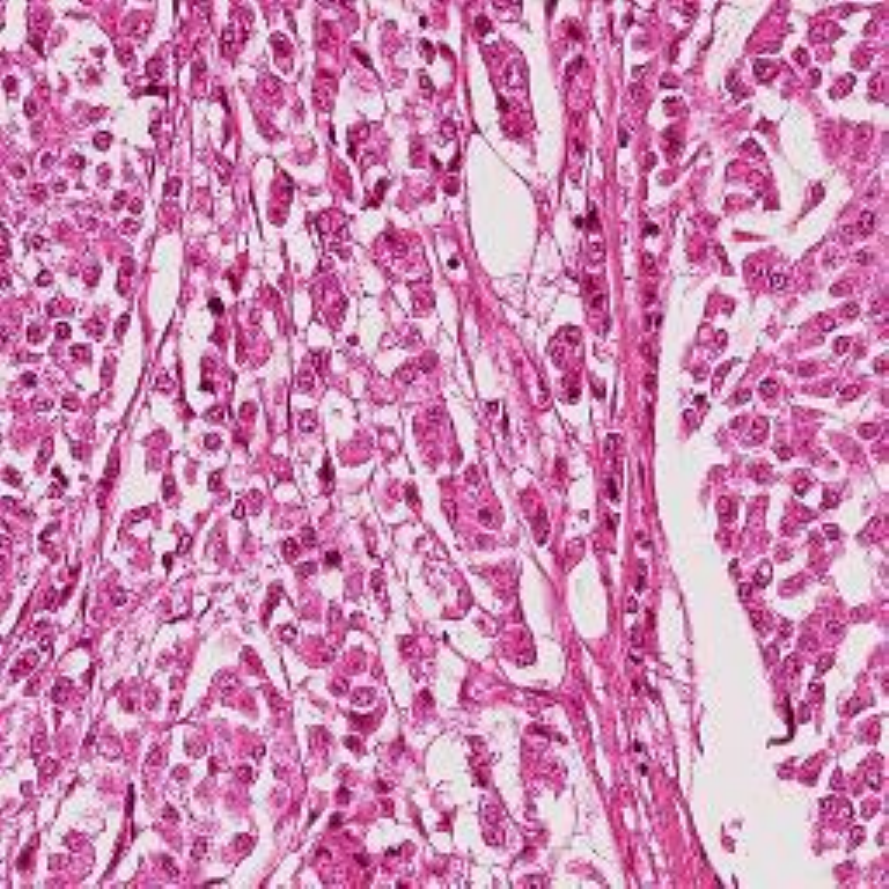}&
{\fboxsep=0mm
\fboxrule=1pt
\fcolorbox{red}{white}{\includegraphics[width=0.11\textwidth]{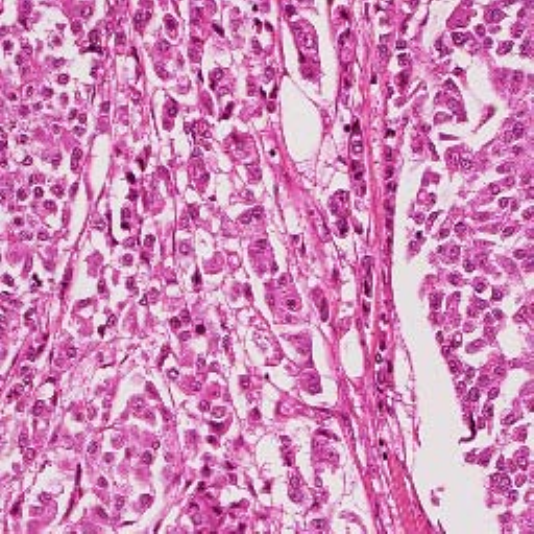}}}\\

\end{tabular}
\end{center}
\caption{Multimarginal Wasserstein Barycenter comparison with state of the art methods on MITOS-ATYPIA'14 challenge dataset. The blue bordered image is the input image and the two red bordered images are the references (with the one above being the intermediate reference). $m$ is the number of intermediate references (in traditional Wasserstein barycenter $m = 0$; there is no intermediate reference).}
\label{fig:visual_comparisons}
\end{figure}

\begin{table}[t!]
\begin{center}
\setlength{\tabcolsep}{8pt}
\caption{Stain Normalization Comparison (Mean $\pm$ Standard Deviation) on MITOS-ATYPIA'14 challenge dataset using Structural Similarity Index (SSIM) \cite{wang2004image}, Feature Similarity Index (FSIM) \cite{zhang2011fsim}. The time is the total time taken for normalizing all 500 images.}
\label{tab:table_visual_comparisons}
\begin{tabular}{l|cc|c}
\hline
Methods & SSIM & FSIM & Time (sec)\\
\hline
Reinhard \cite{reinhard2001color}  & 0.55$\pm$0.13 & 0.63$\pm$0.07 & \textbf{6.76} \\
Macenko \cite{macenko2009method}   & 0.51$\pm$0.08 & 0.62$\pm$0.09 & 59.40 \\
Khan \cite{khan2014nonlinear}      & 0.62$\pm$0.18 & 0.65$\pm$0.08 & 1994.87\\
Vahdane \cite{vahadane2016structure} & 0.63$\pm$0.11 & 0.65$\pm$0.06 & 502.04\\
StainGAN \cite{shaban2019staingan} & 0.68$\pm$0.23 & 0.69$\pm$0.06 & 69.12\\
$m=0$ (Ours)  & 0.59$\pm$0.32 & 0.67$\pm$0.08 & 254.06 \\
$m=1$ (Ours) & \textbf{0.73$\pm$0.06} & \textbf{0.75$\pm$0.11} & 384.28\\
\hline

\end{tabular}
\end{center}
\end{table}

\begin{figure}[ht!]
\begin{center}
\includegraphics[width=0.95\textwidth]{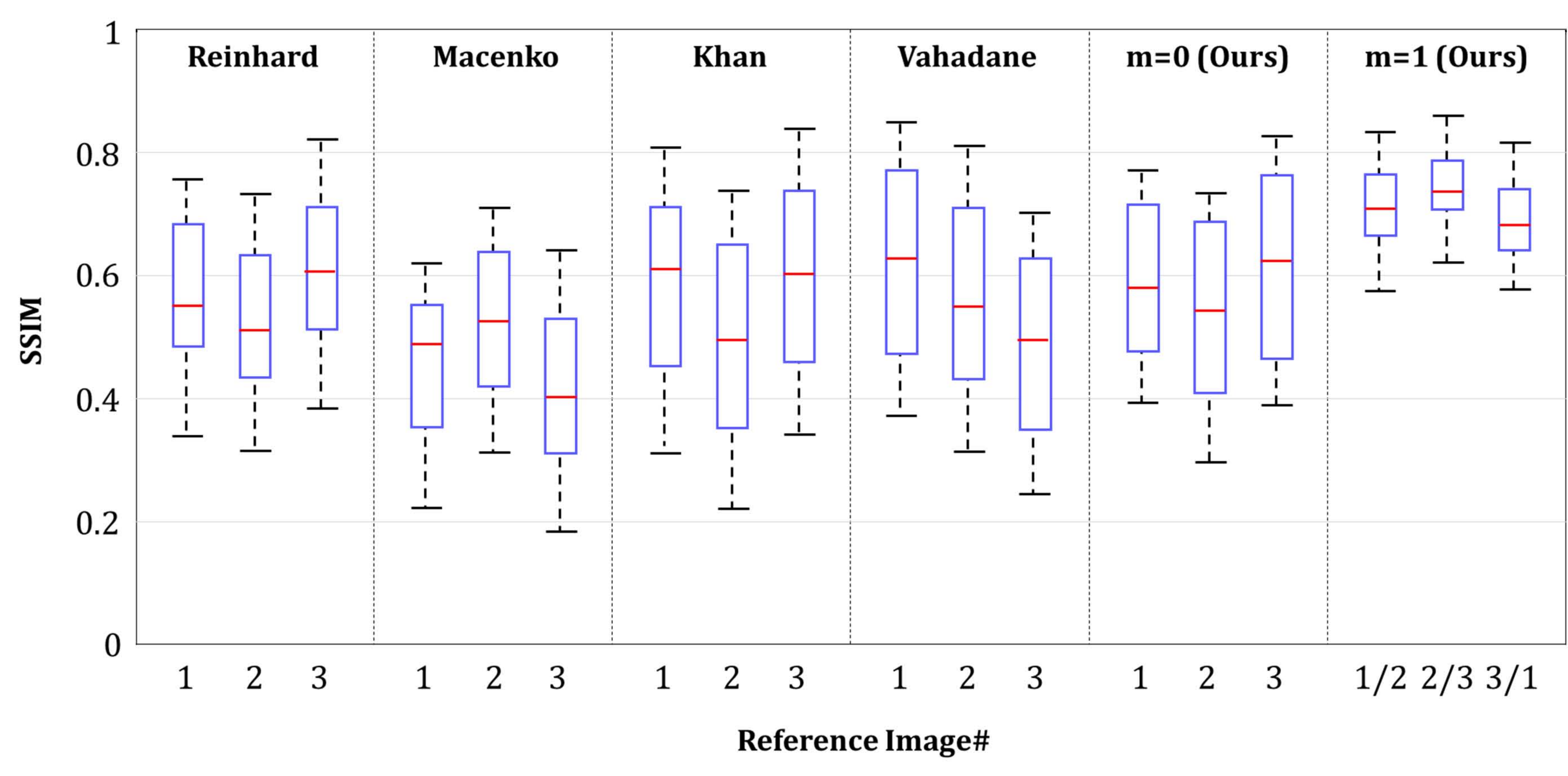}
\end{center}
\caption{Variation of SSIM due to different reference image selection. Three reference images were selected to test the various stain normalization methods. In our multimarginal case (2 references), we picked three combinations from the original three as highlighted under $m=1$.}
\label{fig:visual_comparisons1}
\end{figure}

\subsection{Stain Normalization Evaluation}

We used MITOS-ATYPIA'14 challenge dataset for evaluating our stain normalization. The dataset includes same tissue sections scanned by two different scanners (Aperio-A and Hamamatsu-H) with total 424 \textit{X}20 A-H frame pairs, 300 training and 124 testing. Images from scanner A are normalized and matched against the real corresponding images from H (ground truth). As in StainGAN \cite{shaban2019staingan}, 10,000 random (256$\times$256) patches from 300 training frames were used for training (26 epochs with the regularization parameter $\lambda=10$, learning rate 0.0002, Adam optimizer with a batch size of 4) and 500 patches from 124 testing data used for evaluation. The visual and quantitative comparisons are shown in Figure \ref{fig:visual_comparisons} and Table \ref{tab:table_visual_comparisons}, respectively. For the traditional case (one reference and source), our results are very similar to Reinhard \emph{et al.} \cite{reinhard2001color} since they also do color matching in Lab space, but our results improve drastically given two reference images. The references in our case span patches with different amounts of background visible. We also tested with different reference images and we show that we get a tighter bound as long as the references contain different amounts of background visibility; see Figure \ref{fig:visual_comparisons1} for the box plots of SSIM for different references.

\subsection{Nuclei Segmentation}

We also evaluate our stain augmentation/normalization in the nuclei segmentation settings. Specifically, we test our approach on MoNuSeg challenge dataset \cite{kumar2017dataset}. Previous works \cite{kumar2017dataset,pontalba2019assessing} have already shown that stain normalization improves performance for nuclei segmentation tasks. As mentioned in the Introduction, Tellez \emph{et al.} \cite{tellez2019quantifying} demonstrated that for several key specific tasks, stain augmentation improves performance and robustifies the resultant models. Very importantly, deep learning stain normalization/transfer approaches, e.g., StainGAN \cite{shaban2019staingan}, are not particularly suitable for nuclei segmentation task given the lack of training (source-target image pair) data; see \cite{pontalba2019assessing} for the details. The work of Vahadane \emph{et al.} \cite{vahadane2016structure} has been one of the main methodologies for stain normalization in nuclei segmentation.

Here we explore the effects of using different combinations of stain normalization and augmentation approaches for the same underlying architecture (CNN3), geometric augmentation and post-processing approaches \cite{kumar2017dataset}. We used the same architecture and hyperparameters as reported in Kumar \emph{et al.} \cite{kumar2017dataset}. After training and validation, the Aggregated Jaccard Index was computed on the same test set as in \cite{kumar2017dataset} for direct comparison. To drive the stain normalization and augmentation, in our approach we used 4 reference images, one from each organ present in the training dataset; images from 4 organs were used for training and testing and images from 3 additional organs were included just in the test set.

\begin{table}[t!]
\begin{center}
\setlength{\tabcolsep}{6pt}
\caption{Nuclei Segmentation comparisons using Aggregated Jaccard Index on same MoNuSeg test dataset as reported in \cite{kumar2017dataset} for direct comparison (with images from 4 organs --Breast, Liver, Kidney and Prostate-- also included in the training data and 3 others--Bladder, Colon and Stomach-- not included in training). For the same underlying architecture (CNN3 \cite{kumar2017dataset}) denoted by C, different combinations of the following are explored: Vahadane Stain Normalization (V), geometric augmentation (G) via rigid (rotation and flipping) and affine transformations, color jitter via random HSV/HED shifts (J), stain augmentation via direct perturbations of H\&E color channels (A) as introduced by Tellez et al. \cite{tellez2018whole} ($\alpha_i$ and $\beta_i$ were similarly taken from two uniform distributions), and finally our stain normalization (SN) and augmentation (SA). For SN, we just include the final interpolation and for SA we include all the interpolations.}
\label{tab:table_seg}
\begin{tabular}{|l|c|c|c|c|c|}
\hline
\textbf{Organ} & \textbf{Image} & C+V \cite{kumar2017dataset} & C+SN & C+G+J+A & C+G+SN+SA \\
\hline
\multirow{2}{*}{Breast}    & 1 & 0.4974 & 0.5211 & 0.4532 & 0.5325 \\
\cline{2-6}
                           & 2 & 0.5796 & 0.5726 & 0.4830 & 0.5815\\
\hline
\multirow{2}{*}{Liver}     & 1 & 0.5175 & 0.5462 & 0.6134 & 0.5598\\
\cline{2-6}
                           & 2 & 0.5148 & 0.5829 & 0.5918 & 0.6013\\
\hline
\multirow{2}{*}{Kidney}    & 1 & 0.4792 & 0.4812 & 0.5815 & 0.5648\\
\cline{2-6}
                           & 2 & 0.6672 & 0.7187 & 0.6924 & 0.7414\\
\hline
\multirow{2}{*}{Prostate}  & 1 & 0.4914 & 0.5305 & 0.5491 & 0.6270\\
\cline{2-6}
                           & 2 & 0.3761 & 0.4017 & 0.3191 & 0.5296\\
\hline
\multirow{2}{*}{Bladder}   & 1 & 0.5465 & 0.5634 & 0.5510 & 0.6475\\
\cline{2-6}
                           & 2 & 0.4968 & 0.5016 & 0.4489 & 0.5267\\
\hline
\multirow{2}{*}{Colon}     & 1 & 0.4891 & 0.5108 & 0.4904 & 0.5318\\
\cline{2-6}
                           & 2 & 0.5692 & 0.6179 & 0.5879 & 0.6263\\
\hline
\multirow{2}{*}{Stomach}   & 1 & 0.4538 & 0.5318 & 0.4823 & 0.6408\\
\cline{2-6}
                           & 2 & 0.4378 & 0.4520 & 0.3912 & 0.6551\\

\hline
\hline
\multicolumn{2}{|l|}{Overall} & 0.5083 & 0.5381 & 0.5168 & 0.5976\\
\hline
\end{tabular}
\end{center}
\end{table}

\section{Conclusions and Future Work}

In the paper, we presented a new multimarginal Wasserstein barycenter method for H\&E stain normalization and augmentation given one or more references. The method achieved superior performance in stain normalization and nuclei segmentation tasks because of the use of the intermediate references in the multimarginal setting. This allows one to incorporate additional distributions that can give physically more realistic interpolations (hence augmentation) as well as normalization. Since the normalization is done in the color distribution space and is not dependent on the number of pixels, the method can easily be scaled to whole slide images. In the future, we will also explore incorporating our Wasserstein barycenter formulation as a deep learning loss function.

\section*{Acknowledgements}
This study was supported by AFOSR grants (FA9550-17-1-0435, FA9550-20-1-0029), NIH grant (R01-AG048769), MSK Cancer Center Support Grant/Core Grant (P30 CA008748), and a grant from Breast Cancer Research Foundation (grant BCRF-17-193).


\end{document}